\documentclass[11pt,a4paper]{article}
\usepackage{jheppub}
\usepackage{etoolbox}
\usepackage{multirow}

\makeatletter
\def\@fpheader{\relax}
\makeatother

\usepackage[utf8]{inputenc}
\usepackage{slashed}
\usepackage[czech,english]{babel}
\usepackage{graphicx}
\usepackage{amsmath,amsfonts,amssymb}
\DeclareMathOperator{\MyProd}{\scalebox{1.4}{$\mathrm{I\kern-0.2ex I}$}}

\title{Form Invariance, Topological Fluctuations and Mass Gap of Yang-Mills Theory}

\author[a]{Yachao Qian}

\emailAdd{yachao.qian@stonybrook.edu}

\author[b,\hspace{0.6mm} c]{Jun Nian}

\emailAdd{nian@ihes.fr}

\affiliation[a]{Department of Physics and Astronomy\\
	Stony Brook University\\
	Stony Brook, NY 11794-3800, U.S.A.\\}

\affiliation[b]{Institut des Hautes \'Edutes Scientifiques\\
	Le Bois-Marie, 35 route de Chartres\\
         91440 Bures-sur-Yvette, France\\}

\affiliation[c]{C.N. Yang Institute for Theoretical Physics\\
	Stony Brook University\\
         Stony Brook, NY 11794-3840, U.S.A.\\}

\abstract{In order to have a new perspective on the long-standing problem of the mass gap in Yang-Mills theory, we study the quantum Yang-Mills theory in the presence of topologically nontrivial backgrounds in this paper. The topologically stable gauge fields are constrained by the form invariance condition and the topological properties. Obeying these
constraints, the known classical solutions to the Yang-Mills equation in the 3- and 4-dimensional Euclidean spaces are recovered, and the other allowed configurations form the nontrivial topological fluctuations at quantum level. Together, they constitute the background configurations, upon which the quantum Yang-Mills theory can be constructed. We demonstrate that the theory mimics the Higgs mechanism in a certain limit and develops a mass gap at semi-classical level on a flat space with finite size or on a sphere.}


\arxivnumber{}

\newcommand{\bea}{\begin{eqnarray}}
\newcommand{\eea}{\end{eqnarray}}
\newcommand{\la}{\label}
\newcommand{\be}{\begin{equation}}
\newcommand{\ee}{\end{equation}}

\begin{document}
\maketitle

\section{Introduction}

Yang-Mills theory \cite{YM} has a central position in modern theoretical physics. The successful standard model of particle physics \cite{Glashow, Weinberg, Salam} was formulated in the language of Yang-Mills theory. Historically, each step in understanding the structure and the dynamics of Yang-Mills theory has helped us formulate quantum field theory, which in turn widens and deepens our knowledge of nature. An incomplete list includes the quantization of Yang-Mills theory \cite{Faddeev}, the renormalization of Yang-Mills theory \cite{VeltmantHooft} and the discovery of various nontrivial classical solutions (e.g. \cite{instanton, WuYang, multi-instanton}, for a review see Ref.~\cite{Actor}). More recently, significant progress has been made in understanding the supersymmetric Yang-Mills theory \cite{SW}. There is still a big remaining problem, that is to explain the mass gap of the pure Yang-Mills theory without coupling it to other matter fields \cite{MassGap}. Any progress in resolving this problem will undoubtedly help us study the structure of quantum field theory, which can be generalized and applied to many other branches of physics.

In this paper, we show that the mass gap problem may be tightly related to the nontrivial topological backgrounds, which could be analyzed by introducing two new ingredients, the form invariance condition and the topological fluctuations.  A crucial difference between our approach and some previous works is that the background configurations we study in this paper, though constrained by the form invariance condition and the topological properties, do not necessarily satisfy the field equation. We will see that at quantum level even the pure Yang-Mills theory has topological fluctuations, which in a certain limit have a similar expression of the well-known Higgs mechanism \cite{Anderson, Englert, Higgs, Hagen}.

We study the $SU(2)$ gauge theory in 3- and 4-dimensional Euclidean spaces. We start with a physical Ansatz for the spherically symmetric gauge field, which is assumed to be topologically stable. The main idea is that such Ansatz should be form invariant under Lorentz transformations. By form invariance we mean the following relation. Under a Lorentz transformation, an ordinary vector field $V_\mu$ should satisfy
\be
  (O^{-1})_\mu\,^\nu V_\nu (O x) = V_\mu (x)\, ,
\ee
where $O_\mu\,^\nu$ denotes the Lorentz transformation. A gauge field $A_\mu$ differs from an ordinary vector field in the following way:
\be\label{eq:FIlocal}
  (O^{-1})_\mu\,^\nu \, A_\nu (O\, x) = V^{-1} \, A_\mu (x) \, V + V^{-1} \partial_\mu V\, ,
\ee
where $V$ generates a gauge transformation, i.e., under the Lorentz transformations given on the left-hand side, the gauge field $A_\mu$ is invariant up to a gauge transformation. In this paper, we will focus on the Lorentz transformations with constant parameters, which implies that the gauge transformations on the right-hand side of the equation above should also only have rigid parameters, as we will prove in Appendix~\ref{guchaohao}. Hence, Eq.~\eqref{eq:FIlocal} simplifies to
\be\label{eq:FI}
  \left(O^{-1} \right)_\mu\,^\nu\, A_\nu (O\, x) = V^{-1}\, A_\mu (x)\, V\, ,
\ee
where $O_\mu\,^\nu$ denotes a Lorentz transformation with constant parameters, and $V$ stands for a gauge transformation with rigid parameters. The equation above means that for any constant Lorentz transformation $O_\mu\,^\nu$ there must be at least one constant gauge transformation $V$, such that Eq.~\eqref{eq:FI} is satisfied globally. We refer to Eq.~\eqref{eq:FI} as the form invariance condition throughout the paper. All the topologically stable backgrounds, either solutions to the field equation or not, should satisfy this condition.

In order that the condition above is satisfied, we propose the Ans\"atze of topologically stable $SU(2)$ gauge fields in the 3D and 4D Euclidean spaces in Subsection~\ref{3dansatz} and Subsection~\ref{4dansatz} respectively. They look originally very similar:
\be
  A_\mu = p  \left(U^{-1} \partial_\mu U\right)\, ,\quad U = \textrm{exp} \left[T_a\, \hat{n}^i \, \omega^a\,_i \, \theta  \right]\, ,
\ee
where
\be
  T^a = \frac{\sigma^a}{2i}\, ,\quad \hat{n}^i \equiv \frac{x^i}{|x|}\, , \quad |x| \equiv \sqrt{ {\displaystyle \sum\nolimits_{i=1}^3}  \,  x^i x_i} \, ,
\ee
and $\omega^a\,_i$ is a tensor that connects the gauge indices and the spacetime indices $1, 2, 3$. We prove that, in order for the Ansatz above to be form invariant,   $\omega$ has to be a constant $O(3)$ group element, and for the 3D Yang-Mills theory $p= p(\tau)$ and $\theta = \theta(\tau)$ with $\tau \equiv x^\mu x_\mu = |x|^2$ in 3D. However,  there are additional subtleties due to the form invariance condition in 4D, as we will discuss in Appendix~\ref{sq}, which will impose further constraints on the factors $p$ and $\theta$ in 4D.

In Subsection~\ref{3dtopology} and Subsection~\ref{4dtopology}, we will see that after imposing the form invariance condition, the topological properties will further constrain the Ansatz: In order to have a well-defined topological charge, the  topologically stable Ansatz has to take some fixed values at the boundaries. The constraint can be imposed by introducing a topological term to the original Yang-Mills Lagrangian without modifying the theory at quantum level. This term will have huge impact on the theory.

Based on the constraints from the form invariance condition and the topological properties, we can solve the Yang-Mills equation exactly for the spherically symmetric case, because the classical solutions must be topologically stable and thus satisfy the form invariance condition and the topological properties. This is done in Section~\ref{Solution}. For the 3D case, we recover the Wu-Yang monopole, the pure gauge solution and the trivial vacuum solution, while for the 4D case we find the meron solution, the instanton solution, the anti-instanton solution, the pure gauge solution and the trivial vacuum solution.

Next, we move on to study the quantum Yang-Mills theory, i.e. the path integral of Yang-Mills theory. In principle, we can expand the theory around all the backgrounds satisfying the form invariance condition and the topological properties. Since topological properties only fix the boundaries of the backgrounds, the topological fluctuations with fixed boundaries naturally arise. As a saddle point approximation, we only include the topological fluctuations around classical solutions. The classical solutions together with the topological fluctuations form the background configurations, around which we further turn on quantum fluctuations, that are not constrained by the form invariance condition nor have the topological properties. The exact formalism is discussed in Section~\ref{TF}, in particular, the relations and the differences between topological fluctuations and quantum fluctuations will be discussed in Subsection~\ref{remarks}. Unlike the previous works on background formalism, the background that we consider not only includes classical solutions  but also have topological fluctuations. Correspondingly, we have to study the pseudo zero modes instead of the zero modes.

To simplify the discussion and make the relevant physics more transparent, we adopt an approximation invented by R.~Feynman \cite{Feynman} in Section~\ref{approx} and consider the effective theory after the average over the gauge orientations. As we will see, in a certain limit the effective theory approaches the configuration of quantum fluctuations in the background only consisting of the classical solutions, while in another limit it mimics the Higgs mechanism, where the topological fluctuations play a similar role of the Higgs field. Though the quantum field acquires mass in both cases, the masses acquired through different formalisms are quite different, and we will provide a qualitative analysis.

  Finally, we apply the ideas discussed in the previous sections to the long-standing problem of the Yang-Mills mass gap. As we will see, the quantum phenomena in the IR regime is dominated by the topological fluctuations. By turning off the quantum fluctuations while keeping the topological fluctuations, we analyze the low-energy physics of the quantum Yang-Mills theory, and following the approach by A.~Polyakov \cite{Polyakov} we compute the two-point correlation function of two gauge invariant operators at semi-classical level in Section~\ref{LE}. The exponential decay of the two-point correlation function for a large distance provides a strong evidence for the existence of the mass gap. More rigorous mathematical arguments are presented in Section~\ref{MG}, and the Yang-Mills mass gap problem is simplified to the mass gap problem of a special kind of nonlinear Schr\"odinger equation, which is well-studied. Based on the results from the mathematical literature, we find the mass gap at semi-classical level for quantum $SU(2)$ Yang-Mills theory defined on a flat space with finite size or on a sphere. Some possible directions for the future research are discussed in Section~\ref{Discussion}.

As consistency checks, we observe that our results are supported by many previous works in the literature both from the formal side and from the phenomenological side (including simulations and numerical results). Let us list some of them as follows:
\begin{itemize}
  \item The form invariance condition in this paper can be thought of as one of the axioms by A.~Wightman for a general quantum field theory \cite{Wightman}. For a vector field $j_\mu$ this axiom can be written as
\be
  U(a, A)\, j_\mu (x)\, U(a, A)^{-1} = \Lambda_\mu\,^\nu (A^{-1})\, j_\nu (A x + a)\, ,
\ee
where $\Lambda_\mu\,^\nu$ is a representation of the Lorentz group, and $U(a, L)$ is a unitary or anti-unitary operator on the Hilbert space.

  \item According to Ref.~\cite{Simons}, any weakly stable Yang-Mills field with gauge group $SU(2)$ or $SU(3)$ on the 4-sphere must be self-dual or anti-self-dual, i.e., instantons or anti-instantons. As we will see in Section~\ref{Solution}, these solutions will be automatically singled out from the topologically stable configurations of the gauge field.

  \item The results from the lattice calculations \cite{lattice-1, lattice-2, lattice-3, lattice-4} confirmed that the ghost field decouples from the gauge field and becomes free in the deep infrared, which are consistent with our expectation that there should be some scalar degrees of freedom appearing in the IR regime.

  \item Recent studies on the spectrum of effective strings \cite{Flauger-1, Flauger-2} have found strong evidences for the existence of a massive pseudoscalar on the worldsheet of QCD flux tube, which also supports the emergence of scalar degrees of freedom in the theory.

  \item Ref.~\cite{FeynmanPrinciple} observed that by varying the profile function of instantons one can find a mass gap for glueballs at the classical level. Our results give an explanation to this observation in the sense that varying the profile function of instantons is equivalent to turning on the 4D topological fluctuations allowed by the form invariance condition and the topological properties.
\end{itemize}

\begin{figure}[!htb]
      \begin{center}
        \includegraphics[width=0.8\textwidth]{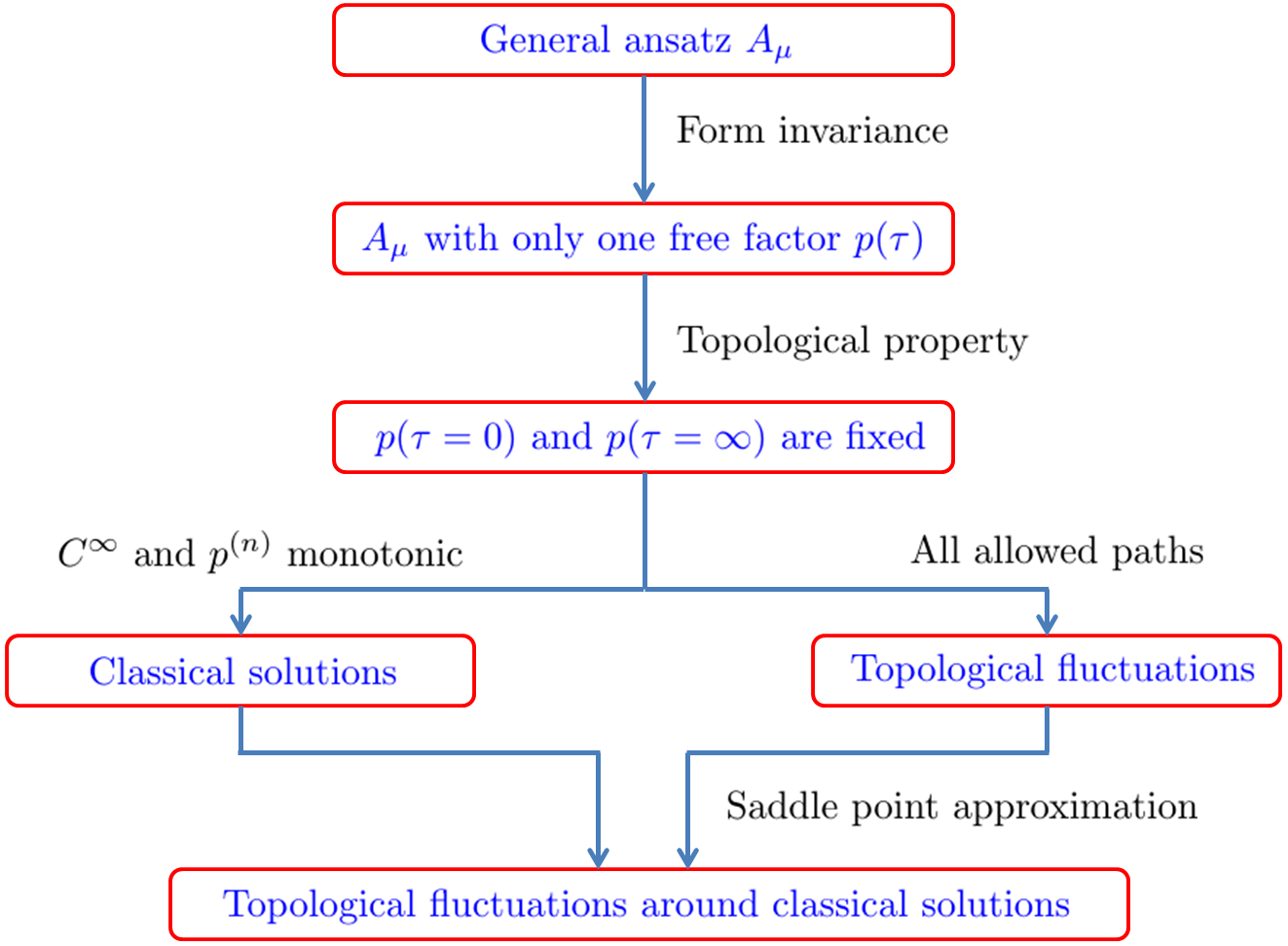}
      \caption{The 4D case as an exmaple.}
      \label{fig:logic}
      \end{center}
    \end{figure}

In the literature, sometimes the terminology ``topological fluctuations'' is used to denote  changes in topological charge, which is different from what we use in this paper. The topological fluctuations considered in this paper preserve the topological charge, because they are restricted by the form invariance condition \eqref{eq:FI} and the topological boundary conditions. To clarify this difference and summarize our approach, we would like to illustrate the logic of the 4D case as an example in Fig.~\ref{fig:logic}.

We add a few appendices to the main text, which contain many details of the derivations. In Appendix~\ref{app:notation} we summarize the notations adopted in this paper. In Appendix~\ref{guchaohao}, a theorem, which is important for our discussion on the form invariance condition, is proved. In Appendix~\ref{omega} we classify the tensor $\omega$, which appears in the Ans\"atze of topologically stable gauge fields. To discuss the topological properties, we have to introduce the topological terms. The one for the 3D case is the Chern-Simons term, and we discuss it in detail in Appendix~\ref{CS}. Unlike the 3D case, the 4D form invariance condition will impose extra constraints on the Ansatz, and we present the details in Appendix~\ref{sq}. The 4D topological term is discussed in Appendix~\ref{FF}. After deriving Ans\"atze for the topologically stable gauge fields, which satisfy both the form invariance condition and the topological properties, we can find the classical solutions to the Yang-Mills equation by applying the Ans\"atze to the equation of motion. The details for the 3D and the 4D case are presented in Appendices~\ref{3dcs} and \ref{4dcs} respectively. After recovering all the known classical solutions, we can move on to the discussions of the quantum Yang-Mills theory. As discussed in the main text, we introduce the concept ``pseudo zero modes''. The corresponding path integral measure for the pseudo zero modes is discussed in Appendix~\ref{topologicalmeasure}. To illustrate the relations of topological fluctuations and quantum fluctuations, we borrow a famous example of the well-understood 1D quantum antiferromagnets from the literature of condensed matter physics, which is briefly summarized in Appendix~\ref{reSpinChain}, while in Appendix~\ref{FeynmanPathInt} we review the Feynman's path integral formulation of the 1D non-relativistic quantum mechanics, in order to show the different roles of topological fluctuations and quantum fluctuations in the configuration space, over which the path integral is defined. Since we have simplified the mass gap problem of the Yang-Mills theory to the mass gap problem of a certain kind of nonlinear Schr\"odinger equation in Section~\ref{MG}, we summarize the relevant results from the mathematical literature and discuss how to map them to our problem in Appendix~\ref{GP}. Although the equations that we are interested are defined in the 3D or the 4D Euclidean space, it turns out that this kind of nonlinear Schr\"odinger equation in 1D has solutions with explicit expressions. In Appendix~\ref{Toy} we study the 1D case as a toy model, which should capture some qualitative features of the 3D and the 4D cases.

We would like to make a remark about the notation used in this paper. Since we focus on the 3D and 4D Euclidean spaces, we do not distinguish the upper and the lower spacetime indices.

\section{\label{FT}Form Invariance and Topological Properties}

\subsection{\label{3df}3D Case}

\subsubsection{\label{3dansatz}Ansatz and Form Invariance}

The Ansatz of a spherically symmetric $SU(2)$ gauge field in the 3-dimensional Euclidean space is given by 
\be\label{eq:3DAnsatz1}
  A_\mu = p(\tau) \left(U^{-1} \partial_\mu U\right)\, , 
\ee
where $\tau \equiv x_\mu x^\mu$, and $U$ is an $SU(2)$ group element.
In general, $U$ can be expressed as
\be\label{eq:3DAnsatz1a}
  U = \textrm{exp} \left[T_a \psi^a (x) \right] = \textrm{exp} \left[T_a \frac{\psi^a (x)}{|\psi (x)|} |\psi (x)| \right] = \textrm{exp} \left[T_a\, \omega^a\,_\mu\, \hat{n}^\mu |\psi (x)| \right] = \textrm{exp} \left[T_a\, \omega^a\,_\mu\, \hat{n}^\mu \theta(\tau) \right]\, ,
\ee
with
\be
  \hat{n}^\mu \equiv \frac{x^\mu}{|x|}\, ,\quad \theta(\tau) \equiv |\psi (x)|\,  ,\quad  T^a = \frac{\sigma^a}{2i}\, .
\ee
We have defined a matrix $\omega^a\,_\mu$ to connect the two unit vectors $\hat{n}^\mu$ and $ \psi^a (x) / |\psi (x)| $ in different spaces. For the 3D case, the indices $\mu$ and $a$ both run over $1, \cdots, 3$.

The Ansatz \eqref{eq:3DAnsatz1} must satisfy the form invariance condition \eqref{eq:FI}:
\begin{displaymath}
  \left(O^{-1} \right)_\mu\,^\nu\, A_\nu (O\, x) = V^{-1}\, A_\mu (x)\, V\, ,
\end{displaymath}
where $O$ is a constant $SO(3)$ group element, and $V$ is a constant $SU(2)$ group element. We will show in Appendix~\ref{omegarestriction} that Eq.~\eqref{eq:FI} restricts $\omega$ to be a constant $O(3)$ group element. In this paper we assume that $\textrm{det}\, \omega = 1$, so the Ansatz \eqref{eq:3DAnsatz1} becomes
\be\label{eq:3DAnsatz}
  A_\mu = p(\tau) \left(U^{-1} \partial_\mu U\right)\, , \quad  U = \textrm{exp} \left[T_a\, \omega^a\,_\mu\, \hat{n}^\mu \theta(\tau) \right] \, ,
\ee
with a constant $SO(3)$ group element $\omega$. With a proper choice of the generators $T_a$, we can write the matrix $\omega$ as   
\bea
\omega = 
\begin{pmatrix}
\,1\, & 0 & 0 \\
0 & \,1\, & 0 \\
0 & 0 & \,1\, \\
\end{pmatrix} \, .
\eea

We would like to emphasize that for the 3D Yang-Mills theory the form invariance condition \eqref{eq:FI} is automatically satisfied by the Ansatz \eqref{eq:3DAnsatz}, hence it does not impose any constraints on the functions $p(\tau)$ and $\theta(\tau)$. For simplicity, in this paper we only consider $p(\tau)$ and $\theta(\tau)$ that are $C^1$-differentiable functions.

\subsubsection{\label{3dtopology}Topology}

Although we consider the pure Yang-Mills theory, the topological properties will become manifest after introducing a topological term.

For the 3-dimensional Euclidean space, the appropriate topological term is the Chern-Simons term:
\be
   S_{CS} = \frac{ik}{4\pi} \int d^3 x\, \textrm{Tr} \left(A \wedge dA + \frac{2}{3} A\wedge A\wedge A \right)\, .
\ee
In general, $S_{CS}$ takes values in $\mathbb{R} / 2 \pi \mathbb{Z}$. If we require that $S_{CS}$ takes values in $2 \pi \mathbb{Z}$, it will not affect the quantum Yang-Mills theory in the path integral.

Plugging the Ansatz \eqref{eq:3DAnsatz} into the Chern-Simons term, we obtain
\be
  A \wedge dA + \frac{2}{3} A\wedge A\wedge A = \left(\frac{2}{3} p^3 - p^2\right) (U^{-1} d U) \wedge (U^{-1} d U) \wedge (U^{-1} d U) - p d p\, d(U^{-1} dU)\, .
\ee
Since the second term in the above expression is proportional to $d(U^{-1} dU)$, which vanishes after taking the trace, the Chern-Simons action now becomes
\be\label{eq:3dcmintegral}
  S_{CS} = \frac{ik}{4\pi} \int d^3 x\, \left(\frac{2}{3} p^3 - p^2 \right) \textrm{Tr} (U^{-1} dU) \wedge (U^{-1} d U) \wedge (U^{-1} d U)\, ,
\ee
which is essentially a Wess-Zumino term. We can define
\be
  S_{CS} = 2\pi i k B\, ,
\ee
where $B$ is the winding number.

\begin{table}[htb!]
\centering
\begin{tabular}{c*{5}{c} }  
 \,  \quad   Winding number $B$  \quad\quad     & \quad    $p|_{\tau=0}$ \quad  \, & \quad    $p|_{\tau=\infty}$ \quad   \,  & \quad   $\theta|_{\tau=0}$ \quad   \,   & \quad   $\theta|_{\tau=\infty}$ \quad  \,    \\
\hline \hline
0    &    0    &   0  &  $\pi$ & $\pi$\\ 
0    &    1/2 & 1/2 & $\pi$ & $\pi$\\  
0    &    1   &  1   & $\pi$ & $\pi$\\  
1    &    0   & 1/2 & $\pi$& $\pi$ \\  
-1   &   1/2 &  0   & $\pi$ & $\pi$\\
$\cdots$    &  $\cdots$ &  $\cdots$ &  $\cdots$ & $ \cdots$ \\
\end{tabular}
\caption{Boundary conditions in 3D.}
\label{table:3dBC}
\end{table}

As shown in Appendix~\ref{CS}, in order that $B$ is an integer, the values of $p$ and $\theta$ at the boundaries $\tau=0$ and $\tau = \infty$ are constrained. We can list the possible boundary conditions in Table~\ref{table:3dBC}.

We would like to emphasize that these boundary values are obtained from the topological constraint, and they do not necessarily lead to solutions to the Yang-Mills equation, but the solutions to the Yang-Mills equation must satisfy these boundary conditions. Eq.~\eqref{eq:3DAnsatz} with the constrained boundary values provides an Ansatz of the topologically stable $SU(2)$ gauge field in the 3-dimensional Euclidean space.

\subsection{\label{4df}4D Case}

The discussions in this section are similar to Section~\ref{3df} for the 3D case. The notations used here are summarized in Appendix~\ref{app:notation}.

\subsubsection{\label{4dansatz}Ansatz and Form Invariance}

Now let us consider the Ansatz for the topologically stable $SU(2)$ gauge field in the 4-dimensional Euclidean space. Similar to Eq.~\eqref{eq:3DAnsatz1}, we can write down a general Ansatz:
\be\label{eq:4DAnsatz}
  A_\mu = p(\tau, x_4) \left(U^{-1} \partial_\mu U\right)\, ,\quad U = \textrm{exp} \left[T_a\, \hat{n}^i \, \omega^a\,_i\, \theta(\tau, x_4) \right]\, ,
\ee  
where  $\tau \equiv x^\mu x_\mu$, and in this case $\mu$ runs from $1$ to $4$, while $i$ runs from $1$ to $3$. The functions $p(\tau, x_4)$ and $\theta(\tau, x_4)$ depend on both $\tau$ and $x_4$, while $\hat{n}^i$ is a unit vector depending only on $x_1$, $x_2$ and $x_3$:
\be
  \hat{n}^i \equiv \frac{x^i}{|x|}\, ,
\ee
where $|x|^2 \equiv \sum_{i =1}^3 x^i x_i$. As we will see in the following, in order that the form invariance condition \eqref{eq:FI}:
\begin{displaymath}
  (\Lambda^{-1})_\mu\,^\nu\, A_\nu (\Lambda\, x) = V^{-1}\, A_\mu (x)\, V
\end{displaymath}
still holds for the 4D case, the expressions of the factors $p(\tau, x_4)$ ,  $\theta(\tau, x_4)$ and $\omega^a\,_i$ have to be fixed, where $\Lambda$ is an $SO(4)$ Lorentz transformation and $V$ is an $SU(2)$ gauge transformation, both of which have parameters independent of $x$.

The Lorentz group of the 4-dimensional Euclidean space is $SO(4)$, which has 6 generators $M_{\alpha\beta}$ $(\alpha,\beta = 1, \cdots, 4)$ with $M_{\alpha\beta} = - M_{\beta\alpha}$. For a fixed value of $x_4$, the rotations in the subspace $(x_1, x_2, x_3)$ are generated by $M_{12}$, $M_{23}$, $M_{31}$, and the form invariance condition restricts $\omega$ to be a constant $O(3)$ group element as for the 3-dimensional case. Again, we assume that $\textrm{det} \, \omega = 1$ in this paper, hence we choose $\omega$ to be a constant $SO(3)$ group element. Moreover, we need to impose the form invariance condition on the Ansatz under the rotations generated by $M_{14}$, $M_{24}$, $M_{34}$ to constrain the functions $p(\tau, x_4)$ and $\theta(\tau, x_4)$. As we will show in Appendix~\ref{thetafix}, this condition constrains $p(\tau, x_4) = p (\tau)$ and fixes the function $\theta(\tau, x_4)$ to be
\be\label{eq:4Dq}
  \textrm{cos} \frac{\theta}{2} = \pm \frac{x_4}{\sqrt{\tau}}\, .
\ee
We choose $\textrm{cos}  (\theta/2 ) = x_4/\sqrt{\tau} $ in this paper. Consequently,  the Ansatz becomes
\be\la{eq:amuansatz4D}
  A_\mu^a\, T^a \,  =  \, 2 \frac{p(\tau)}{\tau}\eta_{a\mu\nu} x^\nu T^a\, ,
\ee
where $\eta_{a\mu\nu}$ are the 't Hooft symbols (see Appendix~\ref{app:notation}). In Appendix~\ref{completecheck}, we prove that this expression is indeed form invariant, while in Appendix~\ref{alternativeapproach} an alternative approach to obtain the form invariant expression \eqref{eq:amuansatz4D} will be discussed, which can be generalized to higher dimensions or curved spacetime.

 \subsubsection{Topology}\label{4dtopology}
Similar to the 3D case, we would like to introduce a topological term, which does not affect the Yang-Mills action at the quantum level. In the 4-dimensional Euclidean space, this term can be
\be
  S = 2 \pi i k
\ee
with the winding number $k$ given by
\begin{align}
  k & = - \frac{1}{16 \pi^2} \int d^4 x\, \textrm{Tr} \left[F^{\mu\nu} (* F_{\mu\nu}) \right] \nonumber\\
  {} & = - \frac{1}{16 \pi^2} \int d^4 x\, 2\, \partial_\mu \epsilon^{\mu\nu\rho\sigma} \, \textrm{Tr} \left[A_\nu \partial_\rho A_\sigma + \frac{2}{3} A_\nu A_\rho A_\rho \right]\, ,
\end{align}
where the second line is an integral over the boundary. For the 4D case, the winding number can be thought of as the second Chern number, which is  topologically invariant. In this paper we focus on the gauge group $SU(2)$, for which only the second Chern class is nonvanishing and all the higher Chern classes vanish.

Plugging the Ansatz \eqref{eq:4DAnsatz} into the topological term above, we obtain
\be
  k = -\frac{1}{8 \pi^2} \oint d\Omega_\mu \, \epsilon^{\mu\nu\rho\sigma} \left(\frac{2}{3} p^3 - p^2 \right)\, \textrm{Tr} \left[ \left( U^{-1} \partial_\nu U \right) \left( U^{-1} \partial_\rho U \right) \left( U^{-1} \partial_\sigma U \right) \right]\, ,
\ee
which is a surface integral evaluated on the boundary of the original 4-dimensional manifold. In the simplest case, there are two boundaries around $\tau=0$ and $\tau = \infty$.

\begin{table}[htb!]
\centering
\begin{tabular}{c*{3}{c} } 
 \, \quad Winding Number $k$   \quad\quad        & \quad  $p|_{\tau=0}$ \quad  \,   & \quad $p|_{\tau=\infty}$ \quad   \,   \\
\hline\hline
0   &   0   &   0 \\
0   &   1/2 & 1/2 \\
0    &    1  &   1\\
1    &   0  &   1  \\
-1    &   1  &  0   \\
$\cdots$    &  $\cdots$ &  $\cdots$  \\
\end{tabular}
\caption{Boundary conditions in 4D.}
\label{table:4dBC}
\end{table}

We list some possible values of $p$ at the boundaries in Table~\ref{table:4dBC}, and more details can be found in Appendix~\ref{FF}. Again, as in the 3D case, the solutions to the Yang-Mills equation have to satisfy these boundary conditions, while the possible boundary values do not always lead to solutions. Eq.~\eqref{eq:amuansatz4D} with the fixed boundary values provides an Ansatz of the topologically stable $SU(2)$ gauge field in 4-dimensional Euclidean space. Moreover, the factor $\theta$ in the 4-dimensional Ansatz is already fixed by the form invariance condition.

\subsection{$C^\infty$-Curve} \label{smoothest}

In the previous subsections, we obtain the Ans\"atze of the topologically stable gauge field for both the 3D and the 4D Yang-Mills equation. We have seen that the boundary values of the factors in the Ans\"atze are fixed by the topological properties. In principle, there can be infinitely many smooth functions that satisfy these boundary conditions, and they provide the candidates for the solutions to the Yang-Mills equation, because the classical solutions must be form invariant and satisfy the boundary values, i.e., they must be topologically stable.

\begin{figure}[!htb]
      \begin{center}
        \includegraphics[width=0.64\textwidth]{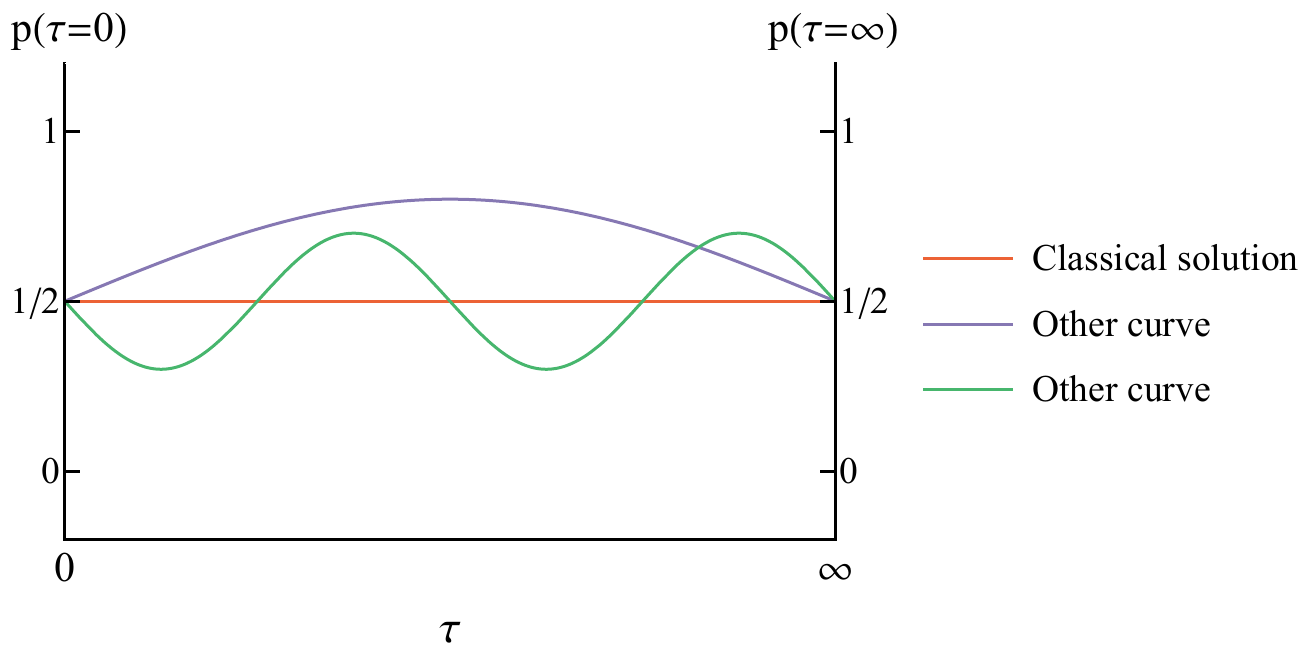}
      \caption{Curves satisfying the boundary conditions for the 4D meron solution.}
      \label{fig:3dmany}
      \end{center}
    \end{figure}

We have made the following observation. Among the possible candidates the true solution is always a monotonic $C^\infty$-curve. For example, for the 4D case there are infinitely many curves that satisfies the boundary conditions for the meron solution (see Fig.~\ref{fig:3dmany}):
\be
  p(\tau = 0) = p(\tau = \infty) = \frac{1}{2}\, .
\ee
The true 4D meron solution is given by $p(\tau) = 1/2$, which is a monotonic $C^\infty$-curve, and its higher derivatives
\be
p^{(n)} \equiv  \partial^n  p(\tau) / \partial \tau^n  = 0
\ee
are also monotonic.

Besides the classical solutions, the other topologically stable configurations are also very important in our following discussions. These configurations are called topological fluctuations, and we will discuss them in great detail in Section~\ref{TF} and \ref{approx}.

\section{\label{Solution}Classical Solutions}

We have seen in the previous section, that the form invariance condition and the topological properties have strongly restricted the expressions of the Ansatz and provided us with the topologically stable gauge fields, which are candidates of the classical solutions. To solve the Yang-Mills  equation in 3D now becomes to solve for the factors $p$ and $\theta$ with the possible boundary values listed in Table~\ref{table:3dBC}, and to solve the Yang-Mills equation in 4D becomes to solve for only the factor $p$ with the possible boundary values listed in Table~\ref{table:4dBC}. We will show the details of the calculations in this section.

\subsection{\label{3ds}3D Case}

In Subsection~\ref{3dansatz}, we have constructed the form-invariant Ansatz to the 3D Yang-Mills equation. Now we choose a gauge fixing condition:
\be\label{eq:3Dgf}
\phi' = (1 - 2 p) {\theta}'\, .
\ee
Consequently, the Ansatz can be written into an equivalent expression with a different parameterization as follows:
\be\label{eq:Ansatz-3D-afterGauge}
  A_{\mu, a}  =   G \left( \frac{\delta_{\mu a}}{|x|} - \frac{x_\mu x_a}{|x|^3} \right) + \left(  H-1 \right) \frac{\epsilon_{\mu a i} x_i}{\tau}\, ,
\ee
where
\be
G \equiv    p_+ \, \textrm{sin}(\theta_+)   +  p_- \, \textrm{sin}(\theta_-)  \, , \quad \quad H \equiv    p_+  \, \cos(\theta_+)   +  p_-  \,  \cos(\theta_-)   
\ee
and
\be
  p_\pm   \equiv \frac{1 \pm \hat{p}}{2}\, ,   \quad \quad   \theta_\pm   \equiv \frac{\phi \pm \theta}{2}\, , \quad \quad   \hat{p} \equiv 2 p -1\, .
\ee
 We will use this gauge through out the calculations in this paper.

The boundary conditions for $p$ and $\theta$, which are discussed in Section~\ref{3df} and listed in Table~\ref{table:3dBC}, are extended here to include the boundary conditions for $G$ and $H$ in Table~\ref{table:3dBC2}.

\begin{table}[h!]
\centering
\begin{tabular}{ccccccccc}
 Winding & \multicolumn{1}{  c  }{\multirow{2}{*}{\,$p|_{0}$\,} }&  \multicolumn{1}{  c  }{\multirow{2}{*}{\,$p|_{\infty}$\,} }   &  \multicolumn{1}{  c  }{\multirow{2}{*}{\,$\theta|_{0}$\,} }    &  \multicolumn{1}{  c  }{\multirow{2}{*}{\,$\theta|_{\infty}$\,} }  & \multicolumn{1}{  c  }{\multirow{2}{*}{$G|_0$} }   & \multicolumn{1}{  c  }{\multirow{2}{*}{ $G|_\infty$} }   & \multicolumn{1}{  c  }{\multirow{2}{*}{   $H|_0$ } }  &  \multicolumn{1}{  c  }{\multirow{2}{*}{   $H|_\infty$ } }   \\
\, number $B$ \quad & {} & {} & {} & {} & {} & {} & {} & {}  \\[1ex]
\hline\hline \\[-2ex]
0   &    0     &   0   &   $\pi$   &   $\pi$   &   $-\cos\left(\frac{\phi}{2}\right)$   &   $-\cos \left(\frac{\phi}{2}\right)$   &   $\quad\sin\left(\frac{\phi}{2}\right)$   &   $\quad\sin\left(\frac{\phi}{2}\right)$\\[1.5ex]
0   &   $\frac{1}{2}$    &  $\frac{1}{2}$     &   $\pi$   &   $\pi$   &   0   &   0   &   0   &   0\\[1.5ex]
0   &   1   &   1   &   $\pi$   &   $\pi$   &   $\quad\cos\left(\frac{\phi}{2}\right)$   &   $\quad\cos \left(\frac{\phi}{2}\right)$   &   $-\sin\left(\frac{\phi}{2}\right)$   &   $-\sin\left(\frac{\phi}{2}\right)$\\[1.5ex]
1   &    0   &   $\frac{1}{2}$     &   $\pi$   &   $\pi$   &   $-\cos\left(\frac{\phi}{2}\right) $   &   0   &   $\quad\sin\left(\frac{\phi}{2}\right)$   &   0\\[1.5ex]
-1   &  $\frac{1}{2}$      &  0  &   $\pi$   &   $\pi$   &   0   &   $-\cos\left(\frac{\phi}{2}\right) $   &   0   &   $\quad\sin\left(\frac{\phi}{2}\right)$ \\[1.5ex]
$\cdots$   &   $\cdots$   &   $\cdots$   &   $\cdots$   &   $ \cdots$   &   $\cdots$   &   $\cdots$   &   $\cdots$   &   $ \cdots$ \\ [1 ex]
\end{tabular}\nonumber
\caption{Boundary conditions for $G$ and $H$ in 3D. $\phi$ is an arbitrary constant.}
\label{table:3dBC2}
\end{table}

The Yang-Mills equation reads
\be\la{eq:3deom}
 D_\mu  F_{\mu\nu}  = 0 \, ,
\ee
where for the 3D Eulidean space we do not distinguish the upper and the lower indices. Plugging Eq.~\eqref{eq:Ansatz-3D-afterGauge} in Eq.~\eqref{eq:3deom}, we obtain
\be
G =  0\, ,\, H = 0 \quad \textrm{or} \quad G = \textrm{cos}\, \Theta\, ,\, H = \textrm{sin}\, \Theta\, ,
\ee
where $\Theta$ is a constant. The details are presented in Appendix~\ref{3dcs}. The solution with $G = \textrm{cos}\, \Theta$ and $H = \textrm{sin}\, \Theta$ is a pure gauge solution, while the solution with $G=H=0$ corresponds to the Wu-Yang monopole, which is also a stright line $p = 1/2$ in the variable $\tau$ hence a monotonic $C^\infty$-curve satisfying the boundary values, as we briefly discussed in Section~\ref{smoothest}. We summarize the 3D solutions with lowest winding numbers in Fig.~\ref{fig:3dsolution}:

    \begin{figure}[!htb]
      \begin{center}
        \includegraphics[width=0.46\textwidth]{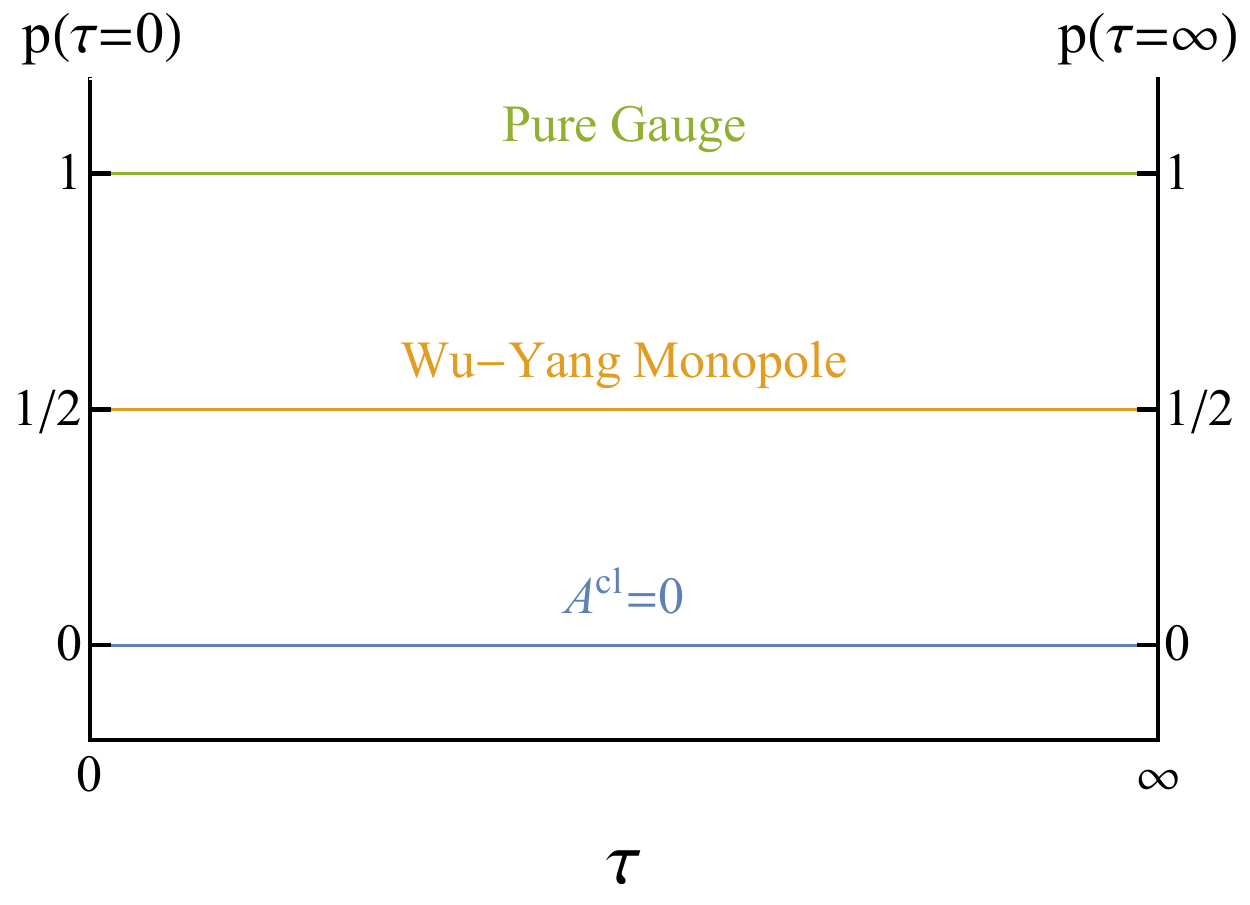}
      \caption{Spherically symmetric solutions to 3D Yang-Mills equation with lowest winding numbers.}
      \label{fig:3dsolution}
      \end{center}
    \end{figure}

\subsection{\label{4ds}4D Case}

We list the results of the 4D classical solutions in this section. The details are given in Appendix~\ref{4dcs}.

First, we have the topologically stable Ansatz to the 4D Yang-Mills equation
\be
  A_{\mu, a} = 2\, \frac{p(\tau)}{\tau}\, \eta_{a\mu\rho} x_\rho\, ,
\ee
where for the 4D Eulidean space we do not distinguish the upper and the lower indices. As we have seen in Appendix~\ref{sq}, the form invariance condition uniquely fixes the factor $\theta$ in the Ansatz and constrains the factor $p = p(\tau)$. The boundary values of $p$ at $\tau=0$ and $\tau=\infty$ are fixed by the topological properties, and some possible choices are listed in Table~\ref{table:4dBC}.

Next, we can compute the field strength and try to solve the 4D Yang-Mills equation.
\be
  F_{\mu\nu}^a = 4 \eta_{a\mu\nu} \left(\frac{p^2}{\tau} - \frac{p}{\tau} \right) + 4 x_\rho (x_\mu \eta_{a\nu\rho} - x_\nu \eta_{a\mu\rho}) \left[\left(\frac{p}{\tau} \right)' + \frac{p^2}{\tau^2} \right]\, ,
\ee
where the prime denotes the derivative with respect to $\tau$. The Yang-Mills equation is
\be
  (D_\mu F_{\mu\nu})^a = 8\, \frac{\eta_{a\nu\rho} x_\rho}{\tau^2}\, (-p + 3 p^2 - 2 p^3 + \tau p' + \tau^2 p'') = 0\, .
\ee
As discussed in Appendix~\ref{4dcs}, we suppose that $p(\tau)$ has the expansions
\be
  p (\tau) = \Bigg\{
  \begin{array}{ll}
    a^{\tau=0}_0 + \sum_{n=1}^\infty a^{\tau=0}_n \, \tau^n\, ,& \textrm{for small} \tau \, ;\nonumber\\
    a^{\tau=\infty}_0 + \sum_{n=1}^\infty a^{\tau=\infty}_n / \tau^n\, ,& \textrm{for large} \tau  \, ,
  \end{array}
\ee 
then there are a few different solutions to this equation:
\begin{enumerate}
  \item $a_0^{\tau=0} = \frac{1}{2}\, ,\, a_0^{\tau=\infty} = \frac{1}{2}$:\quad $p = 1/2$\, ,
  \item $a_0^{\tau=0} = 0\, ,\, a_0^{\tau=\infty} = 1$:\quad $p = \frac{\tau}{\tau + c_1}$\, ,
  \item $a_0^{\tau=0} = 1\, ,\, a_0^{\tau=\infty} = 0$:\quad $p = \frac{c_1}{\tau + c_2}$\, ,
  \item $a_0^{\tau=0} = 1\, ,\, a_0^{\tau=\infty} = 1$:\quad $p = 1$\, ,
  \item $a_0^{\tau=0} = 0\, ,\, a_0^{\tau=\infty} = 0$:\quad $p = 0$\, ,
\end{enumerate}
where $c_1$ and $c_2$ are two positive real constants. The first solution is the meron solution. The second solution and the third solution correspond to the $1$-instanton solution in the regular gauge and the $1$-anti-instanton solution in the singular gauge respectively. The last two solutions can be viewed as the pure gauge solution and the trivial vacuum solution respectively.

We also observe that for these solutions the factors $p$ are all monotonic $C^\infty$-curves satisfying the boundary conditions with $p^{(n)}$ also monotonic. For instance, for the meron solution $p(\tau)$ is just a straight line connecting the boundary values at $\tau=0$ and $\tau=\infty$. For the $1$-instanton solution $p(\tau) = \tau / (\tau + c_1)$ we notice that
\be
\frac{\partial^n p(\tau)}{\partial \tau^n} = \frac{\partial^n  }{\partial \tau^n} \left( \frac{\tau}{\tau + c_1}  \right) = (-1)^{n+1} \frac{c_1 \, n!}{\left( c_1 + \tau  \right)^{n+1}} \, ,
\ee
are also monotonic $C^\infty$-functions.

\begin{figure}[!htb]
      \begin{center}
        \includegraphics[width=0.46\textwidth]{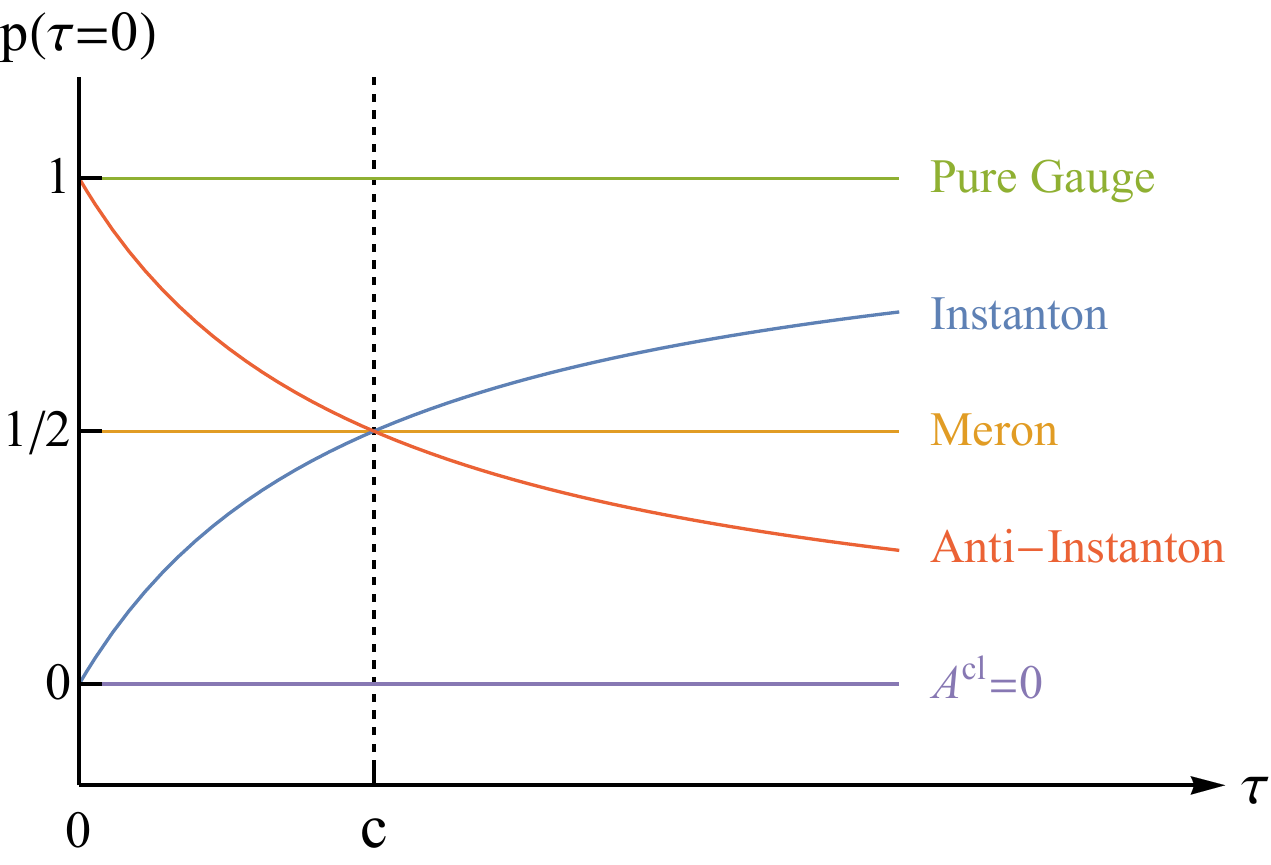}
      \caption{Spherically symmetric solutions to 4D Yang-Mills equation with lowest winding numbers.}
      \label{fig:4dsolution}
      \end{center}
    \end{figure}

 We summarize the 4D solutions with lowest winding numbers in Fig.~\ref{fig:4dsolution}, where for simplicity we assume that $c_1 = c_2 = c$ with a constant $c$.

\begin{figure}[!htb]
      \begin{center}
        \includegraphics[width=0.52\textwidth]{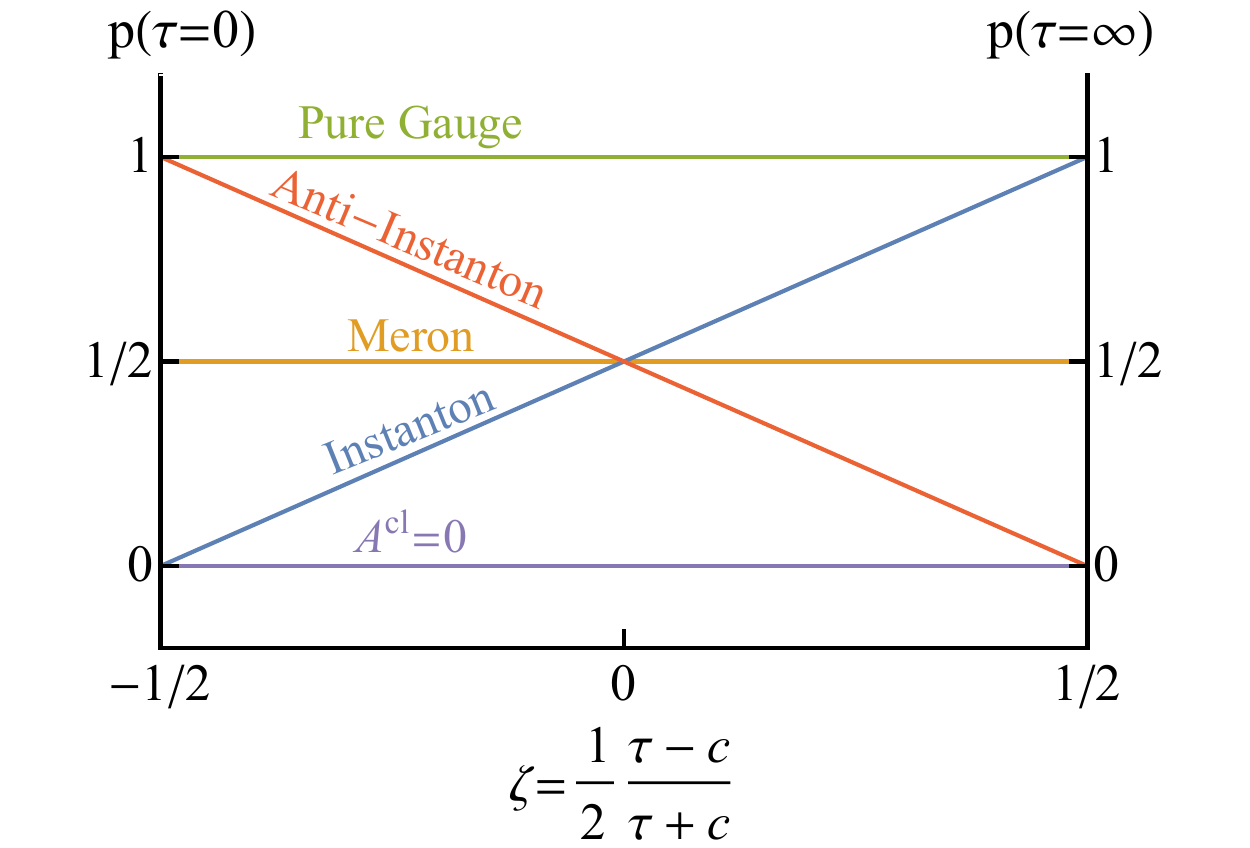}
      \caption{Spherically symmetric solutions to 4D Yang-Mills equation with lowest winding numbers in coordinate $\zeta$.}
      \label{fig:4dconformalsolution}
      \end{center}
    \end{figure}

If we adopt a new coordinate introduced by the conformal transformation
\be
\zeta = \frac{1}{2} \, \frac{\tau - c}{\tau + c} \, ,
\ee
then all the classical solutions shown in Fig.~\ref{fig:4dsolution} can be plotted in the new coordinate shown in Fig.~\ref{fig:4dconformalsolution}. As we can see, for each topologically allowed path there is a corresponding classical solution, which is a monotonic $C^\infty$-curve.

\section{\label{TF}Topological Fluctuations~(General Formalism)}

In this section, we consider the quantum fluctuations of the Yang-Mills field, which we call $Q_\mu$, and the topological fluctuations, which are denoted by $\widetilde{\mathcal{A}}$. The quantum fluctuations $Q_\mu$ are the rapidly varying fields, that generally do not satisfy the form invariance condition or the topological properties. The topological fluctuations $\widetilde{\mathcal{A}}$, on the other hand, are the slowly varying fields that satisfy both the form invariance condition and the topological properties. The expansion considered here is similar to the case of the 1D quantum antiferromagnets, in which we split the spin field $\vec{n}$ into a slowly varying mode $\vec{m}$ and a rapidly varying mode $\vec{l}$ (See Appendix~\ref{reSpinChain}). We will review and explain the legitimacy of such treatment at the end of this section. Let us first set up the general formalism, and then discuss the 3D and the 4D case separately.

From now on we employ the background field formalism, and denote the background Yang-Mills field and the quantum fluctuation by $A_\mu$ and $Q_\mu$ respectively. The full field strength is given by
\be\label{eq:fmunudmuqmu}
  \mathcal{F}_{\mu\nu} = [\mathcal{D}_\mu,\, \mathcal{D}_\nu] = [D_\mu + Q_\mu ,\, D_\nu + Q_\nu]\, ,
\ee
and one can also define a field strength of the background as
\be\label{eq:fmunudmu}
  F_{\mu\nu} = [D_\mu,\, D_\nu]\, ,
\ee
where
\begin{align}
  \mathcal{D}_\mu & \equiv \partial_\mu + A_\mu^{\textrm{top}} + Q_\mu\, ,\nonumber\\
  D_\mu & \equiv \partial_\mu + A_\mu^{\textrm{top}}\, .
\end{align}
The background field $A_\mu^{\textrm{top}}$ not only includes the solution to the classical Yang-Mills equation, which we will call $A_\mu^{\textrm{cl}}$, but it also includes the topological fluctuations around the classical solutions, which are the fields $\widetilde{G}$ and $\widetilde{H}$ for the 3D case and $\widetilde{p}$ for the 4D case, and we will formally denote them as $\widetilde{\mathcal{A}}$. The key point here is that the background field is constrained by the form invariance condition \eqref{eq:FI}:
\begin{displaymath}
 (O^{-1})\, _\mu\,^\nu\, A_\nu^{\textrm{top}} (O \, x) = V^{-1}\, A_\mu^{\textrm{top}} (x)\, V\, .
\end{displaymath}
More explicitly,
\[  A_{\mu, a}^{\textrm{top}} = \left\{ \begin{array}{lll}
          \left(G_0 + \widetilde{G} \right) \left( \frac{\delta_{\mu a}}{|x|} - \frac{x_\mu x_a}{|x|^3} \right) + \left(  H_0 + \widetilde{H}-1 \right) \left(\frac{\epsilon_{\mu a i} x_i}{|x|^2}\right) & \quad \mbox{for 3D Yang-Mills}\, ;\\
& \\
       \Big( p_0 + \widetilde{p}  \Big) \eta_{a \mu\nu}  \left(\frac{x_\nu}{|x|^2}\right) & \quad \mbox{for 4D Yang-Mills}\, ,\end{array} \right. \] 
where we adopt the saddle point approximation and write $G = G_0 + \widetilde{G}$,  $H = H_0 + \widetilde{H}$ and  $p = p_0 + \widetilde{p}$.   $G_0$ and $H_0$ correspond to the classical solutions for the 3D case, while $p_0$ corresponds to the classical solutions for the 4D case.  In order to preserve the topological properties, the topological fluctuations have to vanish at the boundaries, i.e.,
\be
  \widetilde{\mathcal{A}} (\tau = 0) = \widetilde{\mathcal{A}} (\tau = \infty) = 0\, ,
\ee
which has a significant physical meaning  that we will discuss in Subsection~\ref{remarks}.

Expanding the expression of the field strength, we obtain
\be
 \mathcal{F}_{\mu\nu} = F_{\mu\nu} + D_\mu Q_\nu - D_\nu Q_\mu + [Q_\mu,\, Q_\nu]\, ,
\ee
where $F_{\mu\nu}$ is the field strength of the background Yang-Mills field. Consequently, up to some total derivatives
\begin{align}
  \mathcal{L}_{\textrm{YM}} & \equiv - \frac{1}{2}\, \textrm{Tr} \left(\mathcal{F}_{\mu\nu} \mathcal{F}_{\mu\nu}\right) \nonumber\\
  {} & = - \textrm{Tr} \left(\frac{1}{2} F_{\mu\nu} F_{\mu\nu} + 2 Q_\mu \left(D_\nu F_{\mu\nu} \right) - Q_\mu \mathcal{M}_{\mu\nu} Q_\nu + 2 [Q_\mu,\, Q_\nu] (D_\mu Q_\nu) + \frac{1}{2} [Q_\mu,\, Q_\nu]^2 \right)\, ,\label{eq:Lexpansion}
\end{align}
where
\be
  \mathcal{M}_{\mu\nu} \equiv D^2 \delta_{\mu\nu} + F_{\mu\nu} - D_\nu D_\mu\, .
\ee
In order to quantize the theory, we introduce a gauge fixing term and a ghost term:
\begin{align}
  \mathcal{L}_{\textrm{fix}} & =  - \textrm{Tr} \left[ \left(D_\mu Q_\mu \right)^2 \right] \, , \\
  \mathcal{L}_{\textrm{gh}} & =  \textrm{Tr} \left( 2 b  D_\mu D_\mu c   \right)\, ,
\end{align}
i.e., for the quantum fluctuations $Q_\mu$ we choose the background gauge
\be
  D_\mu Q_\mu = 0\, ,
\ee
where
\be
  D_\mu \equiv \partial_\mu + A_\mu^{\textrm{top}}\, .
\ee
The full Lagrangian is
\begin{align}
  \mathcal{L} & = \mathcal{L}_{\textrm{YM}} + \mathcal{L}_{\textrm{fix}} + \mathcal{L}_{\textrm{gh}} \nonumber\\
  {} & = - \textrm{Tr} \left( \frac{1}{2} F_{\mu\nu} F_{\mu\nu} + 2 Q_\mu \left(D_\nu F_{\mu\nu} \right) - Q_\mu M_{\mu\nu} Q_\nu - 2 b M_{\textrm{gh}} c + 2 [Q_\mu,\, Q_\nu] (D_\mu Q_\nu) + \frac{1}{2} [Q_\mu,\, Q_\nu]^2 \right) \label{eq:FullLag}
\end{align}
with
\begin{align}
  M_{\mu\nu} & \equiv \mathcal{M}_{\mu\nu} + D_\mu D_\nu = D^2 \delta_{\mu\nu} + 2 F_{\mu\nu}\, ,\nonumber\\
  M_{\textrm{gh}} & \equiv D^2\, .\label{eq:Mop}
\end{align}
Therefore, the full quantum theory is given by the path integral
\be
  Z   = \int \mathcal{D} A_\mu^{\textrm{top}} \, \mathcal{D} Q_\mu\, \mathcal{D} b\, \mathcal{D} c \,\,  \exp \left(- S [A_\mu^{\textrm{top}},\, Q_\mu,\, b,\, c] \right)  \, ,
\ee
where
\be
A_\mu^{\textrm{top}} = A_\mu^{\textrm{top}} \left(\gamma^{(i)} , A^{\textrm{cl}}, \widetilde{\mathcal{A}}\right) \, ,
\ee
is the topological configuration including the classical solutions $A^{\textrm{cl}}$, their moduli $\gamma^{(i)}$ and the topological fluctuations $\widetilde{\mathcal{A}}$ and
\be
S [A_\mu^{\textrm{top}},\, Q_\mu,\, b,\, c] = \frac{1}{g^2}\, \int d^D x \,   \mathcal{L}  \, .
\ee

For the path integral measure, we have
\be
 \int \mathcal{D} A_\mu^{\textrm{top}} = \int \mathcal{D}  \gamma^{(i)}   \mathcal{D}   A^{\textrm{cl}} \mathcal{D}  \widetilde{\mathcal{A}} \,\,\, [\textrm{Jac}] \, ,
\ee 
where $\gamma^{(i)}$ are the moduli including translations and gauge orientations, and $\widetilde{\mathcal{A}}$ denotes topological fluctuations, which are $\widetilde{G}$ and $\widetilde{H}$ for the 3D case and $\widetilde{p}$ for the 4D case. $A^{\textrm{cl}}$ is the classical solution given by $G_0$ and $H_0$ for the 3D case and by $p_0$ for the 4D case. In 3D, $G_0$ and $H_0$ are constant, hence there is no integral over moduli but only a sum over different classical solutions, which we denote as $\mathcal{A}_0$. For the (anti-)instanton in 4D, $p_0$ has one more modulus which is the size $\rho$, thus for this case there is one more corresponding measure in the path integral:
\be
 \mathcal{D}   A^{\textrm{cl}}_{\textrm{(anti-)instanton}} \quad \supset \quad [\textrm{Jac}]_\rho  \, d\rho \, ,
\ee
For simplicity, we neglect the measure of $\rho$ in the following discussions. In 4D, when we expand the measure $\mathcal{D} A_\mu^{\textrm{top}}$ for the (anti-)instanton, we will recover this term.

Summarizing the discussions above, we obtain
\be
  Z   = \sum_{\{\mathcal{A}_0\}} \int \mathcal{D} \gamma^{(i)}\, \mathcal{D} \widetilde{\mathcal{A}}\, \mathcal{D} Q_\mu\, \mathcal{D} b\, \mathcal{D} c\,\,\, [\textrm{Jac}] \,\, \exp \left(- S [A^{\textrm{cl}}, \, \widetilde{\mathcal{A}},\, Q_\mu,\, b,\, c, \, \gamma_i] \right)\, .\label{eq:genPartFct-1}
\ee
To evaluate the path integral \eqref{eq:genPartFct-1} and derive the effective field theories, we can separate the action into two parts. There are two different ways of separation:
\begin{enumerate}
\item 
\be\la{eq:firstseparation}
S  [A^{\textrm{cl}}, \, \widetilde{\mathcal{A}},\, Q_\mu,\, b,\, c, \, \gamma_i]  =S^{\textrm{qu}}  [A^{\textrm{cl}}, \, \widetilde{\mathcal{A}},\, Q_\mu,\, b,\, c, \, \gamma_i]  +  S^{\textrm{top}} [A^{\textrm{cl}}, \widetilde{\mathcal{A}},\, \gamma_i]\, ,
\ee
where $S^{\textrm{top}}$ is independent of quantum modes $Q_\mu$ and the ghost fields $b$, $c$. When we turn off the quantum modes
\be
S^{\textrm{qu}}  [A^{\textrm{cl}}, \, \widetilde{\mathcal{A}},\, Q_\mu = 0,\, b=0,\, c=0, \, \gamma_i]= 0 \, ,
\ee
i.e.,
\be
S  [A^{\textrm{cl}}, \, \widetilde{\mathcal{A}},\, Q_\mu = 0,\, b= 0,\, c= 0 , \, \gamma_i]  = S^{\textrm{top}} [A_\mu^{\textrm{cl}}, \widetilde{\mathcal{A}},\, \gamma_i] \, .
\ee

\item 
 \be\la{eq:secondseparation}
S  [A^{\textrm{cl}}, \, \widetilde{\mathcal{A}},\, Q_\mu,\, b,\, c, \, \gamma_i]  =S_0 [A^{\textrm{cl}}, \,   Q_\mu,\, b,\, c, \, \gamma_i] +  \Delta S  [A^{\textrm{cl}}, \, \widetilde{\mathcal{A}},\, Q_\mu,\, b,\, c, \, \gamma_i] \, ,
\ee 
where $S_0$ is independent of the topological fluctuations $\widetilde{\mathcal{A}}$. When we turn off the topological fluctuations $\widetilde{\mathcal{A}}$, 
\be
\Delta S  [A^{\textrm{cl}}, \, \widetilde{\mathcal{A}} = 0,\, Q_\mu,\, b,\, c, \, \gamma_i]   = 0\, ,
\ee
i.e.,
\be
S  [A^{\textrm{cl}}, \, \widetilde{\mathcal{A}}=0,\, Q_\mu,\, b,\, c, \, \gamma_i]  =S_0  [A^{\textrm{cl}}, \, Q_\mu,\, b,\, c, \, \gamma_i]  \, .
\ee
\end{enumerate}
Now let us briefly explain the physical reason of having two ways of separation.
\begin{itemize}

\item
The first separation \eqref{eq:firstseparation} is convenient to use, because in Eq.~\eqref{eq:fmunudmuqmu}  we have already separated the gauge field into $A_\mu^{\textrm{top}}$ and $Q_\mu$. Hence, it is natural to calculate the contributions purely from $A_\mu^{\textrm{top}}$ to the action, which is $S^{\textrm{top}}$.

\item
The second separation \eqref{eq:secondseparation}, though inconvenient to use, has more transparent and profound physical meanings, as we will show now. In Appendix~\ref{topologicalmeasure}, we provide the detailed derivations of the Jacobian $\left[\textrm{Jac} \right]$:
\be\label{eq:Jac}
  \left[\textrm{Jac} \right]  = \left[\textrm{Jac} \right]_{\gamma_i} \, \left[\textrm{Jac} \right]_{\widetilde{\mathcal{A}}}\, .
\ee
Then we have
\be
  Z  = \sum_{\{\mathcal{A}_0\}} \int [\textrm{Jac}]_{\widetilde{\mathcal{A}}}\, \mathcal{D} \widetilde{\mathcal{A}}\, \int [\textrm{Jac}]_{\gamma_i}\, \mathcal{D} \gamma^{(i)} \int \mathcal{D} Q_\mu\, \mathcal{D} b\, \mathcal{D} c\,\, e^{-S_0 }\, e^{- \Delta S}\, .
\label{eq:genPartFct-2}
\ee
We emphasize that $[\textrm{Jac}]_{\gamma_i}$ in general depends on the topological fluctuations $\widetilde{\mathcal{A}}$. If we turn off the topological fluctuations $\widetilde{\mathcal{A}}$, we will have
\be
[\textrm{Jac}]_{\gamma_i} \xrightarrow{\quad \textrm{Turn off topological fluctuations } \widetilde{\mathcal{A}} \quad } [\textrm{Jac}]_{\gamma_i}^{\textrm{cl}}\, ,
\ee
and thus reproduce the path integral of quantum fluctuations in classical backgrounds:
\be
Z \longrightarrow  Z_0  = \sum_{\{\mathcal{A}_0\}}  \int [\textrm{Jac}]^{\textrm{cl}}_{\gamma_i}\, \mathcal{D} \gamma^{(i)}\, \int \mathcal{D} Q_\mu\, \mathcal{D} b\, \mathcal{D} c\,\,    e^{-S_0    [A^{\textrm{cl}}, \, Q_\mu,\, b,\, c, \, \gamma_i]}\, ,
\ee
where $[\textrm{Jac}]^{\textrm{cl}}_{\gamma_i}$ is the measure for the classical solutions.

As we can see, $S_0$  corresponds to the quantum fields $Q_\mu$ and the ghosts $b$, $c$ in the classical backgrounds. More importantly, in Section~\ref{approx} we will show that $\Delta S$ in some limit\footnote{The limit will be explained in Section~\ref{approx}.} has an expression similar to the Higgs mechanism, where the topological fluctuations play the similar role of the Higgs field. 
\end{itemize}

In principle, if we could integrate out the moduli $\gamma_i$, we would obtain an effective theory of the topological modes $\widetilde{\mathcal{A}}$ and the quantum modes $Q_\mu$, which is of great interest in physics. However, the calculations can be very involved and usually cannot be performed exactly. In practice the integration over the moduli is left to the end of the computation, after the other part of the path integral is evaluated. In this section, we will first present the exact result formally without evaluating the integration over the moduli, and then in order to make the relevant physics more transparent we will perform the integral over moduli approximately in Section~\ref{approx}.

\subsection{3D Case}\label{3dtf}

Let us recall that for the 3D case the gauge field $A_{\mu, a}$ and the field strength $F_{\mu\nu}^a$ after the gauge fixing \eqref{eq:3Dgf} are (see Appendix~\ref{3dcs}):
\begin{align}
  A_{\mu, a} = & \, G \left( \frac{\delta_{\mu a}}{|x|} - \frac{x_\mu x_a}{|x|^3} \right) + \left(  H-1 \right) \frac{\epsilon_{\mu a i} x_i}{|x|^2}\, ,\\
F_{\mu\nu}^a  = & \, \left( \frac{  x_\mu \delta_{\nu a} - x_\nu \delta_{\mu a}}{|x|^3} \right) \left( 2   \tau G' \right) + \frac{\epsilon_{\mu\nu a}}{|x|^2} \left( G^2 + 2 H     -2 \right) \nonumber\\
& + \frac{x_i \left(  x_\mu \epsilon_{\nu a i} - x_\nu \epsilon_{\mu a i} \right)}{|x|^4} \left( 2 - 2 H + 2 \tau H'   - G^2 \right) + \frac{x_a x_i \epsilon_{\mu\nu i}}{|x|^4} \left( H-1 \right)^2\, ,
\end{align}
where $\tau \equiv x_\mu x^\mu$ and $(\cdots)' \equiv \partial (\cdots) / \partial \tau$. The factors $p$ and $\theta$ in the Ansatz \eqref{eq:3DAnsatz} are encoded in the new factors $G$ and $H$, hence the form invariance allows $G$ and $H$ to have the fluctuations. Therefore, for the 3D case the gauge field with quantum fluctuations now reads
\be
  A_{\mu, a} + Q_{\mu, a} = G \left( \frac{\delta_{\mu a}}{|x|} - \frac{x_\mu x_a}{|x|^3} \right) + \left(  H-1 \right) \frac{\epsilon_{\mu a i} x_i}{|x|^2} + Q_{\mu, a}\, ,
\ee
where $G$ and $H$ are slowly varying fields around the classical backgrounds, and $Q_{\mu, a}$ is a rapidly varying quantum mode. Our discusssion here is similar to the analysis for the 1D antiferromagnetic spin chain, which is reviewed in Appendix~\ref{reSpinChain}.

As shown in Eq.~\eqref{eq:firstseparation}, the whole action can be separated into the topological part and the quantum part. We can first calculate the topological part of the Lagangian
\be
 F_{\mu\nu}^aF_{\mu\nu}^a = \frac{2}{\tau^2} + \left(16 G'^2 - \frac{4}{\tau^2} G^2 + \frac{2}{\tau^2} G^4 \right) + \left(16 H'^2 - \frac{4}{\tau^2} H^2 + \frac{2}{\tau^2} H^4 \right) + \frac{4}{\tau^2} G^2 H^2 \, .
\ee
If we define a  complex vector
\be
\psi \equiv G + i H\, ,
\ee
the topological part of the Lagrangian becomes
\be\label{eq:3DfullTopL}
  \frac{1}{4} F_{\mu\nu}^aF_{\mu\nu}^a = \frac{1}{2 \tau^2} + 4 (\overline{\partial_\tau \psi})  (\partial_\tau \psi) - \frac{1}{\tau^2} \overline{\psi}  \psi + \frac{1}{2 \tau^2} (\overline{\psi}  \psi)^2\, .
\ee
The field $\psi$ contains both the solutions to the Yang-Mills equation and the topological fluctuations around them. As discussed before, $G_0$ and $H_0$ correspond to the classical solutions, while $\widetilde{G}$ and $\widetilde{H}$ correspond to the topological fluctuations, i.e.,
\be
  G = G_0 + \widetilde{G}\, ,\quad H = H_0 + \widetilde{H}\, .
\ee
Equivalently, we can use a complex scalar $\widetilde{\psi}$ to denote the topological fluctuations, i.e.,
\be
  \psi = \psi_0 + \widetilde{\psi}
\ee
with
\be
\psi_0 = G_0 + i H_0 \, ,\quad\quad
\widetilde{\psi} \equiv  
\widetilde{G}  +
i \widetilde{H}
 \, .
\ee

As we have seen in Section~\ref{3ds}, for the 3D case the solutions include
\be
  G_0=0,\, H_0=0 \quad \textrm{and} \quad G_0=\textrm{cos}\, \Theta,\, H_0=\textrm{sin}\, \Theta\, .
\ee
The first case corresponds to the Wu-Yang monopole solution, while the second one corresponds to the pure gauge solution. The trivial vacuum solution is just a speical case of the pure gauge case with $\Theta=\pi/2$.

For the Wu-Yang monopole $G_0=H_0=0$, the topological part of the Lagrangian reads
\be
 \frac{1}{4} F_{\mu\nu}^aF_{\mu\nu}^a \bigg|_{G_0 = H_0 = 0} = \frac{1}{2 \tau^2}  + 4 \big|\partial_\tau \widetilde{\psi}\big|^2 - \frac{1}{\tau^2} \big| \widetilde{\psi} \big|^2 + \frac{1}{2 \tau^2} \big| \widetilde{\psi} \big|^4 \, ,
\ee
where the classical part of the Lagangian equals $1 / (2 \tau^2)$, while the topological fluctuation part is
\be
  4 \big|\partial_\tau \widetilde{\psi}\big|^2 - \frac{1}{\tau^2} \big| \widetilde{\psi} \big|^2 + \frac{1}{2 \tau^2} \big| \widetilde{\psi} \big|^4 \, .
\ee
We see that for the Wu-Yang monopole the classical part and the topological fluctuations are completely separated in the Lagrangian, i.e., there are no mixed terms, which is not true in general.

For the pure gauge case $G_0=\textrm{cos}\, \Theta$, $H_0=\textrm{sin}\, \Theta$, the topological part of the Lagrangian reads
\begin{align}
 \frac{1}{4} F_{\mu\nu}^aF_{\mu\nu}^a \bigg|_{G_0 = \textrm{cos}\, \Theta,\,  H_0 = \textrm{sin}\, \Theta} & = \frac{1}{2 \tau^2} +   4 \big|\partial_\tau \widetilde{\psi}\big|^2 - \frac{1}{\tau^2} \big|\psi_0 +  \widetilde{\psi} \big|^2 + \frac{1}{2 \tau^2} \big| \psi_0 +  \widetilde{\psi} \big|^4 \, ,
\end{align} 
where  
\be
\psi_0 \equiv
G_0  + 
i H_0
=
\textrm{cos}\, \Theta  + 
i \, \textrm{sin}\, \Theta
 = e^{i \Theta} \, .
\ee
This expression contains $\Theta$ explicitly, so it seems to be gauge dependent. However, $e^{i \Theta}$ can always be absorbed by redefining the field $ \widetilde{\psi}$. If we turn off all the topological fluctuations, i.e. $\widetilde{G} = \widetilde{H} = 0$, the topological part of the Lagrangian is equal to the classical part, which vanishes identically for the pure gauge case.

Next, we consider the path integral \eqref{eq:genPartFct-2} for the 3D case, which is just the combination of the topological part and the quantum part of the theory. As discussed before, the path integral should take a sum over the Wu-Yang monopole and the trivial vacuum background. Let us list their contributions separately:
\begin{itemize}
\item For the Wu-Yang monopole:
\be
  \int [\textrm{Jac}]\, d x_0\, d\varphi_0\, \int \mathcal{D} \widetilde{\psi}\, e^{-S^{\mathrm{top}}} \Big|_{\psi_0  = 0}\, \int \mathcal{D} Q_\mu \, \mathcal{D} b\, \mathcal{D} c\, e^{-S^{\textrm{qu}}} \Big|_{\psi_0  = 0}\, ,
\ee
where
\begin{align}
  S^{\textrm{top}} \big|_{\psi_0  = 0} & = \frac{1}{g^2} \int d^3 x\, \Bigg[ \frac{1}{2 \tau^2}  + 4 \big|\partial_\tau \widetilde{\psi}\big|^2 - \frac{1}{\tau^2} \big| \widetilde{\psi} \big|^2 + \frac{1}{2 \tau^2} \big| \widetilde{\psi} \big|^4 \Bigg]\, ,\nonumber\\
  S^{\textrm{qu}} \big|_{\psi_0  = 0} & =  \frac{ 1}{g^2} \int d^3 x\,\, \textrm{Tr} \Bigg[ - 2 Q_\mu \left(D_\nu F_{\mu\nu} \right) + Q_\mu M_{\mu\nu} Q_\nu + 2 b M_{\textrm{gh}} c \nonumber\\ 
  {} & \qquad\qquad\qquad\quad - 2 [Q_\mu,\, Q_\nu] (D_\mu Q_\nu) - \frac{1}{2} [Q_\mu,\, Q_\nu]^2\Bigg]_{\psi_0 = 0}\, .\label{eq:ExactActionWuYang}
\end{align}
The Jacobian $[\textrm{Jac}]$ consists of the contributions from pseudo zero modes, corresponding to the translations~($[\textrm{Jac}]_{x_0}$), the gauge orientations~($[\textrm{Jac}]_\varphi$) and the topological fluctuations~($[\textrm{Jac}]_{\widetilde{\mathcal{A}}}$).  We discuss them in detail in Appendix~\ref{topologicalmeasure}.

\item For the pure gauge solution:
\be
  \int \mathcal{D} \widetilde{\psi}\, e^{-S^{\mathrm{top}}} \Big|_{\psi_0= e^{i \Theta}}\, \int \mathcal{D} Q_\mu \, \mathcal{D} b\, \mathcal{D} c\, e^{-S^{\textrm{qu}}} \Big|_{\psi_0= e^{i \Theta} }\, ,
\ee
where
\begin{align}
  S^{\textrm{top}} \big|_{\psi_0= e^{i \Theta}} & = \frac{1}{g^2} \int d^3 x\, \Bigg[ \frac{1}{2 \tau^2} +   4 \big|\partial_\tau \widetilde{\psi}\big|^2 - \frac{1}{\tau^2} \big| e^{i \Theta} +  \widetilde{\psi} \big|^2 + \frac{1}{2 \tau^2} \big|  e^{i \Theta} +  \widetilde{\psi} \big|^4  \Bigg] \, ,\nonumber\\
  S^{\textrm{qu}} \big|_{\psi_0= e^{i \Theta}} & = \frac{ 1}{g^2}\int d^3 x\, \textrm{Tr} \Bigg[ - 2 Q_\mu \left(D_\nu F_{\mu\nu} \right) + Q_\mu M_{\mu\nu} Q_\nu + 2 b M_{\textrm{gh}} c \nonumber\\ 
  {} & \qquad\qquad\qquad\quad - 2 [Q_\mu,\, Q_\nu] (D_\mu Q_\nu) - \frac{1}{2} [Q_\mu,\, Q_\nu]^2 \Bigg]_{\psi_0 = e^{i \Theta}}\, .
\end{align}
For both the Wu-Yang monopole and the pure gauge case, $M_{\mu\nu}$ and $M_{\textrm{gh}}$ are defined in Eq.~\eqref{eq:Mop}.
\end{itemize}

\subsection{\label{4dtf}4D Case}

Similar to the 3D case, for the 4D case we can also separate the gauge field $A_\mu$ into the topological part and the quantum part, and the topological part includes the classical solutions and the topological fluctuations around the solutions.

We have seen that for the 3D case both the factor $p$ and the factor $\theta$, or equivalently $G$ and $H$, can have topological fluctuations. The 4D case is a little simpler, because the form invariance condition \eqref{eq:FI} has fixed the factor $\theta$ and restricted the factor $p= p (\tau)$. We can write
\be
  p = p_0 + \widetilde{p}\, ,
\ee
where $p_0$ and $\widetilde{p}$ denote the classical background and the topological fluctuations around the classical background respectively. The 4D gauge field now has the form
\be
  A_\mu = p\, U^{-1} \partial_\mu U + Q_\mu\, ,
\ee
where $p$ is a slowly varying field around the classical background, and $Q_\mu$ is a rapidly varying quantum mode. Again, the analysis here is similar to the 1D antiferromagnetic spin chain, which will be reviewed in Appendix~\ref{reSpinChain}.

As we have seen in Section~\ref{4ds}, the 4D classical solutions include
\be
  p_0 = \frac{1}{2}\, ,\quad p_0 = \frac{\tau}{\tau + c_1}\, ,\quad p_0 = \frac{c_2}{\tau + c_2}\, ,\quad p_0 = 1\, ,\quad p_0 = 0\, ,
\ee
where they correspond to the meron solution, the instanton solution in the regular gauge, the anti-instanton solution in the singular gauge, the pure gauge solution and the trivial vacuum solution respectively. For simplicity, we assume that $c_1 = c_2 = c$ in the following. In the new variable $q \equiv p - 1/2$, they read
\be
  q_0 = 0\, ,\quad q_0 = \pm \frac{\tau - c}{2 (\tau + c)}\, ,\quad q_0 = \pm \frac{1}{2}\, .
\ee
  
We can first calculate the topological part of the Lagrangian:
\begin{align}
  \frac{1}{4} F_{\mu\nu}^aF_{\mu\nu}^a & =  \frac{24}{\tau^2} \left[(\tau p')^2 + p^2 (p-1)^2 \right]\nonumber\\
  {} & =  \frac{24}{\tau^2} \left[\frac{1}{16} + (\tau q')^2 - \frac{q^2}{2} + q^4 \right]\, .\label{eq:4DfullTopL}
\end{align}
Then we evaluate this expression by inserting
\be
  q = q_0 + \widetilde{q}
\ee
to take into account the topological fluctuations around the classical background.

For each background with the fixed value $p_0$ or equivalently $q_0$ in the 4D case, the path integral \eqref{eq:genPartFct-2} becomes:
\be
  \int \mathcal{D} \widetilde{\psi}\, e^{-S^{\mathrm{top}}} \Big|_{q_0}\, \int \mathcal{D} Q_\mu \, \mathcal{D} b\, \mathcal{D} c\, e^{-S^{\textrm{qu}}} \Big|_{q_0}\, ,
\ee
where the quantum part takes the following general form
\begin{align}
  S^{\textrm{qu}} \big|_{q_0} & = \frac{ 1}{g^2}\int d^4 x\, \textrm{Tr} \Bigg[ - 2 Q_\mu \left(D_\nu F_{\mu\nu} \right)+ Q_\mu M_{\mu\nu} Q_\nu + 2 b M_{\textrm{gh}} c \nonumber\\ 
  {} & \qquad\qquad\qquad\quad - 2 [Q_\mu,\, Q_\nu] (D_\mu Q_\nu) - \frac{1}{2} [Q_\mu,\, Q_\nu]^2 \Bigg]_{q_0}\, ,\label{eq:4DactionQu}
\end{align}
and $M_{\mu\nu}$ and $M_{\textrm{gh}}$ are again defined in Eq.~\eqref{eq:Mop}.

For different backgrounds, the topological parts of the Lagrangian are slightly different. Let us list them in the following:
\begin{itemize}
\item For the meron solution:
\be
  S^{\textrm{top}} \big|_{q_0 = 0} = \frac{1}{g^2} \int d^4 x\, \frac{24}{\tau^2} \left[\frac{1}{16} + (\tau \widetilde{q}\,')^2 - \frac{\widetilde{q}\,^2}{2} + \widetilde{q}\,^4 \right]\, .\label{eq:4DactionMeron}
\ee

\item For the instanton and the anti-instanton solution:

In this case, since the classical solution $q_0$ itself is not constant, the explicit expressions are relatively complicated, let us leave them in the original form:
\be
  S^{\textrm{top}} \big|_{q_0 = \pm \frac{\tau - c}{2 (\tau + c)}} = \frac{1}{g^2} \int d^4 x\, \frac{24}{\tau^2} \left[\frac{1}{16} + \left(\tau (q_0 + \widetilde{q})' \right)^2 - \frac{(q_0 + \widetilde{q})^2}{2} + (q_0 + \widetilde{q})^4 \right]_{q_0 = \pm \frac{\tau - c}{2 (\tau + c)}}\, .
\ee

\item For the pure gauge solution and the trivial vacuum solution:

The classical part vanishes completely, i.e., the topological part of the Lagrangian is purely the topological fluctuations.
\be
  S^{\textrm{top}} \big|_{q_0 = \pm \frac{1}{2}} = \frac{1}{g^2} \int d^4 x\, \frac{24}{\tau^2} \left[(\tau \widetilde{q}\,')^2 + \widetilde{q}\,^2 (\widetilde{q} \pm 1)^2 \right]\, .
\ee

\end{itemize}
The complete path integral \eqref{eq:genPartFct-1} should be a sum over all the possible backgrounds listed above. Similar to the 3D case, for the 4D case we should also consider the measure in the path integral for different backgrounds, and the details can be found in Appendix~\ref{topologicalmeasure}.

\subsection{Some Remarks}\label{remarks}

In this subsection, we make a few remarks about the discussions in this section.

\begin{itemize}
\item Finiteness:

We would like to emphasize the finiteness of the topological fluctuations. As we have seen before, that the background configurations of Yang-Mills fields are constrained by the topological properties, i.e., they have fixed boundary conditions. It implies that there cannot be any topological fluctuations at the boundary, which should also be persistent under gauge transformations, i.e.,
\begin{align}
  \widetilde{\mathcal{A}} (\tau=0) & = \widetilde{\mathcal{A}} (\tau = \infty) = 0\, ,\\
  \delta \widetilde{\mathcal{A}} (\tau=0) & = \delta \widetilde{\mathcal{A}} (\tau = \infty) = 0\, .
\end{align}
For some integrals we encounter in this paper over $\tau$ with topological fluctuations, they have apparent divergences at $\tau=0$ or $\tau=\infty$, however, due to the vanishing topological fluctuations at the boundaries these integrals are in fact finite. For convenience, in Section~\ref{approx} we will introduce physical cutoffs to explicitly cure the apparent divergences, but we should keep in mind that the exact theory does not have such divergences and they are cured by topological boundary conditions automatically.

\item Topological Fluctuations vs Quantum Fluctuations:

In this section we have encountered two types of quantum modes, topological fluctuations $\widetilde{\mathcal{A}}$ and quantum fluctuations $Q_\mu$. In the following, we would like to discuss their relations and different roles in the path integral.

To demonstrate their relations, we would like to borrow a well-understood example from condensed matter physics, the 1D quantum antiferromagnets, which is also briefly reviewed in Appendix~\ref{reSpinChain}. In that example, initially there is only one field $\vec{n} (j)$ describing the spin configurations, and then one can write $ \vec{n} (j) = \vec{m} (j) + (-1)^j a_0 \, \vec{l} (j)$, where $\vec{m} (j)$  is a slowly varying mode and $\vec{l} (j)$ is a rapidly varying mode. It is interesting to study the effective theory consisting of both $\vec{m} (j)$ and $\vec{l} (j)$, or to integrate out one of them and obtain another effective theory consisting of only one mode. The topological fluctuations $\widetilde{\mathcal{A}}$ and the quantum fluctuations $Q_\mu$ discussed in this paper are similar to the slowly varying mode $\vec{m}$ and the rapidly varying mode $\vec{l}$ in the example of the 1D quantum antiferromagnets.

Now let us turn to the discussion about the path integral. In principle, one should include all the possible configurations in the path integral, however, for most cases such path integrals cannot be evaluated exactly. The best approximation is to start from some core in the configuration space, which we know very well, and then allow fluctuations around these known configurations. Hopefully, one can then probe the whole configuration space without loss of the relevant physics.

There is a subtlety that one has to pay attention to. The core in the configuration space that we start from can be a point, while sometimes it can also be a finite-dimensional or an infinite-dimensional subspace. These two cases are illustrated in Fig.~\ref{fig:sketch}. They are fundamentally the same, but in order to prevent double counting, one should restrict the fluctuations to the orthogonal space of the core.\footnote{We would like to thank Felix G\"unther for helpful discussions on this issue.}

\begin{figure}[!htb]
\hfill
 \minipage{141.5mm}
\minipage{0.5 \linewidth}
\begin{center}
\includegraphics[height=42mm]{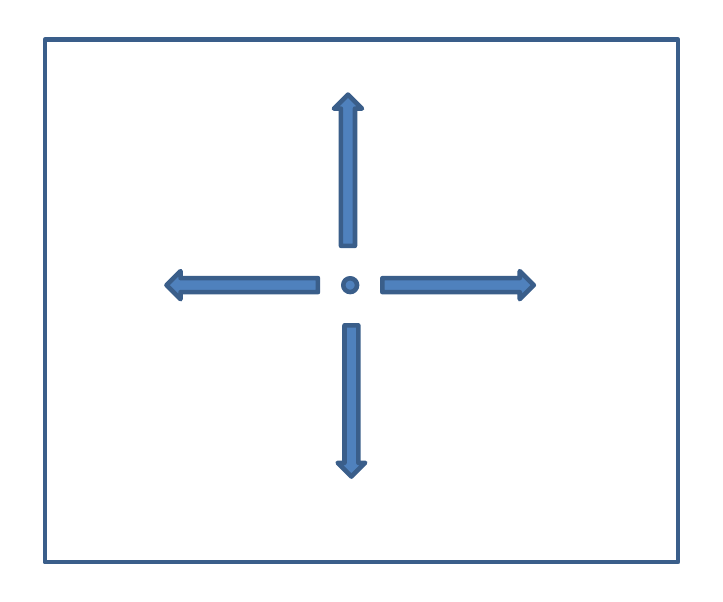}
\end{center}
\endminipage
\minipage{0.5 \linewidth}
\begin{center}
\includegraphics[height=42mm]{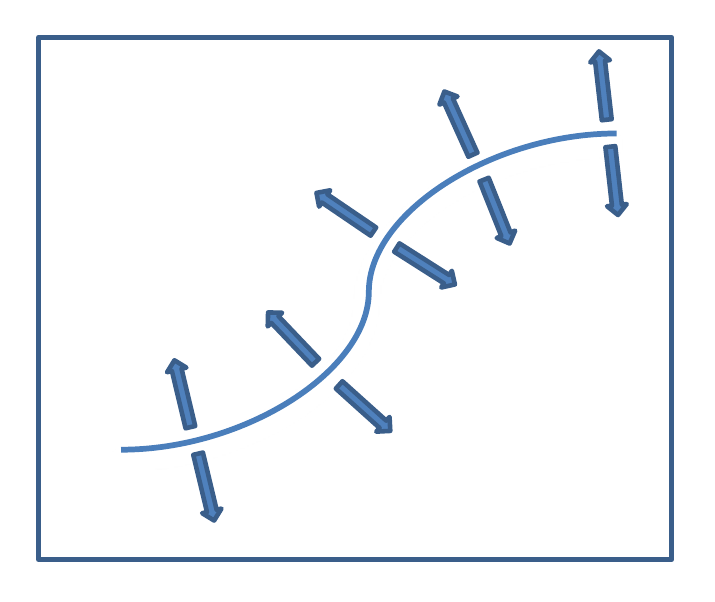}
\end{center}
\endminipage
  \caption{Left:~The configuration space with a single point as the core and fluctuations around it. Right:~The configuration space with a subspace as the core and fluctuations around it.}\label{fig:sketch}
\endminipage
\end{figure}

To study the quantum Yang-Mills theory, there are different choices of the core in the configuration space: One can choose the space of all the classical solutions as in the previous literature, or one can choose the space of all the topologically stable configurations as in this paper. Therefore, the total space of all the configurations, that should be taken into account in the path integral, can be expressed as the following direct sums:
\be
  V = V_{\textrm{sol}} \oplus V_{\textrm{sol}}^\perp  = V_{\textrm{top}} \oplus V_{\textrm{top}}^\perp \, ,
\ee
where $V_{\textrm{sol}}$ and $V_{\textrm{top}}$ stand for the function space of all the solutions and the function space of all the topologically stable configurations respectively, and $V_{\textrm{sol}}^\perp$ and $V_{\textrm{top}}^\perp$ are the orthogonal spaces of $V_{\textrm{sol}}$ and $V_{\textrm{top}}$  respectively. They have the relations:
\be
  V_{\textrm{sol}} \subset V_{\textrm{top}}\, ,\quad V_{\textrm{sol}}^\perp \supset V_{\textrm{top}}^\perp\, .
\ee
Traditionally, one considers the quantum fluctuations around the classical solutions, hence the quantum fluctuations $Q_\mu$ should lie in the space $V_{\textrm{sol}}^\perp$. In this paper, the novelty of our approach is that we consider the quantum fluctuations around the topologically stable configurations including all the classical solutions, therefore, the quantum fluctuations lie in the space $V_{\textrm{top}}^\perp$.

In fact, the formalism discussed here is similar to Feynman's path integral formulation of the 1-dimensional non-relativistic quantum mechanics, which we briefly review in Appendix~\ref{FeynmanPathInt}. In that example, we restrict our discussions to continuous paths without singularities, so $V_{\textrm{top}}$ becomes the space of all the $C^1$-differentiable paths. It is also clear that the space of the classical solutions $V_{\textrm{sol}}$ satisfies
\be
 V_{\textrm{sol}} \subset V_{\textrm{top}} = V_{C^1}\, .
\ee
One can prove \cite{math-1, math-2} that all the continuous but nowhere differentiable paths, i.e. the paths from Brownian motion (random walk), form a space $V_{\textrm{random}}$ that dominates the configuration space. This space also satisfies
\be
  V_{\textrm{random}} \subset V_{\textrm{top}}^\perp\, .
\ee
Hence, we obtain for the 1D non-relativistic quantum mechanics
\be
  V = \Big(V_{\textrm{sol}} \cup (V_{C^1} \backslash V_{\textrm{sol}}) \Big) \oplus \Big(V_{\textrm{random}} \cup \cdots \Big)\, .
\ee
As we discussed in Appendix~\ref{FeynmanPathInt}, this decomposition of the configuration space corresponds to different limits of the theory. When $V_{\textrm{sol}}$ dominates, the theory becomes purely classical. To study the quantum mechanics, one only needs to consider the paths from Brownian motion (random walk) in the space $V_{\textrm{random}}$, as R.~Feynman did in his original paper \cite{FeynmanQM}. However, to probe the IR regime of the quantum theory, both $V_{\textrm{sol}}$ and $V_{C^1} \backslash V_{\textrm{sol}}$ can be important.

One can try to generalize the path integral of the 1D non-relativistic quantum mechanics to other cases such as higher-dimensional relativistic quantum field theory, for instance the 3D and the 4D quantum Yang-Mills theory, but the derivation and the proof become more involved. Nevertheless, the decomposition
\be
  V = \Big(V_{\textrm{sol}} \cup (V_{\textrm{top}} \backslash V_{\textrm{sol}}) \Big) \oplus V_{\textrm{top}}^\perp
\ee
still holds, and the quantum fluctuations $Q_\mu$ should lie in the space $V_{\textrm{top}}^\perp$. For the 1D non-relativistic quantum mechanics, a subspace $V_{\textrm{random}}$ in $V_{\textrm{top}}^\perp$ dominates the configuration space and leads to quantum mechanics, however, in general the quantum fluctuations are not always dominated by continuous but nowhere differentiable functions, because in some cases other configurations in $V_{\textrm{top}}^\perp$, e.g. rough paths, can be more important, in order to lead to the canonical quantization condition.\footnote{We would like to thank J\'ozsef L\"orinczi for helpful discussions on this issue.}

Similar to the example of Feynman's path integral, one can also consider different limits of the more general quantum field theory including the quantum Yang-Mills theory. In the classical limit, all the quantum modes including all the configurations in $V_{\textrm{top}}^\perp$ and $V_{\textrm{top}} \backslash V_{\textrm{sol}}$ are suppressed, and $V_{\textrm{sol}}$ gives us the classical physics in this limit. In the full quantum case, $V_{\textrm{top}}^\perp$ dominates the configuration space and provides the relevant physics, which leads to the ordinary perturbative quantum field theory. To probe the IR regime of the quantum theory, a special limit is relevant, in which one suppresses most quantum modes while still keeping some lowest quantum modes in the space $V_{\textrm{top}} \backslash V_{\textrm{sol}}$, then both $V_{\textrm{sol}}$ and $V_{\textrm{top}} \backslash V_{\textrm{sol}}$ will be important. The transition among different limits will become clearer when we discuss the low-energy physics of the quantum Yang-Mills theory in Section~\ref{LE}.

\end{itemize}

\section{\label{approx}Topological Fluctuations (Approximations)}

As discussed in the previous section, in principle we could integrate out the moduli in the path integral \eqref{eq:genPartFct-2} to obtain an effective theory of the topological modes and the quantum modes. However, in practice it is very hard to obtain the exact analytical result. Hence, in this section we discuss an approximate way of performing the integration over the moduli. Let us illustrate the basic idea of the approximation that we employ.

As we discussed in Subsection~\ref{remarks}, the topological fluctuations in fact do not induce divergences. Hence, it is legitimate to introduce some physical length scales as cutoffs to cure the apparent divergences. We will discuss in Appendix~\ref{topologicalmeasure} that for a careful treatment one has to distinguish the length scale of the classical background, denoted by $\ell_0$, and the averaged length scale of all topological fluctuations, denoted by $\ell_{\textrm{top}}$.

We start with the path integral \eqref{eq:genPartFct-2}:
\begin{displaymath}
  Z = \sum_{\{\mathcal{A}_0\}} \int [\textrm{Jac}]_{\widetilde{\mathcal{A}}}\, \mathcal{D} \widetilde{\mathcal{A}}\, \int [\textrm{Jac}]_{\gamma_i}^{\textrm{top}}\, \mathcal{D} \gamma^{(i)}\, \int \mathcal{D} Q_\mu\, \mathcal{D} b\, \mathcal{D} c\quad e^{- S  [A^{\textrm{cl}}, \, \widetilde{\mathcal{A}},\, Q_\mu,\, b,\, c, \, \gamma_i]} \, ,
\end{displaymath}
where
 \be 
S  [A^{\textrm{cl}}, \, \widetilde{\mathcal{A}},\, Q_\mu,\, b,\, c, \, \gamma_i]  =S_0 [A^{\textrm{cl}}, \,   Q_\mu,\, b,\, c, \, \gamma_i] +  \Delta S  [A^{\textrm{cl}}, \, \widetilde{\mathcal{A}},\, Q_\mu,\, b,\, c, \, \gamma_i] \, ,
\ee 
and $\gamma_i$ denotes translations given by $x_0$ and constant gauge orientations given by $\varphi$. Hence, equivalently,
\be\label{eq:genPartFct-3}
  Z   = \sum_{\{\mathcal{A}_0\}} \int \mathcal{D} Q_\mu\, \mathcal{D} b\, \mathcal{D} c \, \int [\textrm{Jac}]_{\widetilde{\mathcal{A}}}\, \mathcal{D} \widetilde{\mathcal{A}}\,\int   [\textrm{Jac}]_{\gamma_i}^{\textrm{top}}\, d^D x_0  \,  d^3 \varphi \, \, e^{- S  [A^{\textrm{cl}}, \, \widetilde{\mathcal{A}},\, Q_\mu,\, b,\, c, \, \gamma_i]} \, .
\ee

Since in our Ansatz of the topologically stable Yang-Mills field the spacetime indices and the gauge indices are mixed, integrating over all the gauge orientations is equivalent to averaging over all the possible values of the tensor $\omega$, which determines how the indices are mixed. It can be seen as follows. Let us take the 3D case as an example. As shown in Eq.~\eqref{eq:3DRHS}, a constant gauge transformation
\be
  A_\mu \rightarrow V^{-1} A_\mu V
\ee
is equivalent to a rotation of $\omega^a\,_\rho$ by a matrix $U$:
\be
  \omega^a\,_\rho \rightarrow U^a\,_{a'}\, \omega^{a'}\,_\rho\, ,
\ee
where $U$ is defined by
\be
  V^{-1} T_a V = T_b \, U^b\,_a\, .
\ee
Hence, the integration over all the gauge orientations is equivalent to the integration over all the values of $\omega$.

Let us define
\be
  \langle e^{-S} \rangle_\omega \equiv \int   \frac{\sqrt{\textrm{det}\, g (\varphi)}}{2 \pi^2}  d^3 \varphi \, e^{-S}\, ,
\ee
where $ g_{\alpha\beta} (\varphi)$ is the metric on the group manifold.
Then
\be
  \langle e^{-S} \rangle_\omega \geq e^{-\langle S \rangle_\omega}\, .
\ee
Following the argument by R.~Feynman in Ref.~\cite{Feynman}, we can use $e^{-\langle S \rangle_\omega}$ to approximate the original theory, but we would underestimate the free energy, i.e., the approximated free energy will always be greater than or equal to the true free energy. However, we expect that at some values of the parameters the bound in the inequality can be saturated. Hence, in the following we will adopt this approximation and use $e^{-\langle S \rangle_\omega}$ to analyze the theory.

First, we rewrite Eq.~\eqref{eq:genPartFct-3} as
\begin{align}
  Z    & = \sum_{\{\mathcal{A}_0\}} \int \mathcal{D} Q_\mu\, \mathcal{D} b\, \mathcal{D} c \, \int [\textrm{Jac}]_{\widetilde{\mathcal{A}}}\, \mathcal{D} \widetilde{\mathcal{A}}\,\int   \frac{ [\textrm{Jac}]_{\gamma_i}^{\textrm{top}}}{[\textrm{Jac}]_{\gamma_i}^{\textrm{cl}}} \,  [\textrm{Jac}]_{\gamma_i}^{\textrm{cl}}   \, d^D x_0 \, d^3 \varphi \,   e^{- S   }  \nonumber\\
 & \sim \sum_{\{\mathcal{A}_0\}} \int \mathcal{D} Q_\mu\, \mathcal{D} b\, \mathcal{D} c \, \int  \frac{ [\textrm{Jac}]_{\gamma_i}^{\textrm{top}}}{[\textrm{Jac}]_{\gamma_i}^{\textrm{cl}}} \, \mathcal{D} \widetilde{\mathcal{A}}\,\int   [\textrm{Jac}]_{x_0}^{\textrm{cl}}\, d^\textrm{D} x_0 \int   [\textrm{Jac}]_{\varphi}^{\textrm{cl}}\, d^3 \varphi   \,   e^{- S   }  \, ,\label{eq:Approximationbefore}
\end{align}
where we dropped $[\textrm{Jac}]_{\widetilde{\mathcal{A}}}$ for convenience, because it is a constant (see Appendix~\ref{topologicalmeasure}), and we used the fact that
\be
 \frac{ [\textrm{Jac}]_{\gamma_i}^{\textrm{top}}}{[\textrm{Jac}]_{\gamma_i}^{\textrm{cl}}}\,\, \textrm{ is independent of $x_0$ and $\varphi$}\, .
\ee
 We also write
\be
   [\textrm{Jac}]_{\gamma_i}^{\textrm{cl}} =    [\textrm{Jac}]_{x_0}^{\textrm{cl}}     [\textrm{Jac}]_{\varphi}^{\textrm{cl}} \, ,
\ee 
which corresponds to the classical measure for the translations and for the gauge orientations respectively, when the topological fluctuations are turned off. Up to a constant there is
\be
  [\textrm{Jac}]_{\varphi}^{\textrm{cl}} \sim \frac{\sqrt{\textrm{det} \, g (\varphi)}}{2\pi^2}
\ee
except for the instanton and the anti-instanton solution, for which we find that up to a constant
\be
  [\textrm{Jac}]_{\varphi}^{\textrm{cl}} \sim \frac{\sqrt{\textrm{det} \, g (\varphi)}}{2\pi^2} \rho^3 \, ,
\ee
where $\rho$ is the size of the (anti-)instanton. For simplicity, we first consider the 3D Wu-Yang Monopole solution or the 4D meron solution, which does not have the size $\rho$, and the general results for other cases with $\rho$ will be listed in the end.

Applying the approximation by R.~Feynman \cite{Feynman} to Eq.~\eqref{eq:Approximationbefore}, we obtain
\begin{align}
  Z    & \sim  \sum_{\{\mathcal{A}_0\}} \int \mathcal{D} Q_\mu\, \mathcal{D} b\, \mathcal{D} c \, \int  \frac{ [\textrm{Jac}]_{\gamma_i}^{\textrm{top}}}{[\textrm{Jac}]_{\gamma_i}^{\textrm{cl}}} \, \mathcal{D} \widetilde{\mathcal{A}}\,\int   [\textrm{Jac}]_{x_0}^{\textrm{cl}}\, d^\textrm{D} x_0    \,\, \left< e^{- S  [A^{\textrm{cl}}, \, \widetilde{\mathcal{A}},\, Q_\mu,\, b,\, c, \, \gamma_i]}  \right>_\omega\nonumber\\
 & \approx   \sum_{\{\mathcal{A}_0\}} \int \mathcal{D} Q_\mu\, \mathcal{D} b\, \mathcal{D} c \, \int  \frac{ [\textrm{Jac}]_{\gamma_i}^{\textrm{top}}}{[\textrm{Jac}]_{\gamma_i}^{\textrm{cl}}} \, \mathcal{D} \widetilde{\mathcal{A}}\,\int   [\textrm{Jac}]_{x_0}^{\textrm{cl}}\, d^\textrm{D} x_0     \,\, e^{- \left<S  [A^{\textrm{cl}}, \, \widetilde{\mathcal{A}},\, Q_\mu,\, b,\, c, \, \gamma_i] \right>_\omega} \, .\label{eq:Approximation}
\end{align}
The path integral above can be further expressed as (see Eq.~\eqref{eq:secondseparation})
\be
  Z \approx \sum_{\{\mathcal{A}_0\}} \int[\textrm{Jac}]_{\gamma_i}^{\textrm{cl}} \, d^{\textrm{D}} x_0 \, \int \mathcal{D} Q_\mu\, \mathcal{D} b\, \mathcal{D} c\, e^{- \langle S_0   \rangle_{\omega}} \int   \mathcal{D} \widetilde{\mathcal{A}} \,   \frac{[\textrm{Jac}]_{\gamma_i}^{\textrm{top}}}{[\textrm{Jac}]_{\gamma_i}^{\textrm{cl}}}  \, e^{- \langle \Delta S \rangle_\omega } \, ,
\ee
where $S_0$ is the part of the action that is independent of the topological fluctuations, while $\Delta S$ depends on the topological fluctuations.  We can make the following field redefinition:
\be
  \mathcal{D} \widetilde{\mathcal{A}} \, \frac{[\textrm{Jac}]_{\gamma_i}^{\textrm{top}}}{[\textrm{Jac}]_{\gamma_i}^{\textrm{cl}}} = \mathcal{D} \widetilde{\mathcal{A}}'\, ,
\ee
then effectively the path integral can be written as
\be
  Z \approx \sum_{\{\mathcal{A}_0\}} \int [\textrm{Jac}]_{\gamma_i}^{\textrm{cl}}\, d^\textrm{D} x_0 \, \int \mathcal{D} Q_\mu\, \mathcal{D} b\, \mathcal{D} c\, e^{- \langle S_0 [A^{\textrm{cl}}, \,   Q_\mu,\, b,\, c, \, \gamma_i] \rangle_\omega } \int  \mathcal{D} \widetilde{\mathcal{A}}' \, e^{- \langle \Delta S [A^{\textrm{cl}}, \, \mathcal{C} \widetilde{\mathcal{A}}',\, Q_\mu,\, b,\, c, \, \gamma_i] \rangle_\omega } \, ,
\ee
where the change of variable is equivalent to introducing the factor $\mathcal{C}$, which can be thought of as the wave function renormalization and depends on the energy scale. In general, the renormalization factor $\mathcal{C}$ may contain infinities, which can be made finite by adding appropriate counter-terms to cancel the divergence. We will skip the discussion about the counter-term in this paper, and simply consider the finite part of the factor $\mathcal{C}$.

From now on, we drop the prime of $\widetilde{\mathcal{A}}'$ for simplicity.  More explicitly, the path integrals that we encounter in this paper have the following general expressions under the approximation:
\begin{itemize}
\item 3D (Wu-Yang monopole, pure gauge):
\be
  Z \approx \sum_{\{\mathcal{A}_0\}} \int [\textrm{Jac}]_{\gamma_i}^{\textrm{cl}}\, d^3 x_0 \, \int \mathcal{D} Q_\mu\, \mathcal{D} b\, \mathcal{D} c\, e^{- \langle S_0  \rangle_\omega } \int  \mathcal{D} \widetilde{ G } \, \mathcal{D} \widetilde{ H } \, e^{- \langle \Delta S [A^{\textrm{cl}}, \, \mathcal{C}\widetilde{G}, \, \mathcal{C}\widetilde{H},\, Q_\mu,\, b,\, c, \, \gamma_i]  \rangle_\omega } \, .
\ee

\item 4D (trivial solution, pure gauge, meron):
\be
  Z \approx \sum_{\{\mathcal{A}_0\}} \int [\textrm{Jac}]_{\gamma_i}^{\textrm{cl}}\, d^4 x_0 \, \int \mathcal{D} Q_\mu\, \mathcal{D} b\, \mathcal{D} c\, e^{- \langle S_0   \rangle_\omega } \int  \mathcal{D} \widetilde{q} \, e^{- \langle \Delta S   [A^{\textrm{cl}}, \, \mathcal{C}\widetilde{q} ,\, Q_\mu,\, b,\, c, \, \gamma_i]  \rangle_\omega } \, .
\ee

\item 4D (instanton, anti-instanton):
\be
  Z \approx \sum_{\{\mathcal{A}_0\}} \int   [\textrm{Jac}]_{\rho}^{\textrm{cl}} \, \rho^3 d \rho  \int [\textrm{Jac}]_{\gamma_i}^{\textrm{cl}}\, d^4 x_0 \, \int \mathcal{D} Q_\mu\, \mathcal{D} b\, \mathcal{D} c\, e^{- \langle S_0   \rangle_\omega } \int  \mathcal{D} \widetilde{q} \, e^{- \langle \Delta S   [A^{\textrm{cl}}, \,   \mathcal{C}\widetilde{q},\, Q_\mu,\, b,\, c, \, \gamma_i,\, \rho] \rangle_\omega } \, ,
\ee
where we notice that for the 4D (anti-)instanton background there is one more modulus $\rho$ in the classical solution $q_0$. 

\end{itemize}
We will see in the following subsections, that using this approximation to get rid of the moduli of the gauge orientations significantly simplifies the expression of the effective action, which makes the relevant physics more transparent.

\subsection{3D Case}\label{3dapprox}
In this subsection, we discuss the 3D Yang-Mills theory under the approximation \eqref{eq:Approximation}. Following the discussions above, we would like to compute the averaged action under the gauge orientations.

Recall that for the 3D Yang-Mills, the full quantum theory is given by the Lagrangian \eqref{eq:FullLag}:
\begin{displaymath}
  \mathcal{L} = - \textrm{Tr} \left( \frac{1}{2} F_{\mu\nu} F_{\mu\nu} + 2 Q_\mu \left(D_\nu F_{\mu\nu} \right) - Q_\mu M_{\mu\nu} Q_\nu - 2 b M_{\textrm{gh}} c + 2 [Q_\mu,\, Q_\nu] (D_\mu Q_\nu) + \frac{1}{2} [Q_\mu,\, Q_\nu]^2 \right)\, .
\end{displaymath}
To calculate $\langle S \rangle_\omega$, we make use of the following identities proven in Appendix~\ref{omegarotation}:
\begin{align}
  \langle \omega^a\,_\mu \rangle_\omega & = 0\, ,\nonumber\\
  \langle \omega^a\,_{(\mu}\, \omega^b\,_{\nu)} \rangle_\omega & = \frac{1}{3} \delta^{ab} \delta_{\mu\nu}\, ,\label{eq:omegaId}
\end{align}
 where the bracket $(\cdots)$ denotes the symmetrization of the indices. Since all the gauge indices can be traced back to either from $\omega^a\,_\mu$ or from $Q^a\,_\mu$, we can keep track of the gauge indices to see which identity above is needed when we average over $\omega$. The results are
\begin{align}
  \Big\langle \textrm{Tr} \left( \frac{1}{2} F_{\mu\nu} F_{\mu\nu}\right) \Big\rangle_\omega & = \textrm{Tr} \left(   \frac{1}{2} F_{\mu\nu} F_{\mu\nu} \right)\, ,\label{eq:Average-1}\\
  \langle 2 Q_\mu \left(D_\nu F_{\mu\nu} \right) \rangle_\omega & = 0\, ,\\
  \big\langle 2 [Q_\mu,\, Q_\nu] (D_\mu Q_\nu) \big\rangle_\omega & = 2 [Q_\mu,\, Q_\nu] (\partial_\mu Q_\nu)\, ,\\
  \Big\langle \frac{1}{2} [Q_\mu,\, Q_\nu]^2 \Big\rangle_\omega & = \frac{1}{2} [Q_\mu,\, Q_\nu]^2\, .\label{eq:Average-4}
\end{align}
For the terms $- Q_\mu M_{\mu\nu} Q_\nu$ and $2 b M_{\textrm{gh}} c$ let us recall Eq.~\eqref{eq:Mop}:
\begin{align}
  M_{\mu\nu} & \equiv \mathcal{M}_{\mu\nu} + D_\mu D_\nu = D^2 \delta_{\mu\nu} + 2 F_{\mu\nu}\, ,\nonumber\\
  M_{\textrm{gh}} & \equiv D^2\, .\nonumber
\end{align}
Hence,
\begin{align}
\textrm{Tr} \langle Q_\mu M_{\mu\nu} Q_\nu \rangle_\omega & = -\frac{1}{2} \Big\langle Q_\mu^a (D_\rho D_\rho Q_\mu)^a \Big\rangle_\omega-\Big\langle Q_\mu^a \, [F_{\mu\nu}, Q_\nu]^a \Big\rangle_\omega \nonumber\\
  {} & = - \frac{1}{2} Q_\mu^a \partial^2 Q_\mu^a + \frac{1}{3} (Q_\mu^a Q_\mu^a) (A_\nu^b A_\nu^b)\, ,
\end{align}
where we used $\big\langle A_\rho^a A_\rho^b \big\rangle_\omega = (\delta_{a b}/3)  A_\rho^c A_\rho^c$. Similarly,
\be
 \textrm{Tr} \big\langle 2 b M_{\textrm{gh}} c \big\rangle_\omega = -  b^a \partial^2 c^a + \frac{2}{3} (b^a c^a) (A_\nu^b A_\nu^b)\, ,
\ee
where
\be
  A_\nu^b A_\nu^b = \frac{2}{\tau} \left[\left(G_0 +  \mathcal{C}  \widetilde{G} \right)^2 + \left(H_0 +  \mathcal{C}  \widetilde{H} - 1 \right)^2 \right]\, .
\ee
Bringing all the terms after the average together, we obtain
\begin{align}
  \langle S \rangle_\omega & = \frac{ 1}{g^2} \int d^3 x\, \textrm{Tr} \Bigg( - \frac{1}{2} F_{\mu\nu} F_{\mu\nu} + Q_\mu \partial^2 Q_\mu - \frac{4}{3 \tau} (Q_\mu Q_\mu) \left[(G_0 +  \mathcal{C} \widetilde{G})^2 + (H_0 +  \mathcal{C} \widetilde{H} - 1)^2 \right] \nonumber\\
  {} & \qquad\qquad\qquad\quad  + 2 b \partial^2 c  - \frac{8}{3\tau} (bc) \left[(G_0 +  \mathcal{C} \widetilde{G})^2 + (H_0 +  \mathcal{C} \widetilde{H} - 1)^2 \right]  \nonumber\\
  {} & \qquad\qquad\qquad\quad - 2 [Q_\mu, Q_\nu] (\partial_\mu Q_\nu) - \frac{1}{2} [Q_\mu, Q_\nu]^2 \Bigg)\, ,
\label{eq:3DFirstAve}
\end{align}
where the trace is taken over the gauge indices, which the topological fluctuations $\widetilde{G}$ and $\widetilde{H}$ do not have.

The effective action above is still relatively complicated. Since the factor $\mathcal{C}$ is finite, the effective action can be simplified in some special limits of $\mathcal{C}$, which correspond to different corners in the moduli space. In general, we distinguish three cases:
\begin{itemize}

\item $\mathcal{C} \ll 1$:\\
 In this case the topological fluctuations are not important. One only needs to consider the classical backgrounds.
\begin{align}
  \langle S \rangle_\omega & \approx   \langle S_0 \rangle_\omega \nonumber\\
& = \frac{ 1}{g^2} \int d^3 x\, \textrm{Tr} \Bigg( - \frac{1}{2} F_{\mu\nu} F_{\mu\nu} +  Q_\mu \partial^2 Q_\mu - \frac{4}{3 \tau} (Q_\mu Q_\mu) \left[G_0^2 + (H_0   - 1)^2 \right]  + 2 b \partial^2 c \nonumber\\
  {} & \qquad\qquad\qquad\quad   - \frac{8}{3\tau} (bc) \left[G_0^2 + (H_0  - 1)^2 \right]  - 2 [Q_\mu, Q_\nu] (\partial_\mu Q_\nu) - \frac{1}{2} [Q_\mu, Q_\nu]^2 \Bigg)\, ,
\end{align}
where
\[  \frac{1}{4} F_{\mu\nu}^{\textrm{cl}, a} F_{\mu\nu}^{\textrm{cl}, a} = \left\{ \begin{array}{ll}
          \frac{1}{2 \tau^2} & \mbox{for the Wu-Yang monopole};\\
        0 & \mbox{for the trivial and the pure gauge solution}.\end{array} \right. \]
We see that in this limit the topological fluctuations are effectively turned off. To estimate the contributions from the classical background, we can replace  $\sqrt{\tau}$ by the classical background length scale $\ell_{\textrm{cl}} \equiv \ell_0$ in the effective action, as discussed in Appendix~\ref{topologicalmeasure}. Then the quantum fluctuations $Q_\mu$ and the ghosts acquire masses of the order $\sim \ell_0^{-1}$ from the classical background. To carefully calculate the masses, one needs to properly regularize the divergence due to $1/\tau$ and then sum over all the classical backgrounds.

 \item $\mathcal{C} \gg 1$:\\
 In this case the topological fluctuations are important, while the classical backgrounds can be neglected.
\begin{align}
   \langle S_0 \rangle_\omega & \approx \frac{ 1}{g^2} \int d^3 x\, \textrm{Tr} \Bigg(    Q_\mu \partial^2 Q_\mu     + 2 b \partial^2 c  - 2 [Q_\mu, Q_\nu] (\partial_\mu Q_\nu) - \frac{1}{2} [Q_\mu, Q_\nu]^2 \Bigg) \nonumber\\
& \approx \frac{-1}{g^2} \int d^3 x\, \textrm{Tr} \Bigg( \frac{1}{2} G_{\mu\nu}G_{\mu\nu} + (\partial_\mu Q_\mu)^2   - 2 b \partial^2 c    \Bigg) \, ,
\end{align}
where
\be
G_{\mu\nu} = \partial_\mu Q_\nu - \partial_\nu Q_\mu + [Q_\mu, Q_\nu] \, ,
\ee
 and
\begin{align}\label{eq:DeltaSlargeCeq1}
  \langle \Delta S \rangle_\omega & \approx  \frac{- 1}{g^2} \int d^3 x\, \textrm{Tr} \Bigg(  \frac{1}{2} F_{\mu\nu} F_{\mu\nu}  + \frac{4 \mathcal{C}^2 }{3 \tau} (Q_\mu Q_\mu) \big| \widetilde{\psi}\big|^2     + \frac{8\mathcal{C}^2}{3\tau} (bc)\big| \widetilde{\psi}\big|^2 \Bigg) \, ,
\end{align}
where 
\begin{align}\la{eq:fmunueq1}
 \frac{1}{4} F_{\mu\nu}^a F_{\mu\nu}^a   & \approx     4 \mathcal{C}^2 \big| \partial_\tau \widetilde{\psi}\big|^2    - \frac{\mathcal{C}^2 }{\tau^2}  \big| \widetilde{\psi}\big|^2 + \frac{\mathcal{C}^4}{2 \tau^2}\big| \widetilde{\psi}\big|^4 \nonumber\\
 & =    \frac{\mathcal{C}^2}{\tau}    \big| \partial_\mu\widetilde{\psi}\big|^2    - \frac{\mathcal{C}^2 }{\tau^2}  \big| \widetilde{\psi}\big|^2 + \frac{\mathcal{C}^4}{2 \tau^2}\big| \widetilde{\psi}\big|^4  \, ,
\end{align}
and
\be
 \widetilde{\psi} =  
  \widetilde{G}   + 
 i \widetilde{H} \, .
\ee

To analyze the theory in this case, we have to make a detour of discussion. First, we would like to emphasize that Eq.~\eqref{eq:fmunueq1} is well-defined and convergent everywhere. Though $1/\tau$ and $1/\tau^2$ seem to be divergent at $\tau=0$, the topological fluctuations $\widetilde{\psi}$, which we assume to be $C^1$-functions in this paper, vanish at $\tau = 0$ due to the boundary conditions, which will cure the apparent divergence caused by $1/\tau$ or $1/\tau^2$.

$\widetilde{\psi}$ now is a function of $\tau = (x- x_0)^2$, while the quantum mode $Q_\mu(x)$ is a function of $x$ that can be coupled to an external source $J_\mu(x)$, which makes the calculation very difficult. Moreover, the topological fluctuations are part of the backgrounds without the external source, hence they should be translationally invariant for most parts of the space. Therefore, we would like to make the shift:
\be\la{eq:psitranslate}
\widetilde{\psi} \left((x - x_0)^2 \right) \longrightarrow \widetilde{\psi} (x^2)\, .
\ee
However, the topological fluctuations are constrained by the topological boundary conditions, therefore, $\widetilde{\psi}$ is translationally invariant for all the space except the boundaries. Thus, for both before and after the shift \eqref{eq:psitranslate} we can write:
\begin{align}
& \int d^3 x \, \left( \frac{\mathcal{C}^2}{\tau}    \big| \partial_\mu\widetilde{\psi}\big|^2    - \frac{\mathcal{C}^2 }{\tau^2}  \big| \widetilde{\psi}\big|^2 + \frac{\mathcal{C}^4}{2 \tau^2}\big| \widetilde{\psi}\big|^4  \right) \nonumber\\
  = & \left(    \int_{\textrm{near}\,\, x_0} d^3 x   +  \int_{\textrm{near}\,\, 0 } d^3 x  +  \int_{\textrm{else} } d^3 x  \right) \, \left( \frac{\mathcal{C}^2}{\tau}    \big| \partial_\mu\widetilde{\psi} \big|^2    - \frac{\mathcal{C}^2 }{\tau^2}  \big| \widetilde{\psi}\big|^2 + \frac{\mathcal{C}^4}{2 \tau^2}\big| \widetilde{\psi}\big|^4  \right)\, ,\label{eq:psiIntegral}
\end{align}
where the regions near $x_0$ and $0$ can be taken to be the spherical regions centered at $x_0$ and $0$ with the radius $r_0^2$.   We would like to emphasize that this relation is true only inside the path integral, i.e., the integration over the function space of $\widetilde{\psi}$ is implied. For simplicity, we drop the integration over $\widetilde{\psi}$ in the equation above. All the following discussions should be understood with an integration over $\widetilde{\psi}$ assumed.

Away from the regions near $x_0$ and $0$ there is
\begin{align}
& \int_{\textrm{else} } d^3 x \,  \left( \frac{\mathcal{C}^2}{\tau}    \big| \partial_\mu\widetilde{\psi} (x - x_0) \big|^2    - \frac{\mathcal{C}^2 }{\tau^2}  \big| \widetilde{\psi} (x - x_0) \big|^2 + \frac{\mathcal{C}^4}{2 \tau^2 }\big| \widetilde{\psi}(x - x_0)\big|^4  \right)    \nonumber\\
= & \int_{\textrm{else} } d^3 x \,  \left( \frac{\mathcal{C}^2}{\tau}    \big| \partial_\mu\widetilde{\psi} (x ) \big|^2    - \frac{\mathcal{C}^2 }{\tau^2}  \big| \widetilde{\psi} (x  ) \big|^2 + \frac{\mathcal{C}^4}{2 \tau^2 }\big| \widetilde{\psi}(x )\big|^4  \right) \, .
\end{align}
Notice that for $\tau = (x - x_0)^2$ the integrand
\be
 \frac{\mathcal{C}^2}{\tau}    \big| \partial_\mu\widetilde{\psi} (x - x_0) \big|^2    - \frac{\mathcal{C}^2 }{\tau^2}  \big| \widetilde{\psi} (x - x_0) \big|^2 + \frac{\mathcal{C}^4}{2 \tau^2 }\big| \widetilde{\psi}(x - x_0)\big|^4
\ee
is not divergent at $x = 0$ and $x = x_0$. Hence, for a very small region $r_0^2 \to 0$, the integral
\be
 \left(    \int_{\textrm{near}\,\, x_0} d^3 x   +  \int_{\textrm{near}\,\, 0 } d^3 x    \right) \, \left( \frac{\mathcal{C}^2}{\tau}    \big| \partial_\mu\widetilde{\psi} (x - x_0) \big|^2    - \frac{\mathcal{C}^2 }{\tau^2}  \big| \widetilde{\psi} (x - x_0) \big|^2 + \frac{\mathcal{C}^4}{2 \tau^2 }\big| \widetilde{\psi}(x - x_0)\big|^4 \right)
\ee
is negligible. Similarly, when $r_0^2 \to 0$, the integral
\be
  \int_{\textrm{near}\,\, 0} d^3 x  \,  \left( \frac{\mathcal{C}^2}{\tau}    \big| \partial_\mu\widetilde{\psi} (x ) \big|^2    - \frac{\mathcal{C}^2 }{\tau^2}  \big| \widetilde{\psi} (x  ) \big|^2 + \frac{\mathcal{C}^4}{2 \tau^2 }\big| \widetilde{\psi}(x )\big|^4  \right)
\ee 
also vanishes. In sum, we have
\begin{align}
& \frac{1}{g^2} \int  d^3 x \,  \left( \frac{\mathcal{C}^2}{\tau}    \big| \partial_\mu\widetilde{\psi} (x - x_0) \big|^2    - \frac{\mathcal{C}^2 }{\tau^2}  \big| \widetilde{\psi} (x - x_0) \big|^2 + \frac{\mathcal{C}^4}{2 \tau^2 }\big| \widetilde{\psi}(x - x_0)\big|^4  \right)    \nonumber\\
= & \, \frac{1}{g^2} \int  d^3 x \,  \left( \frac{\mathcal{C}^2}{\tau}    \big| \partial_\mu\widetilde{\psi} (x ) \big|^2    - \frac{\mathcal{C}^2 }{\tau^2}  \big| \widetilde{\psi} (x  ) \big|^2 + \frac{\mathcal{C}^4}{2 \tau^2 }\big| \widetilde{\psi}(x )\big|^4  \right)    \nonumber\\
& -  \frac{1}{g^2} \int_{\textrm{near}\,\, x_0} d^3 x  \,  \left( \frac{\mathcal{C}^2}{\tau}    \big| \partial_\mu\widetilde{\psi} (x ) \big|^2    - \frac{\mathcal{C}^2 }{\tau^2}  \big| \widetilde{\psi} (x  ) \big|^2 + \frac{\mathcal{C}^4}{2 \tau^2 }\big| \widetilde{\psi}(x )\big|^4  \right)\, , \label{eq:finiteintegrals}
\end{align}
where $\tau = (x-x_0)^2$. The left-hand side of this equation is finite. The second term on the right-hand side is divergent and gives the difference of the integral \eqref{eq:psiIntegral} near $x_0$ before and after the shift \eqref{eq:psitranslate}, hence it can be viewed as a counter-term, that cancels the divergence of the first term on the right-hand side. In this paper, instead of including the counter-term (i.e. the second term on the right-hand side) explicitly, we will use the first term on the right-hand side as an effective action with some physical cutoffs, that regularize the theory. Nevertheless, the final result is finite and equal to the left-hand side. Because these cutoffs are crucial in the following, we will discuss them in more detail in Subsection~\ref{cutoff}.

As we have shown, the effective action
\be\label{eq:effectiveresult}
  \frac{1}{g^2} \int  d^3 x \,  \left( \frac{\mathcal{C}^2}{\tau}    \big| \partial_\mu\widetilde{\psi} (x ) \big|^2    - \frac{\mathcal{C}^2 }{\tau^2}  \big| \widetilde{\psi} (x  ) \big|^2 + \frac{\mathcal{C}^4}{2 \tau^2 }\big| \widetilde{\psi}(x )\big|^4  \right)
\ee
has divergence, which can be regularized by a UV cutoff $\ell_{\textrm{top}}$ and an IR cutoff $L_{\textrm{top}}$. In a weak-field approximation, we have the expansion $e^{-\langle S\rangle_\omega} \approx 1 - \langle S\rangle_\omega + \cdots$, thus we can perform the integration over $x_0$ directly to $\langle S \rangle_\omega$ in this approximation.  Then we have
\be
\frac{1}{g^2}  \int \frac{d^3 x_0}{V}  \int  d^3 x \,  \left( \frac{\mathcal{C}^2}{\tau}    \big| \partial_\mu\widetilde{\psi} (x ) \big|^2    - \frac{\mathcal{C}^2 }{\tau^2}  \big| \widetilde{\psi} (x  ) \big|^2 + \frac{\mathcal{C}^4}{2 \tau^2 }\big| \widetilde{\psi}(x )\big|^4  \right) \, .\label{eq:integrateoverx0}
\ee
The integration over $x_0$ can be done explicitly, which is equivalent to the following replacements in the limit $L_{\textrm{top}} \gg \ell_{\textrm{top}}$: 
\be
\int \frac{d^3 x_0}{V} \,  \frac{1}{\tau} \rightarrow \frac{3 L_{\textrm{top}}}{L^3} \quad\quad \textrm{and} \quad\quad \int \frac{d^3 x_0}{V} \,  \frac{1}{\tau^2} \rightarrow \frac{3}{L^3 \ell_{\textrm{top}}}\, ,
\ee
where $ \ell_{\textrm{top}}$ is the cutoff at $\tau = 0$, and $L_{\textrm{top}}$  is the cutoff at $\tau = \infty$, both of which are due to the topological boundary conditions. $L$ is the radius of the system, which is of the order of $L_{\textrm{top}}$, but we carefully distinguish them in the paper.  In Appendix~\ref{topologicalmeasure}, we give a qualitative explanation that $ \ell_{\textrm{top}} <  \ell_{0}$ because of the excited topological fluctuation modes.
Defining a dimensionful complex scalar with the unit of mass
\be
\phi \equiv    \frac{\mathcal{C}}{g} \sqrt{\frac{3 L_{\textrm{top}}}{L^3}}  \widetilde{\psi} \, ,
\ee
we obtain
\begin{align}
  \int d^3 x \,    \left[   (\overline{\partial_\mu \phi }) (\partial_\mu \phi ) -  \mu^2_{3D}   \overline{\phi } \phi  +  \lambda_{3D} \left(  \overline{\phi } \phi  \right)^2  \right] \, ,
\end{align}
where 
\be
  \mu^2_{3D} \equiv \frac{1}{L_{\textrm{top}} \,  \ell_{\textrm{top}} }\, ,\quad \textrm{and} \quad \,\lambda_{3D} \equiv \frac{g^2 L^3}{ 6 \, L_{\textrm{top}}^2 \ell_{\textrm{top}}}
\ee
are two constants that can be fixed by comparing with the experimental results or the Monte Carlo simulations.

Let us introduce the parameterization:
\be
\phi  = \frac{\nu_{3D} + \eta (x)}{\sqrt{2}}\, e^{i \gamma(x)}\, ,
\ee
where
\be
\nu^2_{3D} \equiv \mu^2_{3D} / \lambda_{3D}  \, . 
\ee
 
We obtain
\begin{align}
\int d^3 x \,   \left(  \frac{1}{2}  (\partial_\mu \eta)^2  +   \mu^2_{3D} \eta^2  +   \lambda_{3D}\, \nu_{3D}\, \eta^3 +  \frac{\lambda_{3D}}{4}  \eta^4  + \frac{1}{2} (\nu_{3D} + \eta)^2 (\partial_\mu \gamma)^2   - \frac{\lambda_{3D}\, \nu^4_{3D}}{4} \right)\, ,
\end{align}
where $\gamma$ is a Goldstone boson and $\eta$ acts like a Higgs boson with mass $m_\eta = \sqrt{2}\, \mu_{3D}$. 

After neglecting the constant shift $\lambda_{3D}\, \nu^4_{3D} / 4$, Eq.~\eqref{eq:DeltaSlargeCeq1} reads 
\begin{align}
  \langle \Delta S \rangle_\omega & \approx      \int d^3 x\,  \Bigg(    \frac{1}{2}  (\partial_\mu \eta)^2  +   \mu^2_{3D} \eta^2  +   \lambda_{3D}\, \nu_{3D}\, \eta^3 +  \frac{\lambda_{3D}}{4}  \eta^4  + \frac{1}{2} (\nu_{3D} + \eta)^2 (\partial_\mu \gamma)^2  \nonumber\\
  {} & \qquad\qquad\quad + \frac{1}{3  } (Q_\mu^a Q_\mu^a) (\nu_{3D}  + \eta)^2    +  \frac{2 }{3 } (b^ac^a) (\nu_{3D} + \eta)^2 \Bigg) \, ,\label{eq:3dafterinteg}
\end{align}
where the gauge field $Q_\mu$ acquires a mass $ m_Q^{3D} =\sqrt{2/3}\, \nu_{3D} $. This expression is similar to the Higgs mechanism in the Lorentz gauge.   If one chooses a gauge  similar to the unitary gauge before the averaging over the gauge orientations, we believe that the Goldstone boson $\gamma$ can be absorbed into the redefinition of the gauge boson $Q_\mu$, which is the case in the usual Higgs mechanism. Finally, we can extend our result here to the general case and use
\be
1 -   \langle \Delta S \rangle_\omega \approx \exp(- \langle \Delta S \rangle_\omega)\, ,
\ee
to rewrite $\langle \Delta S \rangle_\omega$ back to the exponent.

 \item $\mathcal{C} \sim 1$:\\
 Both topological fluctuations and classical backgrounds are important. One needs to calculate for each background and sum over all of them. So we return to the path integral \eqref{eq:3DFirstAve} without any simplification. Its explicit form could be very complicated, and we will not discuss the details in this paper.


\end{itemize}

\subsection{4D Case}\label{4dapprox}
Following the same spirit of the 3D case, we can work out the 4D case under the approximation. Since in the 4D Ansatz \eqref{eq:4DAnsatz} the $\omega$ tensor mixes the gauge group indices and the spacetime indices of a 3D subspace, the identities \eqref{eq:omegaId} still hold:
\begin{align}
  \langle \omega^a\,_i \rangle_\omega & = 0\, ,\nonumber\\
  \langle \omega^a\,_{(i}\, \omega^b\,_{j)} \rangle_\omega & = \frac{1}{3} \delta^{ab} \delta_{i j}\,,\nonumber
\end{align}
where the indices $i$ and $j$ run from $1$ to $3$. Thus, the results of the averaged terms \eqref{eq:Average-1} $\sim$ \eqref{eq:Average-4} remain valid for the 4D case. We only need to calculate $- Q_\mu M_{\mu\nu} Q_\nu$ and $2 b M_{\textrm{gh}} c$, where again \eqref{eq:Mop}:
\begin{align}
  M_{\mu\nu} & \equiv \mathcal{M}_{\mu\nu} + D_\mu D_\nu = D^2 \delta_{\mu\nu} + 2 F_{\mu\nu}\, ,\nonumber\\
  M_{\textrm{gh}} & \equiv D^2\, .\nonumber
\end{align}
The same results as the 3D case can also be obtained for the 4D case
\begin{align} 
  \textrm{Tr} \langle Q_\mu M_{\mu\nu} Q_\nu \rangle_\omega & =- \frac{1}{2} Q_\mu^a \partial^2 Q_\mu^a + \frac{1}{3} (Q_\mu^a Q_\mu^a) (A_\nu^b A_\nu^b)\, ,\\
  \textrm{Tr} \big\langle 2 b M_{\textrm{gh}} c \big\rangle_\omega & = - b^a \partial^2 c^a + \frac{2}{3} (b^a c^a) (A_\nu^b A_\nu^b)\, ,
\end{align}
except for the 4D case:
\be
  A_\nu^a A_\nu^a = \frac{12\, (q_0 + \mathcal{C} \widetilde{q} + \frac{1}{2})^2}{\tau}\, .
\ee
Hence,
\begin{align}
 \langle S \rangle_\omega & = \frac{1}{g^2}\int d^4 x\, \textrm{Tr} \Bigg( -\frac{1}{2} F_{\mu\nu} F_{\mu\nu} + Q_\mu \partial^2 Q_\mu - \frac{8}{\tau} (Q_\mu Q_\mu) (q_0 + \mathcal{C} \widetilde{q} + \frac{1}{2})^2 + 2 b \partial^2 c \nonumber\\
  {} & \qquad\qquad\qquad\quad - \frac{16}{\tau} (bc) (q_0 + \mathcal{C} \widetilde{q} + \frac{1}{2})^2 - 2 [Q_\mu, Q_\nu] (\partial_\mu Q_\nu) - \frac{1}{2} [Q_\mu, Q_\nu]^2 \Bigg)\, .\label{eq:4DFirstAve}
\end{align}

Similar to the 3D case, we can also discuss the simplified effective action in different limits of the factor $\mathcal{C}$:
\begin{itemize}

\item $\mathcal{C} \ll 1$:\\
  The topological fluctuations are not important. We only need to consider the classical backgrounds.
\begin{align}
 \langle S \rangle_\omega & \approx    \langle S_0 \rangle_\omega \nonumber\\
 & = \frac{ 1}{g^2} \int d^4 x\, \textrm{Tr} \Bigg( - \frac{1}{2} F_{\mu\nu} F_{\mu\nu} + Q_\mu \partial^2 Q_\mu -  \frac{8}{\tau} (Q_\mu Q_\mu) (q_0   + \frac{1}{2})^2 + 2 b \partial^2 c  \nonumber\\
  {} & \qquad\qquad\qquad\quad - \frac{16}{\tau} (bc) (q_0   + \frac{1}{2})^2 - 2 [Q_\mu, Q_\nu] (\partial_\mu Q_\nu) - \frac{1}{2} [Q_\mu, Q_\nu]^2 \Bigg)\, .
\end{align}
Like the corresponding limit in the 3D case, here the topological fluctuations can be neglected, while the quantum fluctuations and the ghosts become massive due to the classical backgrounds. The masses are of the order $\ell_0^{-1}$, where $\ell_0 \equiv \ell_{\textrm{cl}}$ is the classical length scale that can approximately replace $\sqrt{\tau}$, as discussed in Appendix~\ref{topologicalmeasure}.

\item $\mathcal{C} \gg 1$:\\
  The topological fluctuations are important, while the classical backgrounds are negligible.
\begin{align}
   \langle S_0 \rangle_\omega & \approx  \frac{ 1}{g^2}\int d^3 x\, \textrm{Tr} \Bigg(    Q_\mu \partial^2 Q_\mu     + 2 b \partial^2 c  - 2 [Q_\mu, Q_\nu] (\partial_\mu Q_\nu) - \frac{1}{2} [Q_\mu, Q_\nu]^2 \Bigg) \nonumber\\
& \approx  \frac{-1}{g^2} \int d^3 x\, \textrm{Tr} \Bigg( \frac{1}{2} G_{\mu\nu}G_{\mu\nu} + (\partial_\mu Q_\mu)^2   - 2 b \partial^2 c    \Bigg)\, ,
\end{align}
where
\be
G_{\mu\nu} = \partial_\mu Q_\nu - \partial_\nu Q_\mu + [Q_\mu, Q_\nu]\, ,
\ee
and
\begin{align}\label{eq:DeltatildeS4Deq1}
 \langle \Delta S \rangle_\omega & = \frac{-1}{g^2}\int d^4 x\, \textrm{Tr} \Bigg( \frac{1}{2} F_{\mu\nu} F_{\mu\nu}   + \frac{8 \mathcal{C}^2 }{\tau} (Q_\mu Q_\mu)   \widetilde{q}\,^2 +  \frac{16 \mathcal{C}^2 }{\tau} (bc)   \widetilde{q}\,^2      \Bigg)\, ,
\end{align}
where 
\begin{align}
  \frac{1}{4} F_{\mu\nu}^a F_{\mu\nu}^a &\approx   \frac{24 \mathcal{C}^2}{\tau^2} \left( \tau^2 ( \widetilde{q}\,')^2  - \frac{\widetilde{q}\,^2}{2}  +  \mathcal{C}^2 \widetilde{q}\,^4\right) \nonumber\\
  & =   \frac{6 \mathcal{C}^2}{\tau^2} \left( \tau  \left(\partial_\mu \widetilde{q} \right)^2  -  2 \widetilde{q}\,^2   + 4  \mathcal{C}^2 \widetilde{q}\,^4\right) \, . 
\end{align}

In this limit of the 4D case, the effective theory mimics the Higgs mechanism, i.e., the topological fluctuations acquire a vacuum expectation value, which consequently gives masses to the quantum fluctuations and the ghosts. To clearly demonstrate it, we follow the same logic and repeat the same steps as in the limit $\mathcal{C} \gg 1$ of the 3D case. Similarly, after the integration over $x_0$ in the weak-field approximation, the apparent divergences due to the factors $\tau^{-1}$ and $\tau^{-2}$ in the expression above are regularized by an IR cutoff $L_{\textrm{top}}$ and a UV cutoff $\ell_{\textrm{top}}$ with $L_{\textrm{top}} \gg \ell_{\textrm{top}}$, i.e.,
\begin{align}
  {} & \frac{1}{4g^2} \int \frac{d^4 x_0}{V} \, \int d^4 x\, F_{\mu\nu}^a F_{\mu\nu}^a \nonumber\\
  \approx & \, \int d^4 x\, \left[ \frac{ 12 \, \mathcal{C}^2 L_{\textrm{top}}^2}{ g^2 L^4}  \left(\partial_\mu \widetilde{q} (x) \right)^2 - \frac{ 48 \, \mathcal{C}^2 \, \textrm{log}  \left( \frac{L_{\textrm{top}} }{ \ell_{\textrm{top}}} \right) }{g^2 L^4}\, \widetilde{q} (x)^2 + \frac{ 96\, \mathcal{C}^4 \, \textrm{log} \left( \frac{L_{\textrm{top}} }{ \ell_{\textrm{top}}} \right)}{g^2 L^4}\, \widetilde{q} (x)^4 \right]\, ,
\end{align}
where $L$ is the radius of the system, and $\ell_{\textrm{top}}$ can be thought of as the averaged length scale of the topological fluctuations. Again, the length scale of the topological fluctuations is smaller than the length scale of the classical background ($\ell_{\textrm{top}}<\ell_{0}$), because it includes the excited states of the topological fluctuations (see Appendix~\ref{topologicalmeasure} for details). This equation should also be understood as a relation inside a path integral over the topological fluctuations $\widetilde{q}$.

We now define the field
\be
\xi  \equiv \frac{ 2 \sqrt{3}\, \mathcal{C} L_{\textrm{top}} }{ g\, L^2} \widetilde{q} \, ,
\ee
and then obtain after the integration over $x_0$ in the weak-field approximation:
\begin{align}
  \frac{1}{4 g^2} F_{\mu\nu}^a F_{\mu\nu}^a   & = (\partial_\mu \xi)^2 - \mu^2_{4D} \xi^2 + \lambda_{4D} \xi^4 \, ,
\end{align}
where 
\be
\mu^2_{4D} \equiv \frac{4}{L_{\textrm{top}}^2}\, \textrm{log} \left(\frac{L_{\textrm{top}}}{\ell_{\textrm{top}}} \right) \, , \quad \textrm{and} \quad \,\lambda_{4D} \equiv \frac{2g^2 L^4}{3 L_{\textrm{top}}^4} \, \textrm{log} \left(\frac{L_{\textrm{top}}}{\ell_{\textrm{top}}} \right)
\ee
are two constants that can be fixed by comparing with the experiments or the Monte Carlo simulations.

Let us introduce the parameterization:
\be
\xi = \frac{\nu_{4D} + h(x)}{\sqrt{2}} \, ,
\ee 
where
\be
\nu^2_{4D} = \mu^2_{4D} / \lambda_{4D} \, .
\ee

Then we obtain
\begin{align}
  \frac{1}{4g^2} F_{\mu\nu}^a F_{\mu\nu}^a   =  \frac{1}{2} (\partial_\mu h)^2 +\mu^2_{4D} h^2  + \lambda_{4D}\, \nu_{4D}\, h^3 + \frac{\lambda_{4D}}{4}  h^4 - \frac{\lambda_{4D}}{4}  \nu^4_{4D}\, ,
\end{align}
where $h(x)$ acts like a Higgs boson.  Dropping the constant shift $\lambda_{4D} \nu^4_{4D}/4 $, Eq.~\eqref{eq:DeltatildeS4Deq1} reads
\begin{align} 
 \langle \Delta S \rangle_\omega & =   \int d^4 x\,   \Bigg(   \frac{1}{2} (\partial_\mu h)^2 +\mu^2_{4D} h^2  + \lambda_{4D}\, \nu_{4D}\, h^3 + \frac{\lambda_{4D}}{4}  h^4  \nonumber\\
  {} & \qquad\qquad\quad + \frac{1}{3} (Q_\mu^a Q_\mu^a)   (\nu_{4D} + h)^2  + \frac{2}{3} (b^a c^a )   (\nu_{4D} + h)^2       \Bigg)\, ,
\end{align}
where the scalar has the mass $m_h = \sqrt{2}\, \mu_{4D}$ and gauge field $Q_\mu$ acquires a mass $m_{Q}^{4D} =\sqrt{2/3}\, \nu_{4D} $.

 \item $\mathcal{C} \sim 1$:\\
 Both classical backgrounds and topological fluctuations are important. One needs to calculate for each background and sum over all the configurations. Hence, in the scope of this paper the effective action \eqref{eq:4DFirstAve} cannot be further simplified.


\end{itemize}

\subsection{Cutoffs and Topological Boundary Conditions}\label{cutoff}

We would like to discuss an important issue that we have encountered, which is the cutoffs that we introduced in Eq.~\eqref{eq:finiteintegrals}. As we will show, these cutoffs are naturally embedded in the Yang-Mills theory due to the topological properties, which means that they are not just regularizations of the theory, instead they have clear physical meanings and in principle can be determined rigorously.

We take the 3D case as an example. The 4D case follows the same logic and can be done in a similar way. We consider part of the path integral containing only the effective action \eqref{eq:finiteintegrals}, i.e.,
\begin{align}
  {} & \int \frac{d^3 x_0}{V} \, \textrm{exp} \left[ -\frac{1}{g^2} \int  d^3 x \,  \left( \frac{\mathcal{C}^2}{\tau}    \big| \partial_\mu\widetilde{\psi} (x - x_0) \big|^2    - \frac{\mathcal{C}^2 }{\tau^2}  \big| \widetilde{\psi} (x - x_0) \big|^2 + \frac{\mathcal{C}^4}{2 \tau^2 }\big| \widetilde{\psi}(x - x_0)\big|^4  \right) \right]  \nonumber\\
  = & \, \int \frac{d^3 x_0}{V} \, \textrm{exp} \bigg[ -\frac{1}{g^2} \int d^3 x \,  \left( \frac{\mathcal{C}^2}{\tau}    \big| \partial_\mu\widetilde{\psi} (x ) \big|^2    - \frac{\mathcal{C}^2 }{\tau^2}  \big| \widetilde{\psi} (x  ) \big|^2 + \frac{\mathcal{C}^4}{2 \tau^2 }\big| \widetilde{\psi}(x )\big|^4 \right) \nonumber\\
  {} & \qquad\qquad\qquad + \frac{1}{g^2} \int_{\textrm{near}\,\, x_0} d^3 x  \,  \left( \frac{\mathcal{C}^2}{\tau}    \big| \partial_\mu\widetilde{\psi} (x ) \big|^2    - \frac{\mathcal{C}^2 }{\tau^2}  \big| \widetilde{\psi} (x  ) \big|^2 + \frac{\mathcal{C}^4}{2 \tau^2 }\big| \widetilde{\psi}(x )\big|^4  \right) \bigg]\, , \label{eq:finiteintegrals-new}
\end{align}
where again the integration over the function space of $\widetilde{\psi}$ is implied. For simplicity, we drop the integration over $\widetilde{\psi}$ in the equation above. All the following equations should be understood with an integration over $\widetilde{\psi}$ assumed.

As we discussed before, the left-hand side has no divergence due to the topological boundary conditions, while the two terms in the action on the right-hand side are both divergent, but the divergences from two terms cancel each other. Therefore, there are three equivalent ways to analyze this kind of effective theory:
\begin{itemize}
  \item One can directly analyze the finite theory given by the left-hand side of Eq.~\eqref{eq:finiteintegrals-new}. However, it has a non-local expression, which makes it difficult to study in the framework of standard quantum field theory.

  \item One can analyze the theory given by the right-hand side of Eq.~\eqref{eq:finiteintegrals-new}, which should also be finite. In other words, one can view the two terms in the action on the right-hand side as an effective action and its counter-term. In practice, this treatment requires very precise cancellation between two divergent terms, hence it is technically not very feasible.

  \item One can analyze only the first term in the action on the right-hand side of Eq.~\eqref{eq:finiteintegrals-new} as an effective theory, and introduce some physical cutoffs $\ell_{\textrm{top}}$ and $L_{\textrm{top}}$ for $|x-x_0|$, which regularize the theory and give us finite results. This is the approach that we used in this paper.
\end{itemize}

Comparing to standard quantum field theory, we can find the counterparts of these treatments. For example, a UV complete theory usually does not have divergence, hence it can be viewed as the first type above. For most theories that we encounter, they have divergences and need to be regularized by introducing either counter-terms or cutoffs. In principle, different treatments should give us the same result. However, in practice for most theories we know, we are only aware of the part of the effective theory, that corresponds to the first term on the right-hand side of Eq.~\eqref{eq:finiteintegrals-new}, and we have to construct a counter-term by hand to cancel the divergence, which is contrary to the case discussed above, where there is no intrinsic divergence, and the counter-term is known, which is just the second term on the right-hand side of Eq.~\eqref{eq:finiteintegrals-new}.

The cutoffs $\ell_{\textrm{top}}$ and $L_{\textrm{top}}$ have clear physical meanings, and can be determined precisely. For a given configuration $\widetilde{\psi}$, the effective theory on the left-hand side of Eq.~\eqref{eq:finiteintegrals-new} can be evaluated numerically and will give us a finite result, which is formally denoted by $\Upsilon_L [\widetilde{\psi}]$. The first term in the action on the right-hand side of Eq.~\eqref{eq:finiteintegrals-new} with cutoffs $\ell_{\textrm{top}}$ and $L_{\textrm{top}}$ can also be evaluated exactly, which we call $\Upsilon_R \,[\widetilde{\psi},\, \ell_{\textrm{top}},\, L_{\textrm{top}}]$. Therefore, for $\tau = (x-x_0)^2$ we define
\begin{align}
  {} & \Upsilon_L [\widetilde{\psi}] \nonumber\\
  \equiv & \, \int \frac{d^3 x_0}{V} \, \textrm{exp} \left[ - \frac{1}{g^2} \int  d^3 x \,  \left( \frac{\mathcal{C}^2}{\tau}    \big| \partial_\mu\widetilde{\psi} (x - x_0) \big|^2    - \frac{\mathcal{C}^2 }{\tau^2}  \big| \widetilde{\psi} (x - x_0) \big|^2 + \frac{\mathcal{C}^4}{2 \tau^2 }\big| \widetilde{\psi}(x - x_0)\big|^4  \right) \right]\, ,
\end{align}
\begin{align}
  {} & \Upsilon_R\, [\widetilde{\psi},\, \ell_{\textrm{top}},\, L_{\textrm{top}}] \nonumber\\
  \equiv & \, \int \frac{d^3 x_0}{V} \, \textrm{exp} \left[ - \frac{1}{g^2} \int_{\ell_{\textrm{top}} < |x-x_0| < L_{\textrm{top}}} d^3 x \,  \left( \frac{\mathcal{C}^2}{\tau}    \big| \partial_\mu\widetilde{\psi} (x ) \big|^2    - \frac{\mathcal{C}^2 }{\tau^2}  \big| \widetilde{\psi} (x  ) \big|^2 + \frac{\mathcal{C}^4}{2 \tau^2 }\big| \widetilde{\psi}(x )\big|^4  \right) \right]\, ,
\end{align}
and we require that
\be\label{eq:cutoffeq}
  \Upsilon_L [\widetilde{\psi}] = \Upsilon_R\, [\widetilde{\psi},\, \ell_{\textrm{top}},\, L_{\textrm{top}}]\, ,
\ee
which can be viewed as an equation for the cutoffs $\ell_{\textrm{top}}$ and $L_{\textrm{top}}$. By solving this equation numerically, one can always fix the values of $\ell_{\textrm{top}}$ and $L_{\textrm{top}}$. Hence, these cutoffs exist and can be determined at least numerically.  In principle, $\ell_{\textrm{top}}$ and $L_{\textrm{top}}$ depend on the configuration $\widetilde{\psi}$, i.e., they are implicit functionals of $\widetilde{\psi}$. Equivalently, they depend on the energy of the configuration. When we focus on a small range of the energy scale, $\ell_{\textrm{top}}$ and $L_{\textrm{top}}$ can be approximately viewed as constants.

Under the weak-field approximation, the expressions of $\Upsilon_L [\widetilde{\psi}]$ and $\Upsilon_R\, [\widetilde{\psi},\, \ell_{\textrm{top}},\, L_{\textrm{top}}]$ can be expanded to the leading order, and the integrals can be simplied. Consequently, under this approximation Eq.~\eqref{eq:cutoffeq} becomes
\begin{align}\la{eq:weakexpansioncutoffana}
  {} & \int \frac{d^3 x_0}{V} \, \int  d^3 x \,  \left( \frac{\mathcal{C}^2}{\tau}    \big| \partial_\mu\widetilde{\psi} (x - x_0) \big|^2    - \frac{\mathcal{C}^2 }{\tau^2}  \big| \widetilde{\psi} (x - x_0) \big|^2 + \frac{\mathcal{C}^4}{2 \tau^2 }\big| \widetilde{\psi}(x - x_0)\big|^4  \right) \nonumber\\
  = & \, \frac{2\pi}{V}\int_{\ell_{\textrm{top}}^2}^{L_{\textrm{top}}^2} \sqrt{\tau} d \tau  \, \int d^3 x  \,  \left( \frac{\mathcal{C}^2}{\tau}    \big| \partial_\mu\widetilde{\psi} (x ) \big|^2    - \frac{\mathcal{C}^2 }{\tau^2}  \big| \widetilde{\psi} (x  ) \big|^2 + \frac{\mathcal{C}^4}{2 \tau^2 }\big| \widetilde{\psi}(x )\big|^4  \right)\, ,
\end{align}
where we used
\be
\int \frac{d^3 x_0}{V}    \int_{\ell_{\textrm{top}} < |x-x_0|< L_{\textrm{top}}} d^3 x  =  \frac{2\pi}{V}\int_{\ell_{\textrm{top}}^2}^{L_{\textrm{top}}^2} \sqrt{\tau} d \tau  \, \int d^3 x \, .
\ee

If the gauge theory is defined on a sphere $S^n$ instead of the flat space $\mathbb{R}^n$, we can adopt the following stereographic projection:
\be
  \sqrt{\tau} = R \, \textrm{cot} \left(\frac{\theta}{2} \right)\, ,
\ee
where $\theta \in [0,\, \pi]$. For the cutoffs, we can assume the same physical cutoff on $\theta$ near $0$ and $\pi$, i.e.,
\be
  L_{\textrm{top}}  = R \, \textrm{cot} \left(\frac{\theta_0}{2} \right)\, ,\quad  \ell_{\textrm{top}} = R \, \textrm{cot} \left(\frac{\pi - \theta_0}{2} \right)\, ,
\ee
which leads to 
\be
  \ell_{\textrm{top}}\, L_{\textrm{top}} = R^2\, , \quad   \quad \, L_{\textrm{top}} /   \ell_{\textrm{top}} = \textrm{cot}^2 \left( \theta_0/2  \right) \, .
\ee
Consequently, the integral over $\tau$ becomes an integral over $\theta$:
\be
  \int_{\ell_{\textrm{top}}^2}^{L_{\textrm{top}}^2} \sqrt{\tau} d\tau =  \frac{1}{4} R^3 \int_{\theta_0}^{\pi - \theta_0} d\theta \, \frac{\textrm{sin}^2 (\theta)}{\textrm{sin}^6 (\theta/2)}\, .
\ee

When the topological fluctuations are turned off, i.e. $\widetilde{\psi} = 0$, Eq.~\eqref{eq:cutoffeq} is trivially satisfied. When the topological fluctuations are turned on, i.e. $\widetilde{\psi} \neq 0$,   in order for Eq.~\eqref{eq:cutoffeq} to  be solved, the limit $\ell_{\textrm{top}} \to 0$ and $L_{\textrm{top}} \to \infty$ cannot be taken at the same time for a not always vanishing configuration $\widetilde{\psi}$.

As we have seen in this section, the masses acquired by the quantum fluctuation field and the ghost fields can be expressed in terms of the physical cutoffs $\ell_{\textrm{top}}$ and $L_{\textrm{top}}$. In the next section, we will see that the mass gaps of the 3D and the 4D Yang-Mills theory at semi-classical level are also related to these physical cutoffs. According to our discussions above, the finite cutoffs $\ell_{\textrm{top}}$ and $L_{\textrm{top}}$ are naturally embedded in the Yang-Mills theory  on a flat space $\mathbb{R}^n$ of finite size or on a compact space such as a sphere $S^n$ due to the topological properties.

\section{\label{LE}Analysis of Low-Energy Physics}

Based on the discussions in the previous section, we can start addressing the problem of the Yang-Mills mass gap. The rigorous formulation of the problem and the mathematical treatment will be presented in the next section. 

The logic of this section is following. First, in the IR regime the fast varying mode $Q_\mu$ should be less dominant, because it vibrates rapidly and will be suppressed in long distance.  Hence, we can turn off the quantum fluctuations $Q_\mu$ in the IR regime, and derive the low-energy effective action due to the topological fluctuations $\widetilde{\mathcal{A}}$. Since the most relevant background for the mass gap problem is given by the pure gauge solution and the trivial vacuum solution, we consider topological fluctuations around these two solutions. From the asymptotic behavior of the topological fluctuations at semi-classical level, we find the correlation function between two gauge invariant operators, which decays exponentially. This effect implies the existence of a massive mediator that cannot propagate a long distance in the IR regime, hence provides a strong evidence for the existence of the mass gap in the IR regime of the theory.

\subsection{Low-Energy Effective Action}\label{sec:effaction}

In this subsection, let us first analyze the low-energy effective action of the Yang-Mills theory.

\subsubsection{3D Case}\label{sec:3deffaction}

After turning off the quantum fluctuations $Q_\mu$, the path integral in 3-dimensions becomes
\be\label{eq:3dMGPathInt}
  Z = \sum_{\{\mathcal{A}_0\}} \int \mathcal{D} \overline{\Psi}\, \mathcal{D} \Psi\, \int [\textrm{Jac}]_{\gamma_i}^{\textrm{top}} \, \frac{d^3 x_0}{V} \, d^3 \varphi \,   e^{- S}\, ,
\ee
where the action $S$ is actually the topological part of the action $S^{\textrm{top}}$ given by the Lagrangian \eqref{eq:3DfullTopL}:
\be
   S = \frac{1}{g^2} \int d^3 x\, \left[ \frac{1}{2 \tau^2} + 4 (\overline{\partial_\tau \psi})  (\partial_\tau \psi) - \frac{1}{\tau^2} \overline{\psi}  \psi + \frac{1}{2 \tau^2} (\overline{\psi}  \psi)^2 \right]\, ,
\ee
and
\be
  \psi = \psi_0 + \widetilde{\psi}\, .
\ee
$\psi_0$ and $\widetilde{\psi}$ stand for the classical solutions and the topological fluctuations respectively.

In Eq.~\eqref{eq:3dMGPathInt} we work with the exact theory described in Section~\ref{TF}. In principle, one can also apply the approximation introduced in Section~\ref{approx}, which will not change the following discussions. Under the approximation, the Jacobian $[\textrm{Jac}]_{\gamma_i}^{\textrm{top}}$ in Eq.~\eqref{eq:3dMGPathInt} is replaced by $[\textrm{Jac}]_{\gamma_i}^{\textrm{cl}}$, and each topological fluctuation acquires a factor $\mathcal{C}$ from the wave function renormalization. The exact theory described in Section~\ref{TF} results in a complicated $[\textrm{Jac}]_{\gamma_i}^{\textrm{top}}$, which is difficult to analyze, while the approximation introduced in Section~\ref{approx} results in a complicated renormalization factor $\mathcal{C}$, which is also difficult to obtain in general. As we can see, however, we are interested only in the lowest-lying mode between two gauge invariant operators  at semi-classical level, which simplifies the analysis.

As we have seen in the previous sections, for the lowest topological charges in 3D, $\psi_0$ can be the trivial vacuum solution up to a gauge transformation or the Wu-Yang monopole solution. For the mass gap problem the trivial vacuum background including the trivial vacuum solution and the pure gauge solution is the most relevant background, hence in the following we assume $\psi_0$ to be the trivial vacuum background, which is a constant phase. We can absorb the constant phase by redefining the topological fluctuations $\widetilde{\psi}$, then the effective action becomes
\be
  S = \frac{1}{g^2} \int d^3 x\, \left[\frac{1}{\tau} \big| \partial_\mu \widetilde{\psi} \big|^2 + \frac{1}{2 \tau^2} \left( | 1 + \widetilde{\psi} |^2 - 1 \right)^2 \right] \label{eq:3dMGEffAction}
\ee
with the boundary conditions
\be
  \widetilde{\psi} (\tau = 0) = \widetilde{\psi} (\tau = \infty) = 0\, .
\ee
To obtain the regularized effective theory, we apply the procedure discussed in Subsection~\ref{cutoff}, which is to perform the integration over $x_0$ in the weak-field approximation with some counter-terms or physical cutoffs. After some steps the effective action \eqref{eq:3dMGEffAction} becomes
\begin{align}
\left<S\right>_{x_0}=& \frac{1}{g^2} \int \frac{d^3 x_0}{V} \int d^3 x \left[\frac{1}{\tau}  \big| \partial_\mu \widetilde{\psi} (\tau) \big|^2 + \frac{1}{2 \tau^2} \left( | 1 + \widetilde{\psi} (\tau)|^2 - 1 \right)^2 \right]  \nonumber\\
  = & \frac{1}{g^2} \int \frac{d^3 x_0}{V} \int d^3 x \left[\frac{1}{\tau}  \big| \partial_\mu \widetilde{\psi} (\widetilde{\tau}) \big|^2 + \frac{1}{2 \tau^2}\left(   \widetilde{\psi}^\dagger  (\widetilde{\tau}) +  \widetilde{\psi}  (\widetilde{\tau}) + | \widetilde{\psi}   (\widetilde{\tau})|^2   \right)^2 \right] + (\textrm{counter-terms})\nonumber\\
  = & \, \frac{1}{g^2} \int \frac{d^3 x_0}{V} \int_{\ell_{\textrm{top}} < |x-x_0| < L_{\textrm{top}}} d^3 x \left[\frac{1}{\tau}  \big| \partial_\mu \widetilde{\psi} (\widetilde{\tau}) \big|^2 + \frac{1}{2 \tau^2} \left(   \widetilde{\psi}^\dagger  (\widetilde{\tau}) +  \widetilde{\psi}  (\widetilde{\tau}) + | \widetilde{\psi}   (\widetilde{\tau})|^2   \right)^2 \right]\nonumber\\
  \approx & \, \frac{3 L_{\textrm{top}}}{g^2 L^3} \int d^3 x\, \left[ |\partial_{|x|} \widetilde{\psi} (|x|) |^2 + m_{3D}^2 (|1 + \widetilde{\psi} (|x|) |^2 - 1)^2 \right] \nonumber\\
  = & \, \frac{3 L_{\textrm{top}}}{g^2 L^3} \int d^3 x\, \left[ |\partial_{|x|} \Psi (|x|) |^2 + m_{3D}^2 (|\Psi (|x|) |^2 - 1)^2 \right] \, ,\label{eq:3dMGcutoff}
\end{align}
where $\tau = (x - x_0)^2$, and an effective mass is defined by
\be\label{eq:defm3D}
  m_{3D}^2 \equiv \frac{1}{2 \ell_{\textrm{top}} L_{\textrm{top}}} \, .
\ee
Since we focus on a small range of energy scales above the trivial vacuum background, the physical cutoffs $\ell_{\textrm{top}}$ and $L_{\textrm{top}}$ are approximately constant. So is the effective mass $m_{3D}$. In the last expression of Eq.~\eqref{eq:3dMGcutoff} we defined
\be
  \Psi \equiv e^{i \Theta} \left(1 + \widetilde{\psi} \right)\, ,
\ee
where $\Theta$ is a constant phase, and $\Psi$ satisfies the boundary conditions
\be\label{eq:3dMGbc-1}
  |\Psi| (|x| = 0) = |\Psi| (|x| = \infty) = 1\, .
\ee
We also define a new variable $\widetilde{\tau} \equiv x_\mu x^\mu = |x|^2$, which is  $x_0$-independent. Consequently,
\be
  |\partial_\mu \Psi (\widetilde{\tau}) |^2 =   4 \widetilde{\tau} |\partial_{\widetilde{\tau}} \Psi (\widetilde{\tau})|^2 =   \Bigg| \frac{\partial \Psi (|x|) }{\partial |x|} \Bigg|^2 \, .
\ee
The relation between different variables is
\be\label{eq:relbetvariable}
  |x|^2  = \widetilde{\tau} =   \tau(x_0 = 0)\, .
\ee
In the following, the topological field $\Psi$ should be understood as $\Psi (\widetilde{\tau})$ or $\Psi (|x|)$.

As we discussed in Subsection \ref{cutoff}, the second line of Eq.~\eqref{eq:3dMGcutoff} has no intrinsic divergences, and to cure the apparent divergences, we can introduce either counter-terms or physical cutoffs, which correspond to the third and the fourth line of Eq.~\eqref{eq:3dMGcutoff} respectively. From now on we will always use the approach of physical cutoffs.

Let us adopt the parameterization:
\be
  \Psi = |\Psi| \, e^{i \gamma}\, ,
\ee
where $\gamma$ is not well-defined when $\Psi = 0$. This parameterization should be understood as follows. For any interval with $\Psi \neq 0$, we can use the parameterization above, while to pass from one interval to another, which are joint by a point with $\Psi = 0$, we have the freedom to change the phase without affecting Eq.~\eqref{eq:3dMGcutoff}. The field equations for $|\Psi|$ and $\gamma$ are
\begin{align}
  2\, \Delta |\Psi| - 2 |\Psi| (\partial_{|x|} \gamma)^2 - 4 \, m_{3D}^2 \, (|\Psi|^2 - 1) |\Psi| & = 0\, ,\label{eq:3dMGEOM-1}\\
  2\, \partial_{|x|} \left( |x|^2 |\Psi|^2\, \partial_{|x|} \gamma \right) & = 0\, ,\label{eq:3dMGEOM-2}
\end{align}
where for the 3D Euclidean space
\be
  \Delta |\Psi| = \partial_{|x|}^2 |\Psi| + \frac{2}{|x|} \, \partial_{|x|} |\Psi|\, .
\ee
Eq.~\eqref{eq:3dMGEOM-2} implies that $|x|^2 |\Psi|^2 \partial_{|x|} \gamma$ should be a constant, hence for a nonvanishing $\Psi$
\be\la{eq:gammasolution}
  \partial_{|x|} \gamma =  \frac{C}{|x|^2 |\Psi|^2} \, .
\ee
If we plug Eq.~\eqref{eq:gammasolution} into Eq.~\eqref{eq:3dMGEOM-1} and use the boundary condition $|\Psi| = 1$ at $|x|=0$ as well as the fact that $|\Psi|$ is a $C^1$-function, we obtain
\be
C = 0\, .
\ee
Consequently, Eq.~\eqref{eq:3dMGEOM-1} simplifies to
\be
  \partial_{|x|}^2 |\Psi| + \frac{2}{|x|} \, \partial_{|x|} |\Psi| - 2\, m_{3D}^2 \, (|\Psi|^2 - 1) |\Psi| = 0\, .\label{eq:3dMGEOM-3}
\ee

The action \eqref{eq:3dMGcutoff} plays the role of the energy in the 3D Euclidean space, which is always non-negative. Because we are interested in the lowest states of the IR regime, we should consider the solutions to the equations \eqref{eq:3dMGEOM-3} with the boundary condition \eqref{eq:3dMGbc-1}, that have the lowest energies given by Eq.~\eqref{eq:3dMGcutoff}.

\subsubsection{4D Case}\label{sec:4deffaction}

For the 4D case, we can follow the same logic and repeat the same steps in the 3D case. Again, to probe the IR regime of the 4D quantum Yang-Mills theory, we turn off all the quantum fluctuations $Q_\mu$, while keeping the topological fluctuations $\widetilde{\mathcal{A}}$. The path integral then becomes
\be\label{eq:4dMGPathInt}
  Z = \sum_{\{\mathcal{A}_0\}} \int \mathcal{D} q\, \int [\textrm{Jac}]_{\gamma_i}^{\textrm{top}} \, \frac{d^4 x_0}{V} \, d^3 \varphi \,   e^{- S}\, ,
\ee
where $S$ is the topological part of the action $S^{\textrm{top}}$ given by the Lagrangian \eqref{eq:4DfullTopL}:
\begin{align}
   S  = - \frac{1}{g^2} \int d^4 x\, \textrm{Tr} \left(\frac{1}{2} F_{\mu\nu} F_{\mu\nu} \right) & = \frac{1}{g^2} \int d^4 x\, \frac{24}{\tau^2} \left[\frac{1}{16} + (\tau q')^2 - \frac{q^2}{2} + q^4 \right] \nonumber\\
  {} & = \frac{1}{g^2} \int d^4 x\, \left[ \frac{6}{ \tau} \left( \partial_\mu q\right)^2 + \frac{24}{\tau^2} \left(q^2 - \frac{1}{4} \right)^2 \right]\, ,\label{eq:4dMGEffAction}
\end{align}
and
\be
  q = q_0 + \widetilde{q}\, .
\ee
$q_0$ and $\widetilde{q}$ denote the classical solutions and the topological fluctuations around these classical solutions respectively, and in general they are functions of $\tau = (x-x_0)^2$.

Like in the 3D case, we may integrate over $x_0$ in the weak-field approximation as discussed in Subsection~\ref{cutoff} to obtain a regularized effective theory. For the mass gap problem the trivial vacuum background including the trivial vacuum solution and the pure gauge solution is the most relevant background, hence in the following we focus on the trivial vacuum background $q_0 = \pm \frac{1}{2}$ and obtain
\be
  q = \pm \frac{1}{2} + \widetilde{q}\, .
\ee
Following the same steps, we obtain
\begin{align}
\left<S\right>_{x_0}  = & \frac{1}{g^2} \int \frac{d^4 x_0}{V} \, \int d^4 x \left[ \frac{6}{ \tau} \left( \partial_\mu \widetilde{q} (\tau)\right)^2 + \frac{24}{\tau^2} \left( \left( \pm \frac{1}{2} + \widetilde{q} (\tau) \right)^2 - \frac{1}{4} \right)^2 \right]   \nonumber\\
  = & \frac{1}{g^2} \int \frac{d^4 x_0}{V} \, \int d^4 x \left[ \frac{6}{ \tau} \left( \partial_\mu \widetilde{q} ( \widetilde{\tau} )\right)^2+ \frac{24}{\tau^2} \,  \left( \widetilde{q} ( \widetilde{\tau})  \right)^2 \left( \widetilde{q} ( \widetilde{\tau})   \pm 1   \right)^2 \right] + (\textrm{counter-terms}) \nonumber\\
  = & \, \frac{1}{g^2} \int \frac{d^4 x_0}{V} \, \int_{\ell_{\textrm{top}} < |x-x_0| < L_{\textrm{top}}} d^4 x \left[\frac{6}{ \tau} \left( \partial_\mu \widetilde{q} ( \widetilde{\tau} )\right)^2 + \frac{24}{\tau^2} \,  \left( \widetilde{q} ( \widetilde{\tau})  \right)^2 \left( \widetilde{q} ( \widetilde{\tau})   \pm 1   \right)^2 \right] \nonumber\\
  \approx & \, \frac{12 L_{\textrm{top}}^2}{g^2 L^4} \int d^4 x\, \left[ \left(\partial_\mu \widetilde{q} (\widetilde{\tau}) \right)^2 + \frac{ 4 \, \textrm{log}  (L^2_{\textrm{top}} / \ell^2_{\textrm{top}}) }{L_{\textrm{top}}^2}\, \left( \left( \pm \frac{1}{2} + \widetilde{q} (\widetilde{\tau}) \right)^2 - \frac{1}{4} \right)^2 \right] \nonumber\\
  = & \, \frac{12 L_{\textrm{top}}^2}{g^2 L^4} \int d^4 x\, \left[ \left(\partial_{|x|} q (|x|) \right)^2 + 4 \, m_{4D}^2 \left(q^2 (|x|) - \frac{1}{4} \right)^2 \right]\, ,\label{eq:4dMGcutoff}
\end{align}
where $\tau = (x-x_0)^2$. The second line of Eq.~\eqref{eq:4dMGcutoff} has no intrinsic divergences. To cure the apparent divergences, there are different approaches, as we discussed in Subsection~\ref{cutoff}, while the third and the fourth line of Eq.~\eqref{eq:4dMGcutoff} correspond to the approach of counter-terms and the approach of physical cutoffs respectively. In the following, we will always use the latter one. In the last step, we wrote the effective action again in terms of $q = \pm \frac{1}{2} + \widetilde{q}$. Also, we have defined 
\be\label{eq:defm4D}
  \widetilde{\tau} \equiv \tau (x_0 = 0)\, ,\quad m_{4D}^2 \equiv \frac{2\, \textrm{log}  (L_{\textrm{top}} / \ell_{\textrm{top}}) }{L_{\textrm{top}}^2}\, .
\ee
For a small range of energy scales above the trivial vacuum background the physical cutoffs $\ell_{\textrm{top}}$ and $L_{\textrm{top}}$ are approximately constant. Hence, the effective mass $m_{4D}$ is also approximately constant.

To find the lowest states in the spectrum, let us first solve the field equation
\be
  2\, \Delta\, q - 16 \, m_{4D}^2\, q \left(q^2 - \frac{1}{4} \right) = 0\, , \label{eq:4dMGEOM}
\ee
where for the 4D Euclidean space
\be
  \Delta\, q = \partial_{|x|}^2\, q + \frac{3}{|x|} \, \partial_{|x|} q\, .
\ee
The general boundary conditions for the topological fluctuations are
\be
  \widetilde{q} (|x| = 0) = \widetilde{q} (|x| = \infty) = 0\, .
\ee
For the trivial vacuum background, in terms of $q = \pm \frac{1}{2} + \widetilde{q}$ these boundary conditions become
\be\label{eq:4dMGbc-1}
  q (|x| = 0) = q (|x| = \infty) = \pm \frac{1}{2}\, .
\ee
The action \eqref{eq:4dMGcutoff} plays the role of the energy in the 4D Euclidean space, which is always non-negative. Because we are interested in the lowest states of the IR regime, we should consider the solutions to Eq.~\eqref{eq:4dMGEOM} with the boundary conditions \eqref{eq:4dMGbc-1}, which have the lowest energies given by Eq.~\eqref{eq:4dMGcutoff}.

\subsection{Massive Mediator}\label{sec:mediator}

As discussed in Ref.~\cite{Polyakov}, one can consider the two-point correlation function of a gauge invariant operator. If one can only find massive mediators from the correlation function, it implies the existence of the mass gap in the theory.

Similary, to find some physical evidences for the existence of the mass gap in the quantum Yang-Mills theory, we also compute the two-point correlation function of a gauge invariant operator in the trivial vacuum background.  If there is no mass gap, there should be massless propagators, and we expect a power-law decaying behavior for the correlation function at large distance. If there is a mass gap, the mediators should all be massive, and we expect a exponentially decaying behavior for the correlation at large distance.

A natural choice of the gauge invariant operator is
\be\label{eq:opepsilon}
  \epsilon \equiv \frac{1}{4} F_{\mu\nu}^a\, F_{\mu\nu}^a\, ,
\ee
where the field strength $F_{\mu\nu}$ is defined in Eq.~\eqref{eq:fmunudmu}. As discussed in Ref.~\cite{Polyakov}, one can also choose another gauge invariant operator
\be
  \widetilde{\epsilon} \equiv \frac{1}{4} \mathcal{F}_{\mu\nu}^a \mathcal{F}_{\mu\nu}^a\, ,
\ee
where $\mathcal{F}_{\mu\nu}$ is the full field strength defined in Eq.~\eqref{eq:fmunudmuqmu} with quantum fluctuations $Q_\mu$ turned on.  Both $\epsilon$ and $\widetilde{\epsilon}$ are functions of $|x|$ or $\widetilde{\tau}$. As we discussed before, in the IR regime of the theory we turn off all the quantum fluctuations $Q_\mu$, hence we consider the operator $\epsilon$ in the following. We can insert two operators $\epsilon$ at $\vec{x} = \pm \vec{d}$, i.e. $\widetilde{\tau} = d^2$, and consider the two-point correlation function $\langle \epsilon (-\vec{d}\,) \, \epsilon (\vec{d}\,) \rangle$.

\subsubsection{3D Case}\label{sec:3dmediator}

For the 3D case, the operator $\epsilon$ is
\be
  \epsilon = 4 |\partial_{\widetilde{\tau}} \Psi |^2 + \frac{1}{2 \widetilde{\tau}^2} (|\Psi|^2 - 1)^2\, .
\ee
The correlation function $\langle \epsilon (-\vec{d}\,) \, \epsilon (\vec{d}\,) \rangle$ is
\begin{align}\label{eq:3Dcorrelation}
 \langle \epsilon (-\vec{d}\,) \, \epsilon (\vec{d}\,) \rangle & = \frac{1}{Z} \int \mathcal{D} \overline{\Psi}\, \mathcal{D} \Psi\, \int [\textrm{Jac}]_{\gamma_i}^{\textrm{top}} \frac{d^3 x_0}{V} \left( 4 \Big|\partial_{\widetilde{\tau}} \Psi_{\vec{x} = - \vec{d}} \Big|^2 + \frac{1}{2 d^4} \left( \big| \Psi_{\vec{x} = - \vec{d} }\big|^2 - 1 \right)^2 \right) \nonumber\\
  {} & \qquad\qquad\qquad\qquad\qquad\qquad\qquad \cdot \left(4 \Big|\partial_{\widetilde{\tau}} \Psi_{\vec{x} = \vec{d}} \Big|^2 + \frac{1}{2 d^4} \left( \big| \Psi_{\vec{x} = \vec{d}} \big|^2 - 1 \right)^2 \right) \nonumber\\
  {} & \qquad\qquad\qquad\qquad\qquad\qquad\qquad \cdot \textrm{exp} \left(- S \right) \nonumber\\
& = \frac{1}{Z} \int \mathcal{D} \overline{\Psi}\, \mathcal{D} \Psi\, \int [\textrm{Jac}]_{\gamma_i}^{\textrm{top}} \frac{d^3 x_0}{V} \left[ 4 \Big|\partial_{\widetilde{\tau}} \Psi\big|_{\widetilde{\tau} = d^2} \Big|^2 + \frac{1}{2 d^4} \left( \big| \Psi (\widetilde{\tau} = d^2) \big|^2 - 1 \right)^2 \right]^2 \nonumber\\
  {} & \qquad\qquad\qquad\qquad\qquad\qquad\qquad \cdot \textrm{exp} \left(- S \right) \, ,
\end{align}
As discussed above, to study the IR regime of the quantum theory we have turned off the quantum fluctuations $Q_\mu$ and consider the operator $\epsilon$. However, for a full quantumt theory the action $S$ in Eq.~\eqref{eq:3Dcorrelation} in principle is given by $S^{\textrm{top}} + S^{\textrm{qu}}$, which still depends on $Q_\mu$. Since the coupling of the topological fluctuations $\widetilde{\mathcal{A}}$ with the quantum fluctuations $Q_\mu$ only introduces higher order corrections to the correlation function $\langle \epsilon (-\vec{d}\,) \, \epsilon (\vec{d}\,) \rangle$, i.e. at 1-loop or higher order, for the lowest state at the leading order we can replace $S$ in the correlation function by $S^{\textrm{top}}$ discussed in Subsection~\ref{sec:3deffaction}:
\begin{align}
 \langle \epsilon (-\vec{d}\,) \, \epsilon (\vec{d}\,) \rangle_{\textrm{Lowest State}} & \approx \frac{1}{Z} \int \mathcal{D} \overline{\Psi}\, \mathcal{D} \Psi\,  [\textrm{Jac}]_{\gamma_i}^{\textrm{top}} \left[ 4 \Big|\partial_{\widetilde{\tau}} \Psi\big|_{\widetilde{\tau}=d^2} \Big|^2 + \frac{1}{2 d^4} \left( \big| \Psi(\widetilde{\tau} = d^2) \big|^2 - 1 \right)^2 \right]^2 \nonumber\\
  {} &    \quad \cdot \textrm{exp} \left( - \frac{ 3 L_{\textrm{top}}}{g^2 L^3 } \int d^3 x\, \left[ |\partial_{|x|} \Psi (|x|) |^2 + m_{3D}^2 (|\Psi (|x|) |^2 - 1)^2 \right] \right) \, .\label{eq:3dapproxCorr}
\end{align}
 The Jacobian $[\textrm{Jac}]_{\gamma_i}^{\textrm{top}}$ depends on the topological fluctuations, but is independent of $x_0$. At semi-classical level, it cancels out in the the correlation function because of the normalization factor $1/Z$ and does not show up in the final result. Moreover, since the integrand of the correlation function has no $Q_\mu$-dependence, the integral over the gauge orientations $\varphi$ also drops out in the correlation function.

Now let us consider the correlation function \eqref{eq:3dapproxCorr}. In the trivial vacuum background, i.e. $|\Psi| = 1$, the correlation function vanishes identically. If we allow some perturbations around the trivial vacuum background, i.e.,
\be
  |\Psi| = 1 + \phi\, ,
\ee
then at the leading order Eq.~\eqref{eq:3dMGEOM-3} becomes a linear differential equation:
\be
  \partial_{|x|}^2 \phi + \frac{2}{|x|} \, \partial_{|x|} \phi - 4\, m_{3D}^2 \, \phi = 0\, .
\ee
It has the solution
\be
  \phi (|x|) = C_1\, \frac{ e^{-2 \, m_{3D}\, |x|}}{m_{3D}\, |x|} + C_2\,  \frac{e^{2\, m_{3D}\, |x|}}{  m_{3D}\, |x|}\, ,
\ee
where $C_1$ and $C_2$ are two constants. We also expect that $\phi$ should vanish when $|x| \to \infty$, hence we consider only
\be\la{eq:psiexpande}
  \phi (|x|) = C_1\, \frac{ e^{-2 \, m_{3D}\, |x|}}{m_{3D}\, |x|}  \quad \Rightarrow \quad |\Psi| \approx 1 +C_1\, \frac{ e^{-2 \, m_{3D}\, |x|}}{m_{3D}\, |x|}  \, .
\ee
At the semi-classical level, the leading order of the lowest state correlation function for large $d$ becomes
\be
 \langle \epsilon (-\vec{d}\,) \, \epsilon (\vec{d}\,) \rangle_{\textrm{Lowest State}} \approx 16\, C_1^4  \,  \left( \frac{e^{- \, m_{3D}\, d}}{d} \right)^8\, ,
\ee
which implies the existence of a massive mediator and consequently the mass gap in the 3D quantum Yang-Mills theory. To figure out the exact value of the mass gap, one should solve Eq.~\eqref{eq:3dMGEOM-3} exactly as an energy eigenvalue problem with different boundary conditions, and the lowest positive energy in the spectrum corresponds to the mass gap.

As discussed before, we only take into account the leading order contributions and perform the analysis at the semi-classical level. Since we are dealing with a purely bosonic field theory, there are no mechanisms like fermion loops or supersymmetry that can make the mass gap diminished. Hence, we expect that the mass gap found at semi-classical level should retain nonzero, when the full quantum corrections are taken into account.

\subsubsection{4D Case}\label{sec:4dmediator}

For the 4D case, the gauge invariant operator $\epsilon$ has the following expression:
\be
  \epsilon = 24 (\partial_{\widetilde{\tau}} q)^2 +\frac{24}{\widetilde{\tau}^2} \left(q^2 - \frac{1}{4} \right)^2\, ,
\ee
The correlation function reads
\begin{align}
\langle \epsilon (-\vec{d}\,) \, \epsilon (\vec{d}\,) \rangle & = \frac{1}{Z} \int \mathcal{D} q\, \int [\textrm{Jac}]_{\gamma_i}^{\textrm{top}} \frac{d^4 x_0}{V} \left[ 24 \left(\partial_{\widetilde{\tau}} q \big|_{\vec{x} = - \vec{d}} \right)^2 + \frac{24}{d^4} \left(q(\vec{x} = - \vec{d})^2 - \frac{1}{4} \right)^2 \right] \nonumber\\
  {} & \qquad\qquad\qquad\qquad\qquad\quad \cdot \left[ 24 \left(\partial_{\widetilde{\tau}} q \big|_{\vec{x} = \vec{d}} \right)^2 + \frac{24}{d^4} \left(q(\vec{x} = \vec{d})^2 - \frac{1}{4} \right)^2 \right] \nonumber\\
  {} & \qquad\qquad\qquad\qquad\qquad\quad \cdot \textrm{exp} \left(- S \right) \, ,
\end{align}
where in this case the action $S$ should be the sum of the topological action $S^{\textrm{top}}$ and the quantum action $S^{\textrm{qu}}$.

For the lowest state at the leading order, the action $S$ can be replaced by $S^{\textrm{top}}$ discussed in Subsection~\ref{sec:4deffaction}, and the correlation function is
\begin{align}
 \langle \epsilon (-\vec{d}\,) \, \epsilon (\vec{d}\,) \rangle_{\textrm{Lowest State}} & \approx \frac{1}{Z} \int \mathcal{D} q\, [\textrm{Jac}]_{\gamma_i}^{\textrm{top}} \left[ 24 \left(\partial_{\widetilde{\tau}} q \big|_{\widetilde{\tau} = d^2} \right)^2 + \frac{24}{d^4} \left(q^2 (\widetilde{\tau} = d^2) - \frac{1}{4} \right)^2 \right]^2 \nonumber\\
  {} & \quad \cdot \textrm{exp} \left(- \frac{12 L_{\textrm{top}}^2}{g^2 L^4 } \int d^4 x\, \left[ \left(\partial_{|x|} q (|x|) \right)^2 + 4 \, m_{4D}^2 \left(q\,^2 (|x|) - \frac{1}{4} \right)^2 \right] \right) \, .\label{eq:4dapproxCorr}
\end{align}
Again, the Jacobian $[\textrm{Jac}]_{\gamma_i}^{\textrm{top}}$ depends on the topological fluctuations, but is independent of $x_0$. At semi-classical level, it cancels out in the the correlation function because of the normalization factor $1/Z$ and does not show up in the final result. Since the integrand of the correlation function has no $Q_\mu$-dependence, the integral over the gauge orientations $\varphi$ also drops out in the correlation function.

Like in the 3D case, around the 4D trivial vacuum background $q = \pm \frac{1}{2}$ we allow some perturbations:
\be
  q = \pm \frac{1}{2} + \varphi\, ,
\ee
then at the leading order Eq.~\eqref{eq:4dMGEOM} becomes a linear differential equation:
\be\label{eq:lineardiffequa4d}
  \partial_{|x|}^2\, \varphi + \frac{3}{|x|}\, \partial_{|x|}\, \varphi - 4 \, m_{4D}^2\, \varphi = 0\, .
\ee
It has the solution
\be
  \varphi (|x|) = C_1\,  \frac{K_1 (2 \, m_{4D} \, |x|)}{ m_{4D} \, |x|} + C_2\, \frac{ I_1 (2 \, m_{4D} \, |x|)}{ m_{4D} \, |x|}\, ,
\ee
where $C_1$ and $C_2$ are two constants, while $I_1$ and $K_1$ are the modified Bessel functions of the first and the second kind respectively. We also expect that $\varphi$ should vanish when $|x| \to \infty$, hence we consider only
\be\la{eq:psiexpande2}
  \varphi (|x|) = C_1\,  \frac{K_1 (2 \, m_{4D} \, |x|)}{ m_{4D} \, |x|} \quad \Rightarrow \quad q  \approx \frac{1}{2} +C_1\,  \frac{K_1 (2 \, m_{4D} \, |x|)}{ m_{4D} \, |x|}\, .
\ee
At the semi-classical level, the leading order of the lowest state correlation function for large $d$ becomes
\be
\langle \epsilon (-\vec{d}\,) \, \epsilon (\vec{d}\,) \rangle_{\textrm{Lowest State}} \approx \frac{36\, C_1^4\, \pi^2}{ m_{4D}^2  \, d^{2}}\,    \left( \frac{e^{- \, m_{3D}\, d}}{d} \right)^8\, ,
\ee
which implies the existence of a massive mediator and consequently the mass gap in the 4D quantum Yang-Mills theory. To figure out the exact value of the mass gap, one should solve Eq.~\eqref{eq:4dMGEOM} exactly as an energy eigenvalue problem with different boundary conditions, and the lowest positive energy in the spectrum corresponds to the mass gap.

Therefore, for the 4D case our results also support the existence of the mass gap at semi-classical level. Although the quantum corrections can change the value, they cannot make the gap diminished.

\subsection{Field Equation and Nonlinear Schr\"odinger Equation}\label{sec:sec6remark}

We have seen that for both the 3D and the 4D case, after taking the integration over $x_0$, the effective action has a similar expression. For 3D case, we can write Eq.~\eqref{eq:3dMGcutoff} as
\be\label{eq:general3DEffAction}
\left<S\right>_{x_0} =  \frac{3 L_{\textrm{top}}}{g^2 L^3} \int d^3 x\, \left[ \left(\partial_{|x|}  |\Psi|   \right)^2 + m_{3D}^2 \left(|\Psi|^2 - 1\right)^2 \right] \, .
\ee
For the 4D case, after redefining $q \rightarrow q/2$, Eq.~\eqref{eq:4dMGcutoff} reads
\be\label{eq:general4DEffAction}
\left<S\right>_{x_0}  =  \frac{3 L_{\textrm{top}}^2}{g^2 L^4} \int d^4 x\, \left[ \left(\partial_{|x|} q   \right)^2 + m_{4D}^2 \left(q^2   - 1\right)^2 \right]\, .
\ee
If we rescale $x$ as $x/L$ to make it dimensionless, we obtain the following expression for both the 3D and the 4D case
\be\la{eq:rescaleSD}
\left<S\right>_{x_0} =  \frac{3 L_{\textrm{top}}^{D-2} }{g^2 L^2} \int_{|x| \leq 1} d^D x\, \left[ \left(\partial_{|x|} \Phi  \right)^2 + m^2_D \left(\Phi^2 - 1\right)^2 \right] \, .
\ee
where $m_D^2$ is also a dimensionless constant, which is defined as
\be\label{eq:defmD}
  m_D^2 = \bigg\{
  \begin{array}{ll}
    m_{3D}^2 L^2\, , & \quad \textrm{for } D = 3\, ;\\
    m_{4D}^2 L^2\, , & \quad  \textrm{for } D = 4\, .
  \end{array}
\ee
One can replace $m_{3D}^2$ and $m_{4D}^2$ in the expression above with their definitions \eqref{eq:defm3D} \eqref{eq:defm4D}. Because of $L \ge L_{\textrm{top}} \gg \ell_{\textrm{top}}$ there should be $m_D^2 \gg 1$ for both $D=3$ and $D=4$.

The equation of motions from these effective actions are of the same type:
\be
  \Delta u - 2 m^2 u (u^2 - 1) = 0\quad \textrm{with}\quad \Delta u = \partial^2_{|x|} u + \frac{D-1}{|x|} \partial_{|x|} u\, ,
\ee
which is a time-indepedent defocusing nonlinear Schr\"odinger equation.  More precisely, since we consider Euclidean spaces, there is no time in the system, and all the configurations should be time-independent. Therefore, our original problem becomes this special kind of nonlinear Schr\"odinger equation, which has been studied in mathematical literature.

In particular, the defocusing cubic nonlinear Schr\"odinger equation in 3D and 4D
\be
  i \frac{\partial u}{\partial t} + \Delta u = (|u|^2 - 1) u
\ee
has been studied in \cite{Gerard-1, Gerard-2, Visan}. It can be associated with the Ginzburg-Landau energy
\be
  H (u) = \int_{\mathbb{R}^D} d^D x\, \left[\frac{1}{2} |\nabla u(x)|^2 + \frac{1}{4} (|u|^2 - 1)^2 \right]\, ,
\ee
which is essentially the same as Eq.~\eqref{eq:3dMGcutoff} and Eq.~\eqref{eq:4dMGcutoff} in our case. The energy space is defined as
\be
  E = \{u \in H_{\textrm{loc}}^1 (\mathbb{R}^d): \nabla u \in L^2 (\mathbb{R}^D),\, |u|^2 - 1 \in L^2 (\mathbb{R}^D) \}\, .
\ee
For $D \geq 3$ one also defines the homogeneous Sobolev space $\dot{H}^1 (\mathbb{R}^D)$ as
\be
  \dot{H}^1 (\mathbb{R}^D) = \{u \in L^{2^*} (\mathbb{R}^D): \nabla u \in L^2 (\mathbb{R}^D) \} \quad \textrm{with} \quad 2^* = \frac{2 D}{D - 2}\, .
\ee
In particular, G\'erard has proven in Ref.~\cite{Gerard-2} the following theorem:
\begin{flushleft}
  {\bf Theorem} For $D \geq 3$ the energy space $E$ is described as follows:
\end{flushleft}
\be
  E = \{u = c (1 + v): c \in \mathbb{S}^1,\, v \in \dot{H}^1 (\mathbb{R}^D),\, 2 \, \textrm{Re}(v) + |v|^2 \in L^2 (\mathbb{R}^D) \}\, .
\ee
 As a special case, the solution to the time-independent defocusing nonlinear Schr\"odinger equation with a cubic interaction should also lie in this solution space. Hence, our treatment in this section of expanding the topological fluctuations around the trivial vacuum background is consistent with this mathematical theorem.

\section{\label{MG}Mass Gap}

In this section, we would like to discuss the long-standing problem of the mass gap in Yang-Mills theory. According to Ref.~\cite{MassGap}, the mass gap $\Delta$ of a quantum field theory is defined in the following way: The Hamiltonian $H$ has no spectrum in the interval $(0,\, \Delta)$ for some $\Delta > 0$. The spectrum of such $\Delta$ is the mass $m$ with $m < \infty$. The mass gap problem of Yang-Mills theory is formulated as follows \cite{MassGap}:

``{\it Prove that for any compact simple gauge group $G$, a non-trivial quantum Yang-Mills theory exists on $\mathbb{R}^4$ and has a mass gap $\Delta > 0$.}''

In other words, if we could prove that the lowest energy state other than the trivial vacuum in the spectrum has the mass $m > 0$, we would find the mass gap. In Ref.~\cite{Polyakov}, the author explicitly computed all the states for the (2+1)D Georgi-Glashow model, including the lowest-lying state and the excited states, of the mediator exchanged between two gauge invariant operators, and thus not only found the mass gap but also obtained the complete correlation function between two points. In Subsection~\ref{sec:mediator}, we considered the mediator exchanged between two gauge invariant operators, only focusing on the lowest-lying state at semi-classical level. By knowing the lowest-lying state of the mediator, we still cannot compute the complete correlation between two points because the excited states could also be important. However, if the mediator in the lowest-lying state acquires a mass, then the lowest state in the spectrum has a nonzero energy at semi-classical level, which is a very strong evidence for the mass gap.

To more rigorously demonstrate the existence of the mass gap, one should directly analyze the 3D and the 4D effective action, which are given by Eq.~\eqref{eq:3dMGcutoff} and Eq.~\eqref{eq:4dMGcutoff} respectively. As we have seen in the previous section, these effective actions lead to the field equations \eqref{eq:3dMGEOM-3} and \eqref{eq:4dMGEOM} respectively, which are of the same type, i.e. the time-independent defocusing nonlinear Schr\"odinger equation with a cubic interaction, which is also known as the time-independent defocusing Gross-Pitaevskii equation. Hence, in our analysis the mass gap problem of the quantum Yang-Mills theory becomes a mass gap problem of this type of nonlinear Schr\"odinger equation, which generally in $D$-dimensional space is given by
\be\label{eq:generalNLS}
  \Delta u - (|u|^2 - 1) u = 0\, ,
\ee
where for spherically symmetric configurations
\be
  \Delta u = \partial_{|x|}^2 u + \frac{D-1}{|x|}\, \partial_{|x|} u\, ,
\ee
and $x$ takes values in $\mathbb{R}^n$. For the trivial vacuum background, the boundary condition is given by
\be
  \lim_{|x| \to 0} |u(x)| = \lim_{|x| \to \infty} |u(x)| = 1\, .
\ee
This type of nonlinear Schr\"odinger equation has been studied in the mathematical literature, for instance, in Refs.~\cite{Gerard-1, Gerard-2, Visan} mentioned in Subsection~\ref{sec:sec6remark}.

The mass gap problem of the defocusing Gross-Pitaevskii equation was studied in Ref.~\cite{Singapore-1, Lieb, Weinstein} and more recently by Bao and Ruan in Ref.~\cite{Singapore-2} numerically and asymptotically, where the fundamental gap, i.e. the gap between the ground state and the first excited state, of another kind of nonlinear Schr\"odinger equation
\be\label{eq:EqSingapore}
  \left(-\frac{1}{2} \Delta + \beta |\phi|^2 \right) \phi = \mu \phi
\ee
was investigated under various boundary conditions. Here $\beta$ is a positive constant, and $\mu$ is the chemical potential. For the mass gap problem, we focus on the trivial vacuum background, which corresponds to the periodic boundary condition discussed in Ref.~\cite{Singapore-2}. We summarize some relevant results of Ref.~\cite{Singapore-2} for the periodic boundary condition in Appendix~\ref{GP}. To illustrate the other boundary conditions, we present the analytical results of the 1D problem as a toy model in Appendix~\ref{Toy}.

As discussed in Appendix~\ref{GP}, using Eq.~\eqref{eq:mapNLS}:
\begin{displaymath}
  \psi = \sqrt{\frac{\beta}{\mu}} \phi\, ,
\end{displaymath}
we can map Eq.~\eqref{eq:EqSingapore} to the equation derived in Subsection~\ref{sec:sec6remark}:
\be\label{eq:EqOurCase}
  \Delta \psi - 2 m_D^2 (|\psi|^2 - 1) \psi = 0\, ,
\ee
and obtain the relation \eqref{eq:NewE}:
\begin{displaymath}
  E = \mu V - 2 \beta \left(1 - \frac{\widetilde{E}}{\mu} \right)\, ,
\end{displaymath}
where $E$ is the energy for Eq.~\eqref{eq:EqOurCase} in our case defined by
\be
  E = \int_\Omega dx \left[|\nabla \psi|^2 + \mu (|\psi|^2 - 1)^2 \right]\, ,
\ee
and $V$ is the finite volume of the system. The energy $\widetilde{E}$ and the chemical potential $\mu$ for Eq.~\eqref{eq:EqSingapore} are defined by Eqs.~\eqref{eq:SingaporeE} \eqref{eq:Singaporemu} as follows:
\begin{align}
  \widetilde{E} (\phi, \beta) & = \int_\Omega dx \left[\frac{1}{2} |\nabla \phi|^2 + \frac{\beta}{2} |\phi|^4 \right]\, ,\nonumber\\
  \mu (\phi, \beta) & = \widetilde{E}(\phi, \beta) + \frac{\beta}{2} \int_\Omega dx\, |\phi|^4\, .\nonumber
\end{align}
The chemical potential $\mu$ can also be identified with $m_D^2 \gg 1$ appearing in the effective action \eqref{eq:rescaleSD}, and $\Omega$ denotes the finite domain $|x| \leq 1$.

Our case corresponds to the limit $\beta \gg 1$ in the Ref.~\cite{Singapore-2}, which is implied by the condition $\mu = m_D^2 \gg 1$, as we will see in the following. For the trivial vacuum background, we know that the ground state is the trivial vacuum background itself with $E_g=0$, and Ref.~\cite{Singapore-2} has shown that for the periodic boundary condition when $\beta \gg 1$ the energy $\widetilde{E}_1$ and the chemical potential $\mu_1$ for the first excited state are
\begin{align}
  \widetilde{E}_1 & = \frac{1}{2V} \beta + \frac{8}{3 L_1 \sqrt{V}} \beta^{1/2} + \frac{8}{L_1^2} + o(1)\, ,\\
  \mu_1 & = \frac{1}{V} \beta + \frac{4}{L_1 \sqrt{V}} \beta^{1/2} + \frac{8}{L_1^2} + o(1)\, ,\label{eq:mu1}
\end{align}
where $L_1$ is the biggest size of the system. Since the volume $V$ of the domain $|x| \leq 1$ is of the order $1$, hence $\mu_1 = m_D^2 \gg 1$ implies that $\beta \gg 1$, which is also consistent with our assumption.

Therefore, for a large value of $L_1$ Eq.~\eqref{eq:NewE} leads to at leading order
\begin{align}
  E & = \frac{16 \sqrt{V} \beta^{1/2}}{3 L_1} + o (\beta) \nonumber\\
  {} & \approx \frac{16 V \mu_1^{1/2}}{3 L_1} = \frac{16 V m_D}{3 L_1}\, ,
\end{align}
where in the last step we used the relation \eqref{eq:mu1} when $\beta \gg 1$.

Precisely speaking, Ref.~\cite{Singapore-2} considers a system with the rectangular geometry, while in this paper we focus on the spherically symmetric space, hence the results may differ by a constant factor. Nevertheless, for general dimension $D \geq 1$:
\be
  E \propto \frac{  V  }{  L_1}  m_D\,.
\ee
Remember that this result is expressed in terms of the dimensionless variables after the rescaling introduced in Subsection~\ref{sec:sec6remark}. Hence, $V$ and $L_1$ are two dimensionless constants of order $1$.
Combining it with Eq.~\eqref{eq:rescaleSD}, we obtain for $D=3, 4$ the effective action evaluated at the first excited state on the trivial vacuum background up to a constant
\be
  \langle S \rangle_{x_0} \propto \frac{ L_{\textrm{top}}^{D-2} }{g^2 L^2 }\, m_D \, ,
\ee
where $m_D$ is given by Eq.~\eqref{eq:defmD}:
\begin{displaymath}
  m_D^2 = \bigg\{
  \begin{array}{ll}
    m_{3D}^2 L^2\, , & \quad \textrm{for } D = 3\, ;\\
    m_{4D}^2 L^2\, , & \quad \textrm{for } D = 4\, .
  \end{array}
\end{displaymath}
In principle, the coupling $g$ depends on the length scale $L$. Although in the IR regime the coupling $g$ can be very large, as long as $L$ is finite, $g(L)$ should also remain finite. From the result above one can learn that, for a finite $L$ the first excited state on the trivial vacuum background always has a positive finite action compared to the ground state given by the trivial vacuum background with zero energy, therefore, the system has a mass gap.

As an estimate, we can make the assumption $L \approx L_{\textrm{top}}$, and apply the explicit expressions of the effective masses $m_{3D}$, $m_{4D}$ defined by Eq.~\eqref{eq:defm3D} and Eq.~\eqref{eq:defm4D} respectively. The results are following:
\begin{itemize}
\item For the flat space $\mathbb{R}^D$ $(D = 3, 4)$ with finite size:

\be
  \langle S \rangle_{x_0} \propto \Bigg\{
  \begin{array}{ll}
    \frac{1}{g^2} \frac{1}{\sqrt{L_{\textrm{top}} \ell_{\textrm{top}}}} \, , & \quad  \textrm{for 3D}\, ;\\
     \frac{1}{g^2}\, \textrm{log} \left(  \frac{L_{\textrm{top}}}{\ell_{\textrm{top}}} \right) \, , & \quad  \textrm{for 4D}\, .
  \end{array}
\ee

\item For the sphere $S^D$ $(D = 3, 4)$ with a radius $R$:

\be
  \langle S \rangle_{x_0} \propto \Bigg\{
  \begin{array}{ll}
    \frac{1}{g^2 R} \, , &\quad \textrm{for 3D}\, ;\\
     \frac{1}{g^2}\, \textrm{log} \left[  \textrm{cot} \left( \frac{\theta_0}{2} \right) \right] \, ,&\quad \textrm{for 4D}\, ,
  \end{array}
\ee
where $\theta_0$ is the physical cutoff on $\theta$ discussed in Subsection~\ref{cutoff}.

\end{itemize}

In this section we relate the mass gap problem of the quantum Yang-Mills theory to the mass gap problem of two kinds of defocusing cubic nonlinear Schr\"odinger equations, which are related to each other. As we discussed in Section~\ref{LE}, the quantum corrections may change the value of the mass gap, but cannot make it vanish. Hence, for the spherically symmetric configurations at the semi-classical level we can demonstate the existence of the mass gap of the quantum $SU(2)$ Yang-Mills theory on a flat space $R^D$ $(D=3, 4)$ with finite size or on a compact space such as a sphere $S^D$ $(D=3, 4)$.

\section{\label{Discussion}Discussions}

In this paper, we analyzed the quantum Yang-Mills theory. We have explored some key ideas that have been overlooked before. These key ideas include the form invariance, the topological properties and the topological fluctuations. Traditionally, only the solutions to the field equations have been used as the background configurations. However, we have emphasized that the topologically stable configurations, which are not solutions but constrained by the form invariance condition and the topological properties, may also be used as backgrounds. Such kinds of backgrounds bring more abundant structures into play, which can provide us with the missing blocks to help resolve the long-standing mass gap problem in the pure Yang-Mills theory.

Some possible generalizations are being considered, and will be presented in future papers soon. For example, in this paper we discussed the Yang-Mills theory in the 3D and the 4D Euclidean spaces, however, to compare with the physics in the real world we should work in the (3+1)D Minkowski spacetime. It would be more fascinating if some consequences of this new perspective of Yang-Mills theory could be observed in experiments. In this paper, for the classical solutions we focus on the spherically symmetric cases, one can also consider the non-spherically symmetric solutions, e.g. multi-center solutions. Moreover, the Yang-Mills theory defined on a curved spacetime \cite{Nian:2019vbm, Nian:2019ucx}, either Riemannian or Lorentzian, would also be interesting.

Moreover, we would like to apply the low-energy effective theory obtained in this paper to some real physical systems, and compare the results with either experimental or lattice data. In the absence of quarks, we expect that the effective theory should be able to reproduce various glueball masses obtained on lattice (see e.g. \cite{Weingarten, Morningstar:1999rf, Lucini:2004my, Chen:2005mg, Ochs:2013gi}). According to the effective theory, we also conjecture that without invoking the full quantum fluctuations, the topological fluctuations should be enough to account for the existence of mass gap, which in principle can be directly verified on the lattice.

As we suggested in the main text, with this new point of view at hand, people should revisit some old problems and bring new ideas to them, for instance the fine-tuning problem of standard model. We believe that the framework with the topological fluctuations will provide an alternative approach to these problems. Also, the notion of the topological fluctuations in (3+1)-dimensions will provide new interesting ideas for understanding the confinement problem of the Yang-Mills theory. We would like to explore this new perspective in the near future.

Finally, we hope that the new concepts and ideas introduced in this paper can deepen our understanding of gauge theories and widen our view of general quantum field theories. At least, we hope this paper will open up a new way of studying the vacuum structure and the quantum properties of gauge theories.

\section*{Acknowledgements}

We would like to thank Alexander Abanov and Peter van Nieuwenhuizen for many useful discussions. We would like to especially thank Scott Mills for many enlightening discussions and proofreading the manuscript. Y.~Q. wishes to thank Ismail Zahed and Edward Shuryak for helpful discussions, and the Nuclear Theory Group at Stony Brook University for supporting. J.~N. is very grateful to Felix G\"unther, Maxim Kontsevich, Yang Lan, J\'ozsef L\"orinczi and Vasily Pestun for discussions, and would like to thank Institut des Hautes \'Etudes Scientifiques and C.~N. Yang Institute for Theoretical Physics for supporting.

\appendix

\section{\label{app:notation}Convention}

In this appendix we summarize some conventions used in the paper. First, the Lie algebra $\mathfrak{so}(4)$ has the  generators given by
\be\label{eq:SO4gen}
  (M_{\mu\nu})_{mn} \equiv \delta_{\mu m} \delta_{\nu n} - \delta_{\mu n} \delta_{\nu m}\, ,
\ee
which satisfy
\be
  [M_{\mu\nu},\, M_{\rho\sigma}] = \delta_{\nu\rho} M_{\mu\sigma} + \delta_{\mu\sigma} M_{\nu\rho} - \delta_{\mu\rho} M_{\nu\sigma} - \delta_{\nu\sigma} M_{\mu\rho}\, .
\ee
We choose a special representation of the generators as follows:
\be
  M_{23} = \left( \begin{array}{cccc}
    \,0\,& \,0\, & \,0\, & \,0\,\\
    0 & 0 & 1 & 0\\
    0 & -1 & 0 & 0\\
    0 & 0 & 0 & 0
  \end{array} \right) \equiv J_1\, ,\quad
 M_{14} = \left( \begin{array}{cccc}
    \,0\, & \,0\, & \,0\, & 1\\
    0 & 0 & 0 & \,0\,\\
    0 & 0 & 0 & 0\\
    -1 & 0 & 0 & 0
  \end{array} \right) \equiv K_1\, , \nonumber
\ee
\be
M_{31} = \left( \begin{array}{cccc}
    0 & 0 & -1 & 0\\
    \,0\, & \,0\, & \,0\, & \,0\, \\
    1 & 0 & 0 & 0\\
    0 & 0 & 0 & 0
  \end{array} \right) \equiv J_2\, , \quad
M_{24} = \left( \begin{array}{cccc}
    \,0\, & \,0\, & \,0\, & \,0\, \\
    0 & 0 & 0 & 1\\
    0 & 0 & 0 & 0\\
    0 & -1 & 0 & 0
  \end{array} \right) \equiv K_2\, , \nonumber
\ee
\be\label{eq:SO4redef}
M_{12} = \left( \begin{array}{cccc}
    0 & 1 & 0 & 0\\
    -1 & 0 & 0 & 0\\
    \,0\, & \,0\, & \,0\, & \,0\, \\
    0 & 0 & 0 & 0
  \end{array} \right) \equiv J_3\, , \quad
  M_{34} = \left( \begin{array}{cccc}
    \,0\, & \,0\, & \,0\, & \,0\, \\
    0 & 0 & 0 & 0\\
    0 & 0 & 0 & 1\\
    0 & 0 & -1 & 0
  \end{array} \right) \equiv K_3\, .
\ee
They satisfy
\be\label{eq:SO4alg}
  [J_i,\, J_j] = -\epsilon_{ijk} J_k\, ,\quad [K_i,\, K_j] = -\epsilon_{ijk} J_k\, ,\quad [K_i,\, J_j] = -\epsilon_{ijk} K_k\, .
\ee
For convenience, we also define
\be\label{eq:defMN}
  M_i \equiv \frac{1}{2} (J_i + K_i)\, ,\quad N_i \equiv \frac{1}{2} (J_i - K_i)\, ,
\ee
and their (anti-)commutation relations are
\be\label{eq:MNalg}
  [M_i,\, M_j] = -\epsilon_{ijk} M_k\, ,\quad [N_i,\, N_j] = -\epsilon_{ijk} N_k\, ,\quad [M_i,\, N_j] = 0\, ,
\ee
\be
  \{ M_i,\, M_j \} = -\frac{1}{2} \delta_{ij}\, ,\quad \{N_i, \, N_j \} = -\frac{1}{2} \delta_{ij}\, .
\ee

$M_i$ and $N_i$ together form a complete basis of $\mathfrak{so}(4)$ algebra. We can expand $M_{\mu\nu}$ as 
\be\label{eq:etadef}
M_{\mu\nu} \equiv \eta_{i\mu\nu}  M_i + \bar{\eta}_{i\mu\nu}  N_i \, ,
\ee
where $\eta$ and $\bar{\eta}$ are 't Hooft symbols, which can be expressed as
\begin{align}
 \eta_{i\mu\nu}    & = -  \textrm{tr} \left( M_i M_{\mu\nu}  \right) = - (M_i)_{mn} (M_{\mu\nu})_{nm} = 2   (M_i)_{\mu\nu}  =   (\epsilon_{i\mu\nu4} + \delta_{i\mu} \delta_{\nu 4} - \delta_{i\nu} \delta_{\mu 4})  \, ,\nonumber\\
 \bar{\eta}_{i\mu\nu}    & = -  \textrm{tr} \left( N_i M_{\mu\nu}  \right) = - (N_i)_{mn} (M_{\mu\nu})_{nm} = 2   (N_i)_{\mu\nu}  =  (\epsilon_{i\mu\nu 4} - \delta_{i\mu} \delta_{\nu 4} + \delta_{i\nu}\delta_{\mu 4})  ,
\end{align}
where we used
\be 
  \textrm{tr} \left( M_i M_j  \right) = -\delta_{i j} \, , \quad  \textrm{tr} \left( N_i N_j  \right) = -\delta_{i j} \, , \quad \textrm{tr} \left( M_i N_j  \right) =0 \, .
\ee
Some important properties are
\be
\frac{1}{2}\epsilon_{\mu\nu\rho\sigma} \eta_{i\rho\sigma} =  \eta_{i\mu\nu} \, , \quad  \frac{1}{2}\epsilon_{\mu\nu\rho\sigma} \bar{\eta}_{i\rho\sigma}  =  - \bar{\eta}_{i\mu\nu} \, .
\ee

\section{Proof of Eq.~\eqref{eq:FI}}\label{guchaohao}

In this appendix, we review a theorem proven in Ref.~\cite{Gu}. The form invariance condition \eqref{eq:FI} discussed in the text can be proven in the same way.

\begin{flushleft}
  {\bf Theorem:} $\mathscr{F}$ is a spherically symmetric Yang-Mills field strength on $\mathbb{R}^{3+1}$, if and only if after a suitable gauge transformation its gauge potential satisfies
\end{flushleft}
\be\label{appPhase}
  b(Ax, dAx) = (\textrm{ad}\, u_A) \, b(x, dx)
\ee
for each $A\in SO(3)$, where $u_A \in G$ may depend on $A$ and does not depend on $x$, and $b$ is the gauge potential.

\begin{flushleft}
  {\bf \underline{Proof}}
\end{flushleft}

As discussed in Ref.~\cite{Gu}, the field strength $\mathscr{F}$ after a Lorentz transformation $A$, which is denoted by $\mathscr{F}^A$, can be defined by
\be
  \left(\Phi^A \right)_{P^A Q^A} = \Phi_{PQ}\, ,
\ee
where
\be
  \Phi_{PQ} \equiv \mathcal{P}\, \textrm{exp} \left[-\int_Q^P b_\mu (x) \, dx^\mu \right]
\ee
is the phase factor with the endpoints $P$ and $Q$, which is also called the Wilson line. We choose the so-called central gauge, in which the phase factor for a straight line equals the unity $I$. Hence, the phase factor for a path $AB$ equals the phase factor for the loop $OABO$, where the point $O$ is the origin, and $OA$ and $BO$ are straight line segments. For $A\in SO(3)$ and a loop $O Q P O$, where $OP$ and $QO$ are straight line segments, the field strength after the Lorentz transformation is given by
\be\label{appPhase-1}
  \left(\Phi^A \right)_{O Q^A P^A O} = \Phi_{O Q P O}\, .
\ee

A (3+1)D Yang-Mills field given by $\mathscr{F}$ is called spherically symmetric, if for any space rotation denoted by $A$ around the fixed point $O$ (the origin) $\mathscr{F}^A$ is equivalent to $\mathscr{F}$, where the space rotation in $\mathbb{R}^{3+1}$ is given by $A = (a^i\,_j) \in \textrm{SO}(3)$:
\be
  (x')^i = a^i\,_j\, x^j\, ,\quad (x')^0 = x^0\, .
\ee

To have a criterion for the equivalence of Yang-Mills fields, let us also quote Theorem~2.1 in Ref.~\cite{Gu} without proving it here. The theorem says that two Yang-Mills fields are equivalent if and only if their loop phase factors are related by
\be
  \Phi_l' = w \Phi_l w^{-1}\, ,
\ee
where $w \in G$ is independent of $x$ and the loop $l$. Based on this theorem, two Yang-Mills fields are equivalent if and only if their loop phase factors are related by
\be\label{appPhase-2}
  \Phi'_{O Q^A P^A O} = (\textrm{ad}\, u_A) \Phi_{O Q^A P^A O}\, ,
\ee
where $u_A \in G$ is independent of $x$.

Because a (3+1)D spherically symmetric Yang-Mills field is equivalent to the original one after a space rotation, we can combine Eq.~\eqref{appPhase-1} with Eq.~\eqref{appPhase-2} to obtain
\be\label{appPhase-3}
  \Phi_{O Q^A P^A O} = (\textrm{ad}\, u_A) \Phi_{O Q P O}\, .
\ee
In the central gauge, Eq.~\eqref{appPhase-3} implies that
\be
  \Phi_{Q^A P^A} = (\textrm{ad}\, u_A) \Phi_{QP}\, ,
\ee
and consequently,
\be
  b(A x, d A x) = (\textrm{ad}\, u_A) \, b(x, dx)\, ,
\ee
where $u_A \in G$ does not depend on $x$.
{\flushleft $\square$}

We call Eq.~\eqref{appPhase} the form invariance relation in the text. Similarly, one can prove this relation for $\mathbb{R}^3$ or $\mathbb{R}^4$ under some additional constraints on the factors $p$ and $\theta$ in the Ansatz.

\section{\label{omega}Classification of $\omega$}

\subsection{\label{omegarestriction}Form Invariance of 3D Ansatz and Restriction on $\omega$}
In this subsection of Appendix~\ref{omega}, we discuss the constraints on $\omega^a\,_\mu$ due to the form invariance condition \eqref{eq:FI}:
\begin{displaymath}
 O^{-1} \,_\mu\,^\nu\, A_\nu (O\, x) = V^{-1}\, A_\mu (x)\, V\, .
\end{displaymath}

The left-hand side of Eq.~\eqref{eq:FI} equals
\be
 (O^{-1})_\mu\,^\nu\, A_\nu (O x) = p(\tau) \, \textrm{exp} \left[- T_a\, \omega^a\,_\alpha\, O^\alpha\,_\rho\, \hat{n}^\rho\, \theta(\tau) \right] \, \frac{\partial}{\partial x^\mu}\, \textrm{exp} \left[T_b\, \omega^b\,_\beta\, O^\beta\,_\sigma\, \hat{n}^\sigma\, \theta(\tau) \right] \, .\label{eq:3DLHS}
\ee
The right-hand side of Eq.~\eqref{eq:FI} equals
\be
  V^{-1} A_\mu V = p(\tau)\, \textrm{exp} \left[- (T_c U^c\,_a)\, \omega^a\,_\rho \, \hat{n}^\rho\, \theta(\tau) \right] \partial_\mu \, \textrm{exp} \left[(T_d U^d\,_b)\, \omega^b\,_\sigma\, \hat{n}^\sigma\, \theta(\tau) \right]\, ,\label{eq:3DRHS}
\ee
where
\be
\hat{n}'^a \equiv \omega^a\,_\rho \frac{x^\rho}{|x|}\, ,
\ee
and $U_a\,^b$ denotes the group elements of $SO(3)$. We have applied
\be\label{eq:3DSU2SO3Rel}
V^{-1} T_a V = T_c \, U^c\,_a\, ,
\ee
where $U$ is a constant $SO(3)$ matrix, and $V$ is a constant $SU(2)$ matrix.

By comparing the final expressions of Eq.~\eqref{eq:3DLHS} and Eq.~\eqref{eq:3DRHS}, we see that in order for Eq.~\eqref{eq:FI} to hold, there should be
\be \label{eq:3Dcond-1}
T_c\, \omega^c \,_\alpha\, O^\alpha\,_\rho\, \hat{n}^\rho =  T_c \,U^c\,_a \, \omega^a\,_\rho \, \hat{n}^\rho \, ,
\ee 
where both $O$ and $U$ denote the group elements of $SO(3)$. Since the equation above is true for any SU(2) generator $T_c$ and any vector $\hat{n}^\rho$, we obtain
\be\la{eq:omegacond}
\omega O = U \omega\, ,
\ee
where we suppressed the contracted indices. Then we have
\begin{align}
  O^T \omega^T   \omega O  & =   \omega^T U^T   U \omega  \nonumber\\
\Rightarrow\quad  O^T \omega^T \omega O   & =   \omega^T \omega  \, .
\end{align}
The expression above means that $\omega^T \omega $ is invariant under arbitrary $SO(3)$ rotations, and therefore
\be\label{eq:3Dcond-2}
\omega^T \omega \propto I\, ,
\ee
where $I$ is the $3 \times 3$ unit matrix. Let us recall that (see Eq.~\eqref{eq:3DAnsatz1a})
\be
\frac{\psi^a (x)}{|\psi (x)|}  = \omega^a\,_\mu \, \hat{n}^\mu \, ,
\ee
which leads to
\be\label{eq:3Dcond-norm}
1 = \frac{\psi^a (x)}{|\psi (x)|} \frac{\psi^a (x)}{|\psi (x)|} =  \omega^a\,_\mu \, \hat{n}^\mu \, \omega^a\,_\nu \, \hat{n}^\nu = \hat{n}^\mu \left( \omega^T \omega  \right)_{\mu\nu} \hat{n}^\nu \, .
\ee
Combining Eq.~\eqref{eq:3Dcond-2} and Eq.~\eqref{eq:3Dcond-norm}, we obtain
\be\label{eq:3Dcond-2a}
\omega^T \omega =  I \, ,
\ee
and thus $\omega$ is an $O(3)$ group element. Since our Ansatz \eqref{eq:3DAnsatz} is isotropic,  $\omega$ can only depend on $\tau$.

As we will prove now, $\omega$ is in fact a constant $O(3)$ group element. Let us recall Eq.~\eqref{eq:omegacond}:
\begin{displaymath}
\omega O = U \omega  \, ,
\end{displaymath}
which means that for an arbitrary constant $SO(3)$ matrix $O$ there is always a constant $SO(3)$ matrix $U$, such that  Eq.~\eqref{eq:omegacond} is satisfied globally. We notice that $O = U = I$ satisfies Eq.~\eqref{eq:omegacond}, hence we can consider $O$ and $U$ around the unit matrix. Let us expand $O$ and $U$ to the leading order
\be\la{eq:expandaround}
O = I + \delta\lambda^a_O \, \hat{T}_a\, ,\quad U=I + \delta\lambda^a_U \,\hat{T}_a\, ,
\ee
where $\hat{T}_a$ denote the $SO(3)$ generators in the representation $(\hat{T}_a)_{i j} = \epsilon_{a i j}$, and $\delta \lambda^a_{O/U}$ are constants fixed by Eq.~\eqref{eq:omegacond}. Plugging Eq.~\eqref{eq:expandaround} back into Eq.~\eqref{eq:omegacond}, we obtain
\be\la{eq:condeq}
\omega  \left( \delta \lambda^a_O   \hat{T}_a   \right)  =   \left( \delta \lambda^{a'}_U    \hat{T}_{a'} \right) \omega .
\ee
If $\omega$ is a function of $\tau$, we have
\begin{align}
  \omega (\tau)  \left( \delta \lambda^a_O   \hat{T}_a   \right)  & = \left( \delta \lambda^{a'}_U   \hat{T}_{a'} \right)  \omega(\tau) \nonumber\\
  \omega (\tau + \delta \tau)  \left( \delta \lambda^a_O   \hat{T}_a   \right) & = \left( \delta \lambda^{a'}_U  \hat{T}_{a'} \right)  \omega(\tau + \delta \tau) \, .
\end{align}
We can expand $\omega(\tau + \delta \tau)$ to the leading order
\be
\omega(\tau + \delta \tau)  = \omega(\tau) \left(I + \delta \lambda_\omega^a \hat{T}_a\right) \, .
\ee
Combining everything above, we obtain
\be\la{eq:conddiscuss1}
\omega(\tau)  \left(  \delta \lambda_\omega^b \hat{T}_b \right)  \left( \delta \lambda^a_O \hat{T}_a   \right)  = \omega (\tau)  \left( \delta \lambda^{a'}_O \hat{T}_{a'}   \right)   \left( \delta \lambda_\omega^{b'} \hat{T}_{b'}\right) \, .
\ee
Since Eq.~\eqref{eq:conddiscuss1} must be valid for an arbitrary $\delta \lambda_{O}^a$, there should be
\be\la{eq:conddiscuss2}
\left(  \delta \lambda_\omega^b    \right) \omega(\tau)  \hat{T}_b \hat{T}_a = \left(  \delta \lambda_\omega^{b'}   \right)  \omega (\tau)  \hat{T}_a  \hat{T}_{b'} \, .
\ee
Because we suppose that $\omega$ is a function of $\tau$, there should be $\delta \lambda_\omega^b \neq 0$ for at least one value of $b = 1,2,3$. Without loss of generality, we consider $\delta \lambda_\omega^1 \neq 0$, and then obtain
\be\la{eq:unitonly}
\omega(\tau)  \hat{T}_1 \hat{T}_a  = \omega (\tau)  \hat{T}_a  \hat{T}_1 \, .
\ee
Notice that Eq.~\eqref{eq:unitonly} cannot be satisfied for all $a = 1,2,3$. Therefore, $\omega (\tau)$ does not depend on $\tau$. In summary, the matrix $\omega$ can only have constant entries. Together with Eq.~\eqref{eq:3Dcond-2a}, we may conclude that $\omega$ is an element of the rotational group $O(3)$ with constant entries.

If $\textrm{det}\, \omega = 1$, with the definition
\be
\left( T' \right)_a \equiv T_c \, \omega^c\,_a \, .
\ee
Eq.~\eqref{eq:3Dcond-1} becomes 
\begin{align}
T'_a \, O^a\,_\rho\, \hat{n}^\rho & =  T'_a \, (\omega^T)^a\,_c \,U^c\,_a \, \omega^a\,_\rho \, \hat{n}^\rho  \nonumber\\
\Rightarrow \quad T'_a \, O^a\,_\rho\, \hat{n}^\rho & =  T'_a \, \,{U'}^a\,_\rho \,  \hat{n}^\rho \, ,
\end{align} 
where ${U'}^a\,_\rho \equiv (\omega^T)^a\,_c \,U^c\,_a \, \omega^a\,_\rho$. Therefore, in the new choice of generators $\{ T'^a\}$, if we require that
\be\label{eq:3Dcond-3}
U' = O\, ,
\ee
then Eq.~\eqref{eq:3Dcond-1} and consequently Eq.~\eqref{eq:FI} are always true for the 3D Yang-Mills theory.

To see how the gauge transformation parameters and the Lorentz transformation parameters are related, let us assume that the transformations are given by 
\be\label{eq:3Dparameter}
O = \textrm{exp} \left(\varphi_i \hat{T}_i \right)\,\quad \textrm{and}\quad V = \textrm{exp} \left( \alpha_i T_i\right)\, ,
\ee
where $\hat{T}_i$ and $T_i$ denote the $SO(3)$ and $SU(2)$ generators respectively.  Then
\begin{align}
O_{ij} & = \frac{\varphi_i\, \varphi_j}{|\varphi|^2} + \left(\delta_{ij} - \frac{\varphi_i\, \varphi_j}{|\varphi|^2} \right) \, \textrm{cos}\, |\varphi| + \frac{\textrm{sin} \, |\varphi|}{|\varphi|} \, \varphi_k (S_k)_{ij} \nonumber\\
{} & = \frac{\varphi_i \, \varphi_j}{|\varphi|^2} (1 - \textrm{cos}\, |\varphi|) + \delta_{ij} \, \textrm{cos}\, |\varphi| + \textrm{sin} \, |\varphi| \, \frac{\varphi_k}{|\varphi|} \, \epsilon_{kij}\, .
\end{align}
Using the parameters introduced in Eq.~\eqref{eq:3Dparameter} and the relation \eqref{eq:3DSU2SO3Rel}, we obtain
\begin{align}\la{eq:su2so3relation}
U'_{ij} & = -2 \, \textrm{Tr} \left(T_i  V^{-1} T_j V \right) \nonumber\\
{} & = \frac{\alpha_i \, \alpha_j}{|\alpha|^2} (1 - \textrm{cos} \, |\alpha|) + \delta_{ij}\, \textrm{cos}\, |\alpha| + \textrm{sin} \, |\alpha|\, \frac{\alpha_k}{|\alpha|} \, \epsilon_{kij}\, ,
\end{align}
where $\textrm{Tr} \left(  T_i T_j \right) = -  \delta_{i j}/2$. We see that when
\be
\varphi_i = \alpha_i\, ,
\ee
Eq.~\eqref{eq:3Dcond-3} holds automatically. Hence, the gauge transformation parameters can simply be chosen to equal the  Lorentz transformation parameters.

The calculation is similar for  $\textrm{det}\, \omega = -1$. Therefore, with an appropriate choice of $T_a$, we can use
\bea
\omega = 
\begin{pmatrix}
1 & 0 & 0 \\
\,0\, & 1 & \,0\, \\
0 &\,0\, & 1 \\
\end{pmatrix}
\ \ \ \ \ \ 
{\rm or}
\ \ \ \ \ \ 
\omega = 
\begin{pmatrix}
1 & \,0\, & \,0\, \\
\,0\, & 1 & 0 \\
0 & 0 & -1 \\
\end{pmatrix}\, .
\eea
In this paper we assume that $\textrm{det}\, \omega = 1$, hence we choose the first case.

\subsection{\label{omegarotation}Average over All Possible $\omega$'s}
As we discussed in the main text, a gauge transformation can be equivalently formulated as a rotation of $\omega$. The integration over all the gauge orientations is equivalent to the integration over all the values of $\omega$. Since the approximation that we used in Section~\ref{approx} heavily relies on the average of $\omega$, we prove the following important equality for the average:
\be
  \left< \omega(\varphi)^a\,_{(\mu} \, \omega(\varphi)^b\,_{\nu)}  \right>_\varphi = \frac{1}{3} \delta_{a b}\delta_{\mu\nu} \, ,
\ee
where the bracket $(\cdots)$ denotes the symmetrization of the indices.

\begin{flushleft}
  {\bf \underline{Proof}}
\end{flushleft}

Let us start with
\be
\omega(\varphi)^a\,_\mu = O(\varphi)_{a a'} \omega_{a' \mu} = O(\varphi)_{a \mu}   \, ,
\ee
where $O(\varphi)$ is an $SO(3)$ rotation, and we used $\omega_{a' \mu} = \delta_{a' \mu}$. We need to calculate
\be
\left< \omega(\varphi)^a\,_\mu \, \omega(\varphi)^b\,_\nu  \right>_\varphi = \left< O(\varphi)_{a \mu} O(\varphi)_{b \nu}  \right>_\varphi  \, .
\ee

Notice that we only have two tensors that are invariant under $SO(3)$ rotations, $\delta_{ij}$ and 
$\epsilon_{i j k}$. Here there are four indices, and for any tensor with four indices that are $SO(3)$ invariant there should be
\be
T_{a\mu b \nu} = c_1 \delta_{a\mu} \delta_{b \nu} + c_2 \delta_{a b}\delta_{\mu\nu} + c_3 \delta_{a \nu} \delta_{\mu b}\, .
\ee

In our case, we have
\be
 \left< O(\varphi)_{a \mu} O(\varphi)_{b \nu}  \right>_\varphi  \hat{n}_\mu \hat{n}_\nu = \left< \hat{n}_a (\varphi)  \hat{n}_b (\varphi)  \right>_\varphi  =  \frac{1}{3} \delta_{a b}\, ,
\ee
where $\hat{n}_\mu = x^\mu/|x|$, and
\be
T_{a\mu b \nu} \hat{n}_\mu \hat{n}_\nu = (c_1 + c_3) \hat{n}_a \hat{n}_b + c_2 \delta_{a b}.
\ee
Therefore,
\be
c_1 + c_3 = 0 \quad \textrm{and} \quad c_2 = \frac{1}{3}\, .
\ee
Consequently,
\be
  \left< \omega(\varphi)^a\,_{(\mu} \, \omega(\varphi)^b\,_{\nu)}  \right>_\varphi = \frac{1}{3} \delta_{a b}\delta_{\mu\nu} \, .
\ee
 
{\flushleft $\square$}

\section{\label{CS}3D Topological Charge}

In this appendix, we review the topological properties of the Wess-Zumino term induced by the Chern-Simons term and the Ansatz to the 3-dimensional Yang-Mills equation. We mainly follow Appendix A of Ref.~\cite{Zahed}.

The topological charge density is
\be
  B^0 = \frac{1}{8 \pi^2} \, \left(\frac{2}{3} p^3 - p^2 \right) \epsilon^{\nu \alpha \beta} \, \textrm{Tr} \left[L_\nu L_\alpha L_\beta \right]
\ee
with
\be
  L_\mu = U^{-1} \partial_\mu U\quad \textrm{and}\quad U \equiv \textrm{exp} \left[i \vec{\tau} \cdot \hat{\theta} (\vec{x}) \, \theta(\vec{x}) \right]\, .
\ee
The winding number is the topological charge given by
\be
  B = \int d^3 x\, B^0\, ,
\ee
which has integer values, and it can be expressed as an integral over the surfaces around the singular points. To see it, we can use stereographic projection to embed the 3-dimensional space into a 4-dimensional space and define a covariant current
\be
  \widetilde{B}^\mu \equiv \frac{\epsilon^{\mu\nu\alpha\beta}}{8 \pi^2} \, \textrm{Tr} \left[L_\nu L_\alpha L_\beta \right]\, ,
\ee
i.e.,
\be
  \partial_\mu \widetilde{B}^\mu = 0\, .
\ee
Hence,
\be
  B^0 = \left(\frac{2}{3} p^3 - p^2 \right) \widetilde{B}^0\, .
\ee
The integrated conservation law leads to
\be\label{appTopCharge}
  \frac{\partial}{\partial t} \int d^3 x\, B^0 = \int d^3 x\, \left(\frac{2}{3} p^3 - p^2 \right) \frac{\partial \widetilde{B}^0}{\partial t} = - \int d^3 y\, \partial_i \widetilde{B}^i = - \sum_\beta \int d \widetilde{S}_\beta \, \left(\hat{n}_\beta \cdot \vec{\widetilde{B}} \right)\, ,
\ee
where in the intermediate step we changed the variable
\be
  \left(\frac{2}{3} p^3 - p^2 \right)^{\frac{1}{2}} \, d x_i = d y_i\qquad (i = 1, 2, 3)\, ,
\ee
which preserves the unit vector $\hat{n}_\beta$. As long as the factor $\frac{2}{3} p^3 - p^2$ remains nonzero, the measure of the integral is nondegenerate. We will see that except for a few singular points the classical solutions always lie in $0 < p < \frac{3}{2}$ for the whole 3-dimensional space, hence the nondegenerate condition is satisfied. In Eq.~\eqref{appTopCharge}, $\{\beta\}$ denotes the set of the singular points including infinity, where we assume the singular points to be isolated, and $d\widetilde{S}_\beta = dS_\beta \left(2/3\, p^3 - p^2 \right)$ is the surface element around the singular points, on which the factor $p$ is contant, while $\hat{n}_\beta$ is a unit vector orthogonal to the surface $S_\beta$. Inserting the explicit expression
\be
  L_\mu = U^{-1} \partial_\mu U = \textrm{exp} \left[- i (\vec{\sigma}\cdot \hat{\theta}) \theta \right] \partial_\mu\, \textrm{exp} \left[i (\vec{\sigma}\cdot \hat{\theta}) \theta \right]\, ,
\ee
and making use of the identities
\begin{align}
  (\vec{\sigma} \cdot \hat{\theta}) (\vec{\sigma} \cdot \partial_j \hat{\theta}) & = i \vec{\sigma}\cdot (\hat{\theta} \times \partial_j \hat{\theta})\, ,\nonumber\\
  (\vec{\sigma}\cdot \hat{\theta}) (\vec{\sigma}\cdot \hat{\theta}) & = I\, ,
\end{align}
we obtain
\begin{align}\label{appTopCurrent}
  \widetilde{B}^i & = \frac{3 i\, \epsilon^{ijk} \dot{\theta}}{8 \pi^2} \, \textrm{Tr} \Big[\textrm{sin}^2 \theta\, \textrm{cos}^2 \theta (\vec{\sigma}\cdot \hat{\theta}) (\vec{\sigma} \cdot \partial_j \hat{\theta}) (\vec{\sigma}\cdot \partial_k \hat{\theta}) \nonumber\\
 {} & \qquad\qquad\qquad + \textrm{sin}^4 \theta (\vec{\sigma}\cdot \hat{\theta}) (\vec{\sigma} \cdot (\hat{\theta} \times \partial_j \hat{\theta})) (\vec{\sigma} \cdot (\hat{\theta} \times \partial_k \hat{\theta})) \Big]\, .
\end{align}
Choosing an orthogonal comoving coordinate system $\{\hat{e}_k \}$ $(k = 1, 2, 3)$ with $\hat{e}_3 = \hat{n}_\beta$, we can express $\hat{\theta}$ as
\be
  \hat{\theta} = \hat{\theta}_1\, \hat{e}_1 + \hat{\theta}_2 \, \hat{e}_2 + \hat{\theta}_3 \, \hat{e}_3\, .
\ee
Using the Serret-Frenet relations and the zero torsion condition, we can also compute $\partial_1 \hat{\theta}$ and $\partial_2 \hat{\theta}$ in this frame:
\begin{align}
  \partial_1 \hat{\theta} & = (\partial_1 \hat{\theta}_1 - \kappa_1 \hat{\theta}_3) \hat{e}_1 + (\partial_1 \hat{\theta}_2) \hat{e}_2 + (\partial_1 \hat{\theta}_3 + \kappa_1 \hat{\theta}_1) \hat{e}_3\, ,\nonumber\\
  \partial_2 \hat{\theta} & = (\partial_2 \hat{\theta}_1) \hat{e}_1 + (\partial_2 \hat{\theta}_2 - \kappa_2 \hat{\theta}_3) \hat{e}_2 + (\partial_2 \hat{\theta}_3 + \kappa_2 \hat{\theta}_3) \hat{e}_3\, ,
\end{align}
where $\kappa_i$'s are the curvature tensions. Applying some vector analysis to Eq.~\eqref{appTopCurrent}, one can simplify Eq.~\eqref{appTopCharge} as follows:
\be
  \frac{d B}{d t} = \frac{3}{\pi^2} \sum_\beta \int dS_\beta\, \left(\frac{2}{3} p_\beta^3 - p_\beta^2 \right) \dot{\theta} \, \textrm{sin}^2 \theta \left[\hat{\theta}\cdot (\partial_1 \hat{\theta} \times \partial_2 \hat{\theta}) \right]\, ,
\ee
where $p_\beta$ is the value of $p$ on the surface around the singular point $\beta$. Consequently, the winding number is
\be
  B = \frac{3}{2\pi^2} \sum_\beta \int dS_\beta \left(\frac{2}{3} p_\beta^3 - p_\beta^2 \right) \left(\theta - \frac{1}{2} \textrm{sin} 2 \theta \right)_\beta \, \hat{\theta}\cdot (\partial_1 \hat{\theta} \times \partial_2 \hat{\theta})\, ,
\ee
If the surface around the singular points shrinks to zero, the factors defined on the surface become constant, so we can put them outside the integral, i.e.
\be\label{app:3DWinding}
  B = \frac{3}{2\pi^2} \sum_\beta \left(\frac{2}{3} p_\beta^3 - p_\beta^2 \right) \left(\theta_\beta - \frac{1}{2} \textrm{sin} 2 \theta_\beta \right) \int dS_\beta \, \hat{\theta}\cdot (\partial_1 \hat{\theta} \times \partial_2 \hat{\theta})\, ,
\ee
where $p_\beta$ and $\theta_\beta$ are the values of $p$ and $\theta$ at the boundary of the surface around the singular point $\beta$. Since the winding number $B$ has to be an integer, additional constraints are imposed on the factors $p$ and $\theta$.

The expression above is valid for arbitrary numbers of singular points. In this paper, we focus on the spherically symmetric configurations of the Yang-Mills fields, hence we only consider two singular points at $\tau=0$ and $\tau=\infty$ in the main text. We would like to emphasize that due to the opposite boundary orientations at $\tau=0$ and $\tau=\infty$, the contribution from the surface integral in Eq.~\eqref{app:3DWinding} differs by a sign for $\tau=0$ and $\tau=\infty$.

\section{\label{sq}Form Invariance of 4D Ansatz}

\subsection{Restriction on $p$ and $\theta$}\label{thetafix}

We have seen that for the 3D case the Lorentz transformation acting on the Ansatz is equivalent to an $SU(2)$ gauge transformation, hence the Ansatz is form invariant. For the 4D case, since $SO(4) \cong SU(2) \times SU(2) \cong SO(3) \times SO(3)$, a Lorentz transformation can be decomposed into a rotation in the $(1, 2, 3)$-subspace which is generated by $M_{12}$, $M_{23}$ and $M_{31}$, and a transformation generated by $M_{14}$, $M_{24}$ and $M_{34}$. One can show as before that the rotation in the $(1, 2, 3)$-subspace restricts the matrix $\omega$ to be a constant $O(3)$ matrix. In this paper, we assume that $\textrm{det}\, \omega = 1$, therefore, we fix $\omega$ to be a constant $SO(3)$ group element.  Furthermore, we need to consider the transformation generated by $M_{14}$, $M_{24}$ and $M_{34}$. To maintain the form invariance, this transformation has to have the same expression as an $SU(2)$ gauge transformation. In this appendix, we will calculate $p(\tau, x^4)$ and $\theta(\tau, x^4)$ obeying
\be\la{eq:FI4D}
  (\Lambda^{-1})\, _\mu\,^\nu\, A_\nu (\Lambda \, x) = V^{-1}\, A_\mu (x)\, V\, ,
\ee
where $\Lambda$ is the Lorentz transformation generated by $M_{i4}$,  and $V$ is an $SU(2)$ gauge transformation.

A general Lorentz transformation generated by $M_{i4}$ is given by
\begin{align}
\Lambda & = e^{\varphi_{14} \, M_{14} + \varphi_{24} \, M_{24} + \varphi_{34} \, M_{34}} \nonumber\\
{} & = I \, \textrm{cos}^2 \left(  \frac{|\varphi|}{2} \right) + \frac{\varphi_i M_i}{|\varphi|} \, \textrm{sin} |\varphi|  -   \frac{\varphi_i N_i}{|\varphi|} \, \textrm{sin} |\varphi| - 4 \frac{\varphi_i \varphi_j}{|\varphi|^2} M_i N_j \, \textrm{sin}^2  \left(  \frac{|\varphi|}{2} \right) \, ,
\end{align}
where $M_i$ and $N_i$ are defined in Appendix~\ref{app:notation}. Using the properties of $M_i$ and $N_i$,  we obtain
\begin{align}
\Lambda_{\mu\nu} & = \left(e^{\varphi_{14} \, M_{14} + \varphi_{24} \, M_{24} + \varphi_{34} \, M_{34}} \right)_{\mu\nu} \nonumber\\
{} & = \delta_{\mu\nu} + \frac{\varphi_\mu}{|\varphi|} \delta_{\nu 4} \, \textrm{sin} |\varphi| - \frac{\varphi_\nu}{|\varphi|} \delta_{\mu 4} \, \textrm{sin} |\varphi| -  2 \frac{\varphi_\mu \varphi_\nu}{|\varphi|^2}\,  \textrm{sin}^2  \left(  \frac{|\varphi|}{2} \right) - 2 \delta_{\mu 4} \delta_{\nu 4} \, \textrm{sin}^2  \left(  \frac{|\varphi|}{2} \right)\, ,\label{eq:LorentzTrafo}
\end{align}
where $\varphi_\mu \equiv (\varphi_i,\, \varphi_4 = 0)$ and $|\varphi| \equiv \sqrt{(\varphi_i)^2}$. More explicitly,
\begin{align}\label{eq:lorentztransformationformula}
\Lambda_{ij} & = \delta_{ij} - \frac{2 \varphi_i \varphi_j}{|\varphi|^2} \, \textrm{sin}^2 \frac{|\varphi|}{2} \, ,\nonumber\\
\Lambda_{i4} & = \frac{\varphi_i}{|\varphi|} \, \textrm{sin} |\varphi| \, ,\nonumber\\
\Lambda_{4i} & = - \frac{\varphi_i}{|\varphi|} \, \textrm{sin} |\varphi| \, ,\nonumber\\
\Lambda_{44} & = \textrm{cos} |\varphi| \, .
\end{align} 
 
Now we consider how the components $A_\mu^a$ transform under the Lorentz transformations. First, $A_\mu$ has the following  expression:
\begin{align}
A_\mu & = p \left[\textrm{cos}  \left( \frac{\theta}{2} \right) - 2 (\vec{T}\cdot \hat{n}) \, \textrm{sin}  \left( \frac{\theta}{2} \right) \right] \nonumber\\
{} & \quad  \cdot \left[-\frac{1}{2} \, \textrm{sin} \left( \frac{\theta}{2} \right) (\partial_\mu \theta) + (\vec{T} \cdot \hat{n}) \, \textrm{cos}  \left( \frac{\theta}{2} \right) (\partial_\mu \theta) + 2 T_a (\partial_\mu \hat{n}^a) \, \textrm{sin}  \left( \frac{\theta}{2} \right) \right]\, ,
\end{align}
where $\hat{n}^a \equiv \omega^a\,_{i} \frac{x^i}{|x|}$. Then the components $A_\mu^a$ are given by
\begin{align}
A_\mu^a & = -2\, \textrm{Tr} (A_\mu T^a) \nonumber\\
{} & = p  \left[\hat{n}^a (\partial_\mu \theta) + \textrm{sin}\, \theta (\partial_\mu \hat{n}^a) \right] - p \, \epsilon^{abc} (1 - \textrm{cos}\, \theta) \hat{n}_b (\partial_\mu \hat{n}_c)\, .\label{eq:Comp-1}
\end{align}

We have proven that in 3D the Lorentz transformation $\Lambda^{-1} A_\mu (\Lambda x)$ is equivalent to a gauge transformation. Now we consider $\Lambda^{-1} A_\mu (\Lambda x)$, where $\Lambda$ denotes a Lorentz transformation in the $14$, $24$ and $34$ directions. Since our Ansatz has the form
\be
A_\mu = p (\tau, x_4) \, \textrm{exp} \left[- T_a \omega^a\,_i \frac{x^i}{|x|} \theta(\tau, x_4) \right] \frac{\partial}{\partial x^\mu} \, \textrm{exp} \left[T_b\, \omega^b\,_j \frac{x^j}{|x|} \theta(\tau, x_4) \right]\, ,
\ee
under a Lorentz transformation it becomes
\begin{align} 
\Lambda^{-1} A_\mu (\Lambda x) & = p \left(\tau, \left(\Lambda x\right)_4\right)  \, \textrm{exp} \left[- T_a \omega^a\,_i \frac{(\Lambda x)^i}{|\Lambda x|} \theta \left(\tau, (\Lambda x)_4\right) \right] \frac{\partial}{\partial x^\mu} \, \textrm{exp} \left[T_b\, \omega^b\,_j \frac{(\Lambda x)^j}{|\Lambda x|} \theta \left(\tau, (\Lambda x)_4\right) \right] \nonumber\\
{} & =p \left(\tau, \left(\Lambda x\right)_4\right) \, \textrm{exp} \left[- T_a\, \hat{n}'_a\, \theta\left(\tau, (\Lambda x)_4\right) \right] \frac{\partial}{\partial x^\mu} \, \textrm{exp} \left[T_b\, \hat{n}'_b\, \theta\left(\tau, (\Lambda x)_4\right) \right]\, ,
\end{align}
where $\tau \equiv x^\mu x_\mu$, $\hat{n}'_a \equiv \omega_a\,^i \frac{(\Lambda x)_i}{|\Lambda x|}$ and $|\Lambda x| \equiv \sqrt{(\Lambda x)_i\, (\Lambda x)_i}$.

According to Eq.~\eqref{eq:lorentztransformationformula}, $(\Lambda x)_i$ and $(\Lambda x)_4$ have the following expressions:
\begin{align}
(\Lambda x)_i & = x_i - \frac{2 \varphi_i \varphi_j x_j}{|\varphi|^2} \, \textrm{sin}^2 \frac{|\varphi|}{2} + \frac{\varphi_i x_4}{|\varphi|} \, \textrm{sin} |\varphi| \, ,\nonumber\\
(\Lambda x)_4 & = -\frac{\varphi_i x_i}{|\varphi|} \, \textrm{sin} |\varphi| + \textrm{cos} |\varphi| \, x_4\, .
\end{align}
Following the same steps, we can derive a similar expression for the components $A_{\mu}^{'a}$ after the transformation $\Lambda^{-1} A_\mu (\Lambda x)$:
\be
A_\mu^{'a} =p'  \left[\hat{n}'_a (\partial_\mu {\theta}') + \textrm{sin}\, {\theta}' (\partial_\mu \hat{n}'_a) \right] - p'    \, \epsilon^{abc} (1 - \textrm{cos}\, {\theta}') \hat{n}'_b (\partial_\mu \hat{n}'_c)\, ,\label{eq:Comp-2}
\ee
where $\theta' \equiv \theta\left(\tau, (\Lambda x)_4\right)$ and $p' \equiv p \left(\tau, \left(\Lambda x\right)_4\right) $. As we mentioned before, we expect that the Lorentz transformation $\Lambda^{-1} A_\mu (\Lambda x)$ is equivalent to a gauge transformation, which may possibly restrict the form of $p$ and $\theta$.

To see how the form invariance restricts the factors $p$ and $\theta$, let us consider a special case $\mu = 4$. Then Eq.~\eqref{eq:Comp-1} becomes 
\be
A_{\mu=4}^a = p(\tau, \, x_4)  \, \hat{n}_a \partial_{4} \theta\, .
\ee
After a gauge transformation, it has the expression
\be\label{eq:Comp-3}
V^{-1}A_{\mu=4}^a T_a V = p  \,
\left[\frac{\psi_a \psi_b}{|\psi|^2} (1 - \textrm{cos} |\psi|) + \delta_{ab} \, \textrm{cos} |\psi| - \epsilon_{abc} \, \textrm{sin} |\psi| \frac{\psi_c}{|\psi|} \right]    \, T_b \hat{n}_a \partial_{4} \theta\, ,
\ee
where 
\be
V = \exp \left(  \psi^a \, T_a  \right) \, ,
\ee
and we used Eq.~\eqref{eq:3DSU2SO3Rel} and Eq.~\eqref{eq:su2so3relation}. After a Lorentz transformation in the $14$, $24$ and $34$ directions the components $A_\mu^{'a}$ are given by Eq.~\eqref{eq:Comp-2}:
\begin{align}\label{eq:Comp-4}
\Lambda^{-1} A_{\mu=4}^a (\Lambda x) T_a &=
p' \left[\hat{n}'_a (\partial_4 \theta') + \textrm{sin} \theta' (\partial_4 \hat{n}'_a) \right] T_a   - p'  \, T_a \, \epsilon_{abc} \, \omega^b\,_i  \, \omega^c\,_j  \,(1 - \textrm{cos} \theta') \frac{x^i}{|\Lambda x|} \frac{\frac{\varphi^j \, \textrm{sin} |\varphi|}{|\varphi|}}{|\Lambda x|}\, .
\end{align}
The form invariance requires that the expression \eqref{eq:Comp-3} equals \eqref{eq:Comp-4}. By comparing the terms $\sim \epsilon_{abc}$, one obtains
\be\la{eq:solvetheta1}
\psi^c = \pm \omega^c\,_i \varphi^i\, ,\quad  - (1 - \textrm{cos} {\theta}') \frac{ p' }{|\Lambda x|^2} = \pm  \frac{ p }{|x|} \partial_4 \theta\, .
\ee
For the special case $\Lambda = 1$ the second equation above becomes a differential equation:
\begin{align}\label{eq:thetasolution1}
{} & \quad - (1 - \textrm{cos}\, \theta) \frac{1}{|x|^2} = \pm \frac{1}{|x|} \partial_4 \theta \nonumber\\
\Rightarrow & \quad \frac{1}{|x|} = \pm \partial_4\, \textrm{cot} \left(\frac{\theta}{2} \right) \nonumber\\
\Rightarrow & \quad \textrm{cot} \left(\frac{\theta}{2} \right) = \pm \frac{x_4}{|x|} \pm  f(|x|)\, ,
\end{align}
where $f$ is an arbitrary smooth function. The sign $\pm$ in Eq.~\eqref{eq:thetasolution1} corresponds to the choice $U = \exp \left( \pm T_a \,  \omega^a\,_j \, n^j \, \theta  \right)$.  

The left-hand side of Eq.~\eqref{eq:solvetheta1} is invariant under Lorentz transformations:
\begin{align}\la{eq:tofixpandtheta1}
 (1 - \textrm{cos} {\theta}')  \frac{p'}{|\Lambda x|^2} & = (1 - \textrm{cos} {\theta})  \frac{p}{|x|^2}  \nonumber\\
\Rightarrow\quad \frac{1}{1 + \cot^2 \left( \frac{ {\theta}'}{2} \right)} \frac{p'}{|\Lambda x|^2} & = \frac{1}{1 + \cot^2 \left( \frac{ {\theta}}{2} \right)} \frac{p}{|x|^2} \, .
\end{align}
For our Ansatz $A_\mu = p \, U^{-1} \partial_\mu U$, let us consider the following gauge transformation:
\be
U A_\mu U^{-1} + U \partial_\mu U^{-1} = (1 - p) U \partial_\mu U^{-1} \, .
\ee
It is easy to show that if $p \, U^{-1} \partial_\mu U$ is form invariant, then $(1 - p) U \partial_\mu U^{-1}$ is also form invariant. Instead of using the Ansatz  $A_\mu = p \, U^{-1} \partial_\mu U$, if we use $A_\mu = (1-p) \, U \partial_\mu U^{-1}$, we obtain
\be \la{eq:tofixpandtheta2}
\frac{1}{1 + \cot^2 \left( \frac{ {\theta}'}{2} \right)} \frac{(1- p')}{|\Lambda x|^2} =  \frac{1}{1 + \cot^2 \left( \frac{ {\theta}}{2} \right)} \frac{1- p}{|x|^2} \, .
\ee
Comparing Eq.~\eqref{eq:tofixpandtheta1} and Eq.~\eqref{eq:tofixpandtheta2}, we can read off
\be
\frac{1 -p \left(\tau, \left(\Lambda x\right)_4\right) }{ p \left(\tau, \left(\Lambda x\right)_4\right) } = \frac{1 - p \left(\tau,  x_4 \right)}{p \left(\tau,  x_4 \right)} \, .
\ee
Thus,
\begin{align}
  p \left(\tau, \left(\Lambda x\right)_4\right) & = p \left(\tau,  x_4 \right) \nonumber\\
  \Rightarrow\quad  p & =  p(\tau) \, ,
\end{align}
i.e., the factor $p$ is invariant under Lorentz transformations. It should be expected a priori, because $p$ appears as a multiplicative factor, which should not transform under Lorentz transformations when we consider spherically symmetric configurations.

Plugging Eq.~\eqref{eq:thetasolution1} into Eq.~\eqref{eq:tofixpandtheta2} with $p = p(\tau)$, we obtain
\be\la{eq:thetasolution2}
  2 \left(\Lambda x \right)_4 |\Lambda x| f(|\Lambda x|) + |\Lambda x|^2 f^2(|\Lambda x|)   =  2   x_4  |  x| f(|  x|) + |  x|^2 f^2(| x|) \, .
\ee
Notice that the left-hand side and the right-hand side of Eq.~\eqref{eq:thetasolution2} is linear in $x_{\mu=4}$, and therefore cannot be invariant under Lorentz transformations unless
\be
f(|x|) = 0\, .
\ee

Therefore, the form invariance in the 4-dimensional Euclidean space imposes the constraints
\be
\textrm{cot} \left(\frac{\theta}{2} \right) =\pm   \frac{x_4}{|x|} \, \quad \textrm{and} \quad p = p(\tau)\, .
\ee 
For simplicity, we choose $\textrm{cot} \left( \theta / 2 \right) =   x_4/|x|$ and $\omega^a\,_i = \delta_{a i}$.  The form invariant Ansatz $A_\mu$ is given by  
\begin{align}\la{eq:4dforminvariantresult}
A_\mu & = p (\tau) \left[ \frac{x_4 - 2 \left(T^a x_a\right) }{\sqrt{\tau}}   \right] \partial_\mu \left[ \frac{x_4 + 2 \left(T^b x_b\right) }{\sqrt{\tau}}   \right]\nonumber\\
{} & = 2 \frac{p(\tau) }{\tau}\eta_{a\mu\nu} x_\nu T^a\, ,
\end{align}
where $\eta_{a\mu\nu}$ is the 't Hooft symbol~(see Appendix~\ref{app:notation}). To satisfy the form invariance condition \eqref{eq:FI4D}, the gauge transformation parameters can simply be chosen to equal the Lorentz transformation parameters according to Eq.~\eqref{eq:solvetheta1}.

\subsection{\label{completecheck}Complete Check}
 In this subsection, we prove that the Ansatz \eqref{eq:4dforminvariantresult} indeed satisfies the form invariance condition \eqref{eq:FI4D}. First, let us consider
\be
A_\mu^a\, T^a \, = \, 2 \frac{p(\tau) }{\tau}\eta_{a\mu\nu} x_\nu T^a\, .
\ee
The form invariance condition \eqref{eq:FI4D}:
\begin{displaymath}
  (\Lambda^{-1})\,_\mu\,^\nu\, A_\nu (\Lambda\, x) = V^{-1}\, A_\mu (x)\, V
\end{displaymath}
in this case is equivalent to
\be\label{eq:4Dcond}
  (\Lambda^{-1})\,_{\mu\mu'} \,  \eta_{a\mu'\nu'} \,  \Lambda_{\nu' \nu} \, T^a = \eta_{a\mu\nu} \, V^{-1} \, T^a \, V\, ,
\ee
where $\Lambda$ denotes a $4D$ Lorentz transformation, and $V$ stands for an $SU(2)$ gauge transformation. 
For an $SO(4)$ element,
\be
\exp \left( \frac{1}{2} \varphi_{\mu\nu} M_{\mu\nu} \right) = \exp \left[ \varphi^+_i \frac{(J_i + K_i)}{2} + \varphi^-_i \frac{(J_i - K_i)}{2} \right] \, ,
\ee
where
\be
\varphi^\pm_i = \frac{1}{2} \epsilon_{i j k}\, \varphi_{j k} \pm \varphi_{i 4} \, .
\ee
Hence, the general form of a 4D Lorentz transformation $\Lambda$ is given by
\begin{align}
\Lambda & = \exp \left( \frac{1}{2} \varphi_{\mu\nu} M_{\mu\nu} \right) \nonumber\\
&= \left[ \cos \left( \frac{ |\varphi^+| }{2} \right) + \frac{2 (\varphi^+ \cdot M) }{|\varphi^+|}\sin \left( \frac{ |\varphi^+| }{2} \right) \right] \left[ \cos \left( \frac{ |\varphi^-| }{2} \right) + \frac{2 (\varphi^- \cdot N) }{|\varphi^-|}\sin \left( \frac{ |\varphi^-| }{2} \right) \right]  \, . 
\end{align}

The left-hand side of Eq.~\eqref{eq:4Dcond} contains
\be
  (\Lambda^{-1})\,_{\mu\mu'} \eta_{a\mu'\nu'} \Lambda_{\nu' \nu} =  2 (\Lambda^{-1})\,_{\mu\mu'}   \left(M_a\right)_{\mu'\nu'} \Lambda_{\nu' \nu} \, ,
\ee
where we used $\eta_{a\mu'\nu'} = 2 (M_a)_{\mu'\nu'}$. Then $\Lambda^{-1} M_a \Lambda$ is given by
\begin{align}
  &\quad  \left[ \cos \left( \frac{ |\varphi^+| }{2} \right) - \frac{2 (\varphi^+ \cdot M) }{|\varphi^+|}\sin \left( \frac{ |\varphi^+| }{2} \right) \right] M_a \left[ \cos \left( \frac{ |\varphi^+| }{2} \right) + \frac{2 (\varphi^+ \cdot M) }{|\varphi^+|}\sin \left( \frac{ |\varphi^+| }{2} \right) \right] \nonumber\\
&= M_c \left[ \cos \left( |\varphi^+| \right) \left( \delta_{a c} - \frac{\varphi^+_a \varphi^+_c}{|\varphi^+|^2} \right) + \frac{\varphi^+_a \varphi^+_c}{|\varphi^+|^2} - \frac{ \epsilon_{a b c} \varphi^+_b }{|\varphi^+|}\sin \left( |\varphi^+| \right) \right] \, ,
\end{align}
where we made use of the fact that $M_i$ commute with $N_i$. Therefore, the left-hand side of Eq.~\eqref{eq:4Dcond} becomes
\be
\Lambda^{-1}\,_{\mu\mu'} \eta_{a\mu'\nu'} \Lambda_{\nu' \nu} T^a = \eta_{c\mu\nu} \left[ \cos \left( |\varphi^+| \right) \left( \delta_{a c} - \frac{\varphi^+_a \varphi^+_c}{|\varphi^+|^2} \right) + \frac{\varphi^+_a \varphi^+_c}{|\varphi^+|^2} - \frac{ \epsilon_{a b c} \varphi^+_b }{|\varphi^+|}\sin \left( |\varphi^+| \right) \right] T^a \, .
\ee
 
The right-hand side of Eq.~\eqref{eq:4Dcond} reads
\be
\eta_{a\mu\nu}\, V^{-1} \, T_a \, V = \eta_{c\mu\nu} \, T_a \, U_{a c}  \, .
\ee
If we choose
\be\la{eq:4DVparametrize}
V = \exp \left(  \psi^a \, T_a  \right) \, ,
\ee
then
\be
U_{a c} = \cos \left( |\psi| \right) \left( \delta_{a c} - \frac{\psi_a \psi_c}{|\psi|^2} \right) + \frac{\psi_a \psi_c}{|\psi|^2} - \frac{ \epsilon_{a b c} \psi_b }{|\psi|}\sin \left( |\psi| \right) \, .
\ee

We notice that when
\be
\psi_a = \varphi^+_a \, ,
\ee
the equivalent form invariance condition \eqref{eq:4Dcond} is satisfied.

\subsection{An Alternative Approach\label{alternativeapproach}}
In this subsection, we present an alternative approach to construct the Ansatz of the solution to the 4D Yang-Mills equation. This approach can easily be generalized to higher dimensions or larger gauge groups and also curved spacetime \cite{Korean}.

Let us start with a 4-dimensional Yang-Mills field with a gauge group $SO(4)$. Before writing down the Ansatz, we recall some facts from Appendix~\ref{app:notation}. The generators of the Lie algebra $\mathfrak{so}(4)$ are given by Eq.~\eqref{eq:SO4gen}:
\begin{displaymath}
  (M_{\mu\nu})_{mn} \equiv \delta_{\mu m} \delta_{\nu n} - \delta_{\mu n} \delta_{\nu m}\, ,
\end{displaymath}
After the redefinition of the generators \eqref{eq:SO4redef}:
\begin{displaymath}
  J_i \equiv \frac{1}{2} \epsilon_{ijk} M_{jk}\, ; \quad K_i \equiv M_{i4}\, \quad (i = 1,\, 2,\, 3) \, ,
\end{displaymath}
we obtain the commutation relations \eqref{eq:SO4alg}:
\begin{displaymath}
  [J_i,\, J_j] = -\epsilon_{ijk} J_k\, ,\quad [K_i,\, K_j] = -\epsilon_{ijk} J_k\, ,\quad [K_i,\, J_j] = -\epsilon_{ijk} K_k\, .\end{displaymath}
If we define \eqref{eq:defMN}:
\begin{displaymath}
  M_i \equiv \frac{1}{2} (J_i + K_i)\, ,\quad N_i \equiv \frac{1}{2} (J_i - K_i)\, ,
\end{displaymath}
then they satisfy \eqref{eq:MNalg}:
\begin{displaymath}
  [M_i,\, M_j] = -\epsilon_{ijk} M_k\, ,\quad [N_i,\, N_j] = -\epsilon_{ijk} N_k\, ,\quad [M_i,\, N_j] = 0\, .
\end{displaymath}
We see that $-i M_i$ and $-i N_i$ generate two independent $SU(2)$'s respectively, i.e., $SO(4) \cong SU(2) \times SU(2)$.

Next, we claim that the following Ansatz for the 4-dimensional Yang-Mills theory with an $SO(4)$ gauge group \cite{Ma:1983vt, Ma:1984wm} is form invariant.
\be\label{eq:AltAnsatz}
  A_{\mu, ab} = \frac{ p_1(\tau)}{\tau}\, (M_{\mu\nu})_{ab} \, x_\nu  +  \frac{1}{2} \frac{ p_2(\tau)}{  \tau}\, \epsilon_{\mu\nu\rho\sigma} \, (M_{\rho\sigma})_{ab} \, x_\nu   \,.
\ee
In this case, both the Lorentz group and the gauge group are $SO(4)$. The Lorentz group element with the parameters $\varphi_{\mu\nu}$ is
\be
\Lambda_{mn} = \textrm{exp} \big[(M_{\mu\nu}) \varphi_{\mu\nu} \big]_{mn}\, ,
\ee
while the gauge group element with the parameters $\psi_{\mu\nu}$ is
\be
  V_{ab} = \textrm{exp} \big[(- M_{\mu\nu}) \psi_{\mu\nu} \big]_{ab}\, .
\ee

Let us rewrite the form invariance condition Eq.~\eqref{eq:FI}  as
\be\label{eq:FIforalternative}
 V_{a a'} \,  \Lambda^{-1}_{\mu\nu} \, A_{\nu, a' b'} (\Lambda\, x)  \,  V^{-1}_{b' b} =   A_{\mu , a b} (x) \, .
\ee
First,
\be
 \Lambda_{\mu\mu'}^{-1}\, A_{\mu', ab} (\Lambda  x) =  \frac{ p_1 (\tau)}{\tau}\, \Lambda^{-1}_{\mu\mu'} (M_{\mu' \nu'})_{ab}\, \Lambda_{\nu'\nu}\, x_\nu\, +  \frac{1}{2} \frac{ p_2(\tau)}{  \tau}\, \epsilon_{\mu\nu\rho\sigma} \,  \Lambda^{-1}_{\rho\rho'}  (M_{\rho' \sigma'})_{ab} \Lambda_{\sigma' \sigma}  \, x_\nu   \, ,
\ee
where we used
\be
\Lambda_{\mu\mu'} \Lambda_{\nu\nu'} \Lambda_{\rho\rho'} \Lambda_{\sigma\sigma'} \epsilon_{\mu'\nu'\rho'\sigma'} = \epsilon_{\mu\nu\rho\sigma}\, .
\ee
A subsequent rigid gauge transformation leads to
\be
  V_{aa'} \, \Lambda_{\mu\mu'}^{-1}\, (M_{\mu' \nu'})_{a'b'}\, \Lambda_{\nu' \nu}\, V^{-1}_{b'b} = ( V \Lambda  )_{a \mu }\, (V \Lambda )_{b \nu } - (V \Lambda  )_{ a \nu}\, ( V \Lambda  )_{b \mu}  \, ,
\ee
where
\be
  V^{-1}_{b' b} = V_{b b'}\, ,\quad \Lambda^{-1}_{\mu a'} = \Lambda_{a' \mu}\, .
\ee
If we choose the parameters in the Lorentz transformation and the gauge transformation to be
\be\label{eq:chooseParam}
  \varphi_{\mu\nu} =  \psi_{\mu\nu}\, ,
\ee
then
\be
  (  V \Lambda)_{ a \mu} = \delta_{  a \mu}\, .
\ee
Therefore,  the form invariance condition Eq.~\eqref{eq:FIforalternative} is satisfied.

According to \eqref{eq:etadef}:
\begin{displaymath}
M_{\mu\nu} \equiv \eta_{i\mu\nu} M_i + \bar{\eta}_{i\mu\nu} N_i \, ,
\end{displaymath}
we can rewrite \eqref{eq:AltAnsatz} as
\be
 A_{\mu} = \frac{p_1(\tau) + p_2(\tau)}{\tau} \, \eta_{i \mu\nu}  \, M_i \,  x_\nu + \frac{p_1(\tau) -  p_2(\tau)}{\tau} \bar{\eta}_{i \mu\nu}  \, N_i \,   x_\nu \, ,
\ee
where we used
\be
\frac{1}{2}\epsilon_{\mu\nu\rho\sigma} \eta_{i \rho\sigma} =  \eta_{i\mu\nu} \, , \quad  \frac{1}{2}\epsilon_{\mu\nu\rho\sigma} \bar{\eta}_{i\rho\sigma}  =  - \bar{\eta}_{i\mu\nu} \, .
\ee
Since $M_i$ and $N_i$ form two independent $SU(2)$ groups, we can choose the parameters to be $p_1(\tau) = p_2(\tau) = p(\tau)$, such that the Ansatz for the 4D Yang-Mills field with the gauge group $SU(2)$ is
\be\label{eq:AltAnsatz-2}
  A_{\mu, a} = 2 \frac{p(\tau)}{\tau} \, \eta_{a\mu\nu} \, x_\nu\, .
\ee
which is the same as the Ansatz given by Eq.~\eqref{eq:amuansatz4D}. One can also solve the Yang-Mills equation directly using the Ansatz \eqref{eq:AltAnsatz} with an $SO(4)$ gauge group, and the same solutions presented in Appendix~\ref{4dcs} can be obtained.

\section{\label{FF}4D Topological Charge}

In this appendix, we show that for the 4D Yang-Mills theory
\be\label{eq:app4Dwinding}
  k = - \frac{1}{16 \pi^2} \int d^4 x\, \textrm{Tr} \left[F^{\mu\nu} (* F_{\mu\nu}) \right]  
\ee
is  an integer-valued quantity, which can be interpreted as the winding number, and only the singular points of the intergrand contribute to it. We follow closely Appendix~A of Ref.~\cite{LectureonInst}.

First, the integrand in Eq.~\eqref{eq:app4Dwinding} is a total derivative:
\be
  \textrm{Tr} \left[F^{\mu\nu} (* F_{\mu\nu}) \right] = 2 \epsilon^{\mu\nu\rho\sigma} \partial_\mu \textrm{Tr} \left(A_\nu \partial_\rho A_\sigma + \frac{2}{3} A_\nu A_\rho A_\sigma \right)\, .
\ee
Hence, the integral in Eq.~\eqref{eq:app4Dwinding} becomes a surface integral, and only boundaries contribute to it. For a smooth manifold like $\mathbb{R}^4$ or $S^4$, the boundaries can be thought of as the singularities of the integrand. Supposing that the singluarities are isolated singular points, one can wrap a surface $S^3$ with radius $R\to 0$ around each of them, and all these small spheres together form the boundary.

Let us focus on one of the singular points. With our Ansatz Eq.~\eqref{eq:app4Dwinding} becomes
\be
  k = -\frac{1}{8 \pi^2} \oint_{S_\beta^3} d\Omega_\mu \, \epsilon^{\mu\nu\rho\sigma} \left(\frac{2}{3} p^3 - p^2 \right)\, \textrm{Tr} \left[ (U^{-1} \partial_\nu U)\, (U^{-1} \partial_\rho U)\, (U^{-1} \partial_\sigma U) \right]\, ,
\ee
where the surface $S_\beta^3$ surrounds the singular point $\beta$, and the radius of the sphere can be taken to be very small. We assume that the function $p$ itself has no singularites, hence, the factor $\frac{2}{3} p^3 - p^2$ has a constant value $\left(\frac{2}{3} p^3 - p^2\right)_\beta$ in the small sphere and can be brought outside the integration. We use $x^\mu$ and $\xi^i (x)$ $(i = 1, 2, 3)$ to denote the spacetime coordinates and the group coordinates respectively. Using
\be
  \textrm{Tr} \left[ (U^{-1} \partial_\nu U)\, (U^{-1} \partial_\rho U)\, (U^{-1} \partial_\sigma U) \right] = \frac{\partial \xi^i}{\partial x^\nu} \frac{\partial \xi^j}{\partial x^\rho} \frac{\partial \xi^k}{\partial x^\sigma} \, \textrm{Tr} \left[ (U^{-1} \partial_i U)\, (U^{-1} \partial_j U)\, (U^{-1} \partial_k U) \right]\, ,
\ee
and expressing the volume element as
\be
  d \Omega_\mu = \frac{1}{6} \epsilon_{\mu\alpha\beta\gamma}\, d x_\alpha\, d x_\beta\, d x_\gamma\, ,
\ee
we can express the integrand on a surface element of $S_\beta^3$ as
\begin{align}
  d k & = -\frac{1}{8 \pi^2} \epsilon^{ijk} \left(\frac{2}{3} p^3 - p^2 \right)_\beta \, \textrm{Tr} \left[(U^{-1} \partial_i U) (U^{-1} \partial_j U) (U^{-1} \partial_k U)  \right] \, d^3 \xi \nonumber\\
  {} & = \frac{3}{16 \pi^2} \left(\frac{2}{3} p^3 - p^2 \right)_\beta \, (\textrm{det}\, e)\, d^3 \xi\, ,
\end{align}
where
\be
  U^{-1} \partial_i U = e^a_i (\xi)\, T_a\, ,
\ee
and $(\textrm{det}\, e) d^3 \xi$ is the Haar measure on the group manifold. Hence,
\be
  k = \int dk
\ee
takes values in $\mathbb{Z}$, which impose additional constraints on the values of $p$ at the boundary.

An explicit calculation shows that for the $SU(2)$ gauge group and the surface $S^3$ at $|x| \to \infty$ the integral gives
\be
  \frac{1}{16 \pi^2} \int_{S^3_{|x| \to \infty} } (\textrm{det}\, e)\, d^3 \xi = 1\, .
\ee 
One can use this result to evaluate the winding numbers for other cases. The only point that one has pay attention to is the orientation of the surface. For instance, the surfaces around $|x| = 0$ and around $|x| \to \infty$ have opposite orientations, which will consequently differ by a sign in the contribution to the winding number.

\section{\label{3dcs}3D Classical Solutions}

In Section~\ref{3ds}, the Ansatz to the 3D Yang-Mills equation reads 
\be 
  A_{\mu, a}  =   G \left( \frac{\delta_{\mu a}}{|x|} - \frac{x_\mu x_a}{|x|^3} \right) + \left(  H-1 \right) \frac{\epsilon_{\mu a i} x_i}{|x|^2}\, .
\ee
After  simple algebra we obtain
\begin{align}
F_{\mu\nu}^a  = & \, \left( \frac{  x_\mu \delta_{\nu a} - x_\nu \delta_{\mu a}}{|x|^3} \right) \left( 2   \tau G' \right) + \frac{\epsilon_{\mu\nu a}}{|x|^2} \left( G^2 + 2 H     -2 \right) \nonumber\\
& + \frac{x_i \left(  x_\mu \epsilon_{\nu a i} - x_\nu \epsilon_{\mu a i} \right)}{|x|^4} \left( 2 - 2 H + 2 \tau H'   - G^2 \right) + \frac{x_a x_i \epsilon_{\mu\nu i}}{|x|^4} \left( H-1 \right)^2\, ,
\end{align}
where $\tau \equiv x_\mu x^\mu$ and $(\cdots)' \equiv \partial (\cdots) / \partial \tau$.

We also have
\begin{align}
\left( D_\mu F_{\mu\nu}  \right)^a  & =  \partial_\mu F_{\mu\nu}^a + \epsilon_{a b c}A_\mu^b F_{\mu\nu}^c  \nonumber\\ 
{} & = \frac{x_a x_\nu}{|x|^5} \left( - G + G^3 + G H^2 - 2 \tau G' + 4 \tau H G' - 4 \tau G H' - 4 \tau^2 G'' \right) \nonumber\\
 & \quad + \frac{\epsilon_{\nu a i} x_i}{|x|^4} \left( H - G^2 H - H^3 + 2 \tau H' + 4 \tau^2 H'' \right) \nonumber\\
& \quad + \frac{\delta_{\nu a}}{|x|^3} \left( G - G^3 - G H^2  + 2 \tau G' + 4 \tau^2 G''\right)\, .
\end{align}
The Yang-Mills equation
\be
\left( D_\mu F_{\mu\nu}  \right)^a  = 0  
\ee
now becomes a system of three independent equations
\begin{align}
- G + G^3 + G H^2 - 2 \tau G' + 4 \tau H G' - 4 \tau G H' - 4 \tau^2 G'' & = 0\, ,\la{eq:solve3deqc1}\\
G - G^3 - G H^2  + 2 \tau G' + 4 \tau^2 G'' & = 0\, ,\la{eq:solve3deqc2}\\
H - G^2 H - H^3 + 2 \tau H' + 4 \tau^2 H'' & = 0\, .\la{eq:solve3deqc3}
\end{align}
To solve this equation system, we first observe that Eq.~\eqref{eq:solve3deqc1} plus Eq.~\eqref{eq:solve3deqc2} leads to
\be\la{eq:solve3deqc4}
  4 \tau H G' = 4 \tau G H'\, .
\ee
For $G\neq 0$ and $H\neq 0$ the equation above implies that
\be
  G = c\, H\, ,
\ee
where $c$ is a nonzero constant. Then all three equations become the same:
\be\label{eq:app3DEOM}
  H - (1+c^2) H^3 + 2 \tau H' + 4 \tau^2 H'' = 0\, .
\ee
According to the topological constraints discussed in Appendix~\ref{CS}, $G$ and $H$ should have fixed values at the boundaries (See Table~\ref{table:3dBC2}), which are $\tau=0$ and $\tau=\infty$ for the spherically symmetric solutions. We can expand $H(\tau)$ in the following way:
\be
  H (\tau) = \Bigg\{
  \begin{array}{ll}
    a^{\tau=0}_0 + \sum_{n=1}^\infty a^{\tau=0}_n \, \tau^n\, ,& \textrm{for small } \tau\, ;\nonumber\\
    a^{\tau=\infty}_0 + \sum_{n=1}^\infty a^{\tau=\infty}_n / \tau^n\, ,&  \textrm{for large } \tau\, .
  \end{array}
\ee
Plugging the expansions into Eq.~\eqref{eq:app3DEOM}, we obtain at the leading order
\be
  a_0^{\tau = 0\, , \infty} = 0 \,\, \textrm{or} \,\, \pm\frac{1}{\sqrt{1+c^2}}\, .
\ee
For both cases, the higher order terms of the expansion for large value of $\tau$ lead to
\be
  a_n^{\tau = \infty} = 0\quad \textrm{for}\quad n\geq 1\, .
\ee
Hence, for $G\neq 0$ and $H\neq 0$ we obtain
\be
  H = \pm\frac{1}{\sqrt{1+c^2}}\, ,\quad G = \pm \frac{c}{\sqrt{1+c^2}}\, .
\ee
Since the constant $c$ can be either positive or negative, it is possible to define
\be
  H = \textrm{sin}\, \Theta\, ,\quad G = \textrm{cos}\, \Theta\, ,
\ee
where $\Theta$ is a constant. Actually the results above also hold for $G\neq 0$, $H = 0$ or $G=0$, $H\neq 0$, since for both cases the 3D Yang-Mills equation can be reduced to an equation similar to Eq.~\eqref{eq:app3DEOM} with $c=0$. The case $G\neq 0$, $H=0$ corresponds to $\Theta = \pm \pi/2$, while the case $G=0$, $H\neq 0$ corresponds to $\Theta = 0 \,\, \textrm{or} \,\, \pi$. Finally, $G=H=0$ is also a solution, which is in fact the Wu-Yang monopole solution and corresponds to $p = 1/2$.

In fact, one can obtain the equations for $G$ and $H$ by varying an action. First,
\be
 F_{\mu\nu}^aF_{\mu\nu}^a = \frac{2}{\tau^2} + \left( \frac{4}{\tau}  \left(  \partial_\mu G \right)^2 - \frac{4}{\tau^2} G^2 + \frac{2}{\tau^2} G^4 \right) + \left(\frac{4}{\tau}  \left(  \partial_\mu H \right)^2 - \frac{4}{\tau^2} H^2 + \frac{2}{\tau^2} H^4 \right) + \frac{4}{\tau^2} G^2 H^2\, ,
\ee
and then the action can be defined as
\be
S = \frac{1}{4 g^2} \int d^3 \, x \,   F_{\mu\nu}^aF_{\mu\nu}^a \, .
\ee
The variation of this action results in
\begin{align}
- \partial_\mu \left(  \frac{8}{\tau}   \partial_\mu G  \right) -  \frac{8}{\tau^2} G +  \frac{8}{\tau^2} G^3 +  \frac{8}{\tau^2} G H^2  & = 0 \, ,  \nonumber\\
- \partial_\mu \left(  \frac{8}{\tau}   \partial_\mu H  \right) -  \frac{8}{\tau^2}H +  \frac{8}{\tau^2} H^3 +  \frac{8}{\tau^2} H G^2  & = 0  \, ,
\end{align}
which can be simplified to
\begin{align}
- 4 \tau^2 G'' - 2 \tau G' -    G +   G^3 +   G H^2  & = 0 \, ,  \nonumber\\
- 4 \tau^2 H'' - 2 \tau H' -    H +   H^3 +   H G^2  & = 0  \, .
\end{align}
These equations are just Eq.~\eqref{eq:solve3deqc2} and Eq.~\eqref{eq:solve3deqc3}.

\section{\label{4dcs}4D Classical Solutions}
Similar to the 3D case, we analyze the classical solutions to the 4D Yang-Mills equation in this appendix. The form invariance condition \eqref{eq:FI} has restricted the 4D Ansatz to be
\be
  A_{\mu, a} = 2\, \frac{p(\tau)}{\tau}\, \eta_{a\mu\nu} x_\nu\, .
\ee
Again, for the Eulidean case we do not distinguish the upper and lower indices. Then, consequently
\be
  F_{\mu\nu}^a = 4 \eta_{a\mu\nu} \left(\frac{p^2}{\tau} - \frac{p}{\tau} \right) + 4 x_\rho (x_\mu \eta_{a\nu\rho} - x_\nu \eta_{a\mu\rho}) \left[\left(\frac{p}{\tau} \right)' + \frac{p^2}{\tau^2} \right]\, ,
\ee
where the prime denotes the derivative with respect to $\tau$. The Yang-Mills equation now appears as
\be
  (D_\mu F_{\mu\nu})^a = 8\, \frac{\eta_{a\nu\rho} x_\rho}{\tau^2}\, (-p + 3 p^2 - 2 p^3 + \tau p' + \tau^2 p'') = 0\, ,
\ee
and we only need to solve the differential equation
\be\label{eq:app4DEOM}
  - p + 3 p^2 - 2 p^3 + \tau p' + \tau^2 p'' = 0\, .
\ee
As seen from Table~\ref{table:4dBC}, the boundary values of $p$ are fixed by the topological properties. Hence, we can expand $p$ near the boundaries, which for the spherically symmetric case are just $\tau=0$ and $\tau = \infty$. The series is given as follows:
\be
  p (\tau) = \Bigg\{
  \begin{array}{ll}
    a^{\tau=0}_0 + \sum_{n=1}^\infty a^{\tau=0}_n \, \tau^n\, ,& \textrm{for small } \tau\, ;\nonumber\\
    a^{\tau=\infty}_0 + \sum_{n=1}^\infty a^{\tau=\infty}_n / \tau^n\, ,& \textrm{for large } \tau\, .
  \end{array}
\ee
Plugging these expressions into Eq.~\eqref{eq:app4DEOM}, at the leading order they both lead to
\be
  a_0^{\tau = 0, \infty} = 0\,\, \textrm{or}\,\, \frac{1}{2}\,\, \textrm{or}\,\, 1\, .
\ee
Combining the boundary values of $a_0$ at $\tau=0$ and $\tau=\infty$, there are $3\times3 = 9$  combinations. However, the topological properties rule out some combinations, for example $a_0^{\tau = 0}=0$ and $a_0^{\tau = \infty}=1/2$. We can easily show that there are only 5 possible boundary values that respect the topological properties: 
\begin{enumerate}
  \item
  $a_0^{\tau=0} = \frac{1}{2}\, ,\quad a_0^{\tau=\infty} = \frac{1}{2}$ :

  We plug the expansion of $p$ for large $\tau$ into Eq.~\eqref{eq:app4DEOM}. By requiring that $p(\tau)$ is infinitely differentiable, all the higher order terms should vanish, hence
\be
  p = a_0 = \frac{1}{2}
\ee
in this case. It corresponds to the 4D meron solution.

  \item
  $a_0^{\tau=0} = 0\, ,\quad a_0^{\tau=\infty} = 1$ :

  By plugging the expansion of $p$ for large $\tau$ into Eq.~\eqref{eq:app4DEOM}, we obtain
\be
  a_n^{\tau = \infty} = (a_1^{\tau=\infty})^n\, ,
\ee
which leads to
\be
  p(\tau) = 1 + \sum_{n=1}^\infty \frac{(a_1^{\tau=\infty})^n}{\tau^n} = \frac{\tau}{\tau - a_1^{\tau = \infty}}\, ,
\ee
which is also consistent with $p(\tau=0)=0$ for $a_1^{\tau=\infty} \neq 0$. If we require that $p(\tau)$ is a smooth function without singularities, then $a_1^{\tau=\infty}$ should be a real negative constant in this case.

  \item
  $a_0^{\tau=0} = 1\, ,\quad a_0^{\tau=\infty} = 1$ :

  This case can be viewed as the previous case with
\be
  a_1^{\tau=\infty} = 0\, .
\ee
Hence,
\be
  a_n^{\tau=\infty} = 0\quad \textrm{for}\quad n \geq 1\, ,
\ee
and
\be
  p(\tau) = 1\quad \textrm{for}\quad 0 \leq \tau \leq 1\, .
\ee

  \item
  $a_0^{\tau=0} = 1\, ,\quad a_0^{\tau=\infty} = 0$ :

  Similarly we obtain
\be
  a_n^{\tau = \infty} = (-1)^{n-1} (a_1^{\tau=\infty})^n\, .
\ee
Therefore,
\be
  p (\tau) = \sum_{n=1}^\infty (-1)^{n-1} \frac{(a_1^{\tau=\infty})^n}{\tau^n} = \frac{a_1^{\tau=\infty}}{\tau + a_1^{\tau=\infty}}\, ,
\ee
which also satisfies the boundary value $p(\tau=0) = 1$ for $a_1^{\tau=\infty} \neq 0$. Again, we require that $p(\tau)$ is a smooth function without singularities, then $a_1^{\tau=\infty}$ should be a real positive constant in this case.

  \item
  $a_0^{\tau=0} = 0\, ,\quad a_0^{\tau=\infty} = 0$ :

  This case can be viewed as the previous case with
\be
  a_1^{\tau=\infty} = 0\, .
\ee
Hence,
\be
  a_n^{\tau=\infty} = 0\quad \textrm{for}\quad n \geq 1\, ,
\ee
and
\be
  p(\tau) = 0\quad \textrm{for}\quad 0 \leq \tau \leq 1\, .
\ee
\end{enumerate}

Interestingly, for each topologically allowed boundary value, there is a corresponding classical solution. In all, there are five classical solutions as listed in Section~\ref{4ds}.

Similar to Appendix~\ref{3dcs}, to obtain the equation for $p$ one can define an action
\be
S = \frac{1}{4 g^2} \int d^4 \, x \,   F_{\mu\nu}^aF_{\mu\nu}^a \, ,
\ee
where
\be
\frac{1}{4} F_{\mu\nu}^aF_{\mu\nu}^a =  \frac{6}{\tau} \left( \partial_\mu p \right)^2 +  \frac{24}{\tau^2} p^2 (p-1)^2    \, .
\ee 
Variation of this action gives us
\be
- \partial_\mu \left( \frac{1}{\tau} \partial_\mu  p \right) + \frac{4}{\tau^2} \left( p - 3 p^2 + 2 p^3  \right) = 0 \, ,
\ee
which can be simplified to
\be
- \tau^2 p'' - \tau p' +p - 3 p^2 + 2 p^3 = 0 \, .
\ee
This equation  is the same as Eq.~\eqref{eq:app4DEOM}.

\section{\label{topologicalmeasure}Complete Measure for Pseudo Zero Modes}

In this appendix, we discuss how to calculate the path integral measure from pseudo zero modes in the presence of topological fluctuations.

Let us start with
\be
  \delta A_\mu^{\textrm{top}} = \left(\frac{\partial A_\mu^{\textrm{top}}}{\partial \gamma^{(i)}} + D_\mu \Lambda^{(i)} \right) \, \delta \gamma^{(i)} +  \left( \frac{\partial A_\mu^{\textrm{top}}}{\partial \widetilde{\mathcal{A}}} \right) \delta  \widetilde{\mathcal{A}} \, ,
\ee
where the translations and gauge orientations $\gamma^{(i)}$ leave the action invariant, and therefore their gauges need to be fixed. The topological fluctuations $ \widetilde{\mathcal{A}}$ preserve the topology but not the action, hence no gauge fixing is needed for $\widetilde{\mathcal{A}}$. We will denote the first term in the equation above as $\delta_{\gamma^{(i)}} A_\mu^{\textrm{top}}$.

  The gauge of $A_\mu^{\textrm{top}}$ was already fixed, when we considered the topologically stable Ansatz that satisfies the form invariance condition and the topological properties. $\delta_{\gamma^{(i)}} A_\mu^{\textrm{top}}$ has its own gauge choice, and the only condition is that $\delta_{\gamma^{(i)}} A_\mu^{\textrm{top}}$ should not be a gauge transformation. This condition can be achieved by requiring that $\delta_{\gamma^{(i)}} A_\mu^{\textrm{top}}$ is orthogonal to a gauge transformation, i.e.,
\be
  \int d^D x\, (D_\mu \Lambda) \, \delta_{\gamma^{(i)}} A_\mu^{\textrm{top}} = 0\, ,
\ee
which after a partial integration leads to the gauge condition for $\delta_{\gamma^{(i)}} A_\mu^{\textrm{top}}$:
\be\label{eq:gaugedA}
  D_\mu \left(\delta_{\gamma^{(i)}} A_\mu^{\textrm{top}}\right) = 0\, ,
\ee
where
\be
  D_\mu \equiv \partial_\mu + A_\mu^{\textrm{top}}\, .
\ee

  Let us define
\be
  Z_\mu^{(i)} \equiv \frac{\partial A_\mu^{\textrm{top}}}{\partial \gamma^{(i)}} + D_\mu \Lambda^{(i)}
\ee
to be the pseudo zero modes, which will become the zero modes if all the topological fluctuations are turned off, but generally they are not zero modes of  $\mathcal{M}_{\mu\nu}$. We require that $Z_\mu^{(i)}$  satisfy the gauge condition:
\be
  D_\mu Z_\mu^{(i)} = 0\quad \textrm{with}\quad D_\mu \equiv \partial + A_\mu^{\textrm{top}}\, .
\ee
Similarly, for the topological fluctuations $\widetilde{\mathcal{A}}$, one can also define the corresponding pseudo zero modes:
\be
  Z_\mu^{\widetilde{\mathcal{A}}} \equiv \frac{\partial A_\mu^{\textrm{top}}}{\partial \widetilde{\mathcal{A}}}\, ,
\ee
whose explicit forms depend on the dimension, as we will show later in this section. These pseudo zero modes differ from the previous ones, in the sense that they will vanish identically when the topological fluctuations are turned off.

To see the relation between the pseudo zero modes and the Jacobian, we rewrite it in the quantum state language
\be\la{eq:quantumnormeq1}
\left| A_\mu^{\textrm{top}} \right> = Z_\mu^{(1)} \left| \gamma^{(1)} \right> \otimes   Z_\mu^{(2)} \left| \gamma^{(2)} \right> \otimes \cdots \,\, .
\ee 
If $\left|  \gamma^{(i)} \right>$ are properly normalized, the norm of $\left| A_\mu^{\textrm{top}} \right>$ reads
\be
\left< A_\mu^{\textrm{top}}  | A_\mu^{\textrm{top}} \right> = \det \left| U^{ij}  \right|\, ,
\ee
where
\be
  U^{ij} \equiv \langle Z_\mu^{(i)} \big| Z_\mu^{(j)} \rangle = - \frac{2}{g^2} \int d^D x\, \textrm{Tr} \left[Z_\mu^{(i)} Z_\mu^{(j)} \right]\, .
\ee
If $\left|  \gamma^{(i)} \right>$ are orthogonal to each other, then $U^{ij}$ is a diagonal matrix. In this paper, we adopt the following approximation
\be
\left| A_\mu^{\textrm{top}} \right> \approx  \sqrt{\det \left| U^{ij}  \right|} \,  \left| \gamma^{(1)} \right> \otimes    \left| \gamma^{(2)} \right> \otimes \cdots \, ,
\ee
where it has the same norm as Eq.~\eqref{eq:quantumnormeq1}. Hence,
\be
[\textrm{Jac}]_{\gamma^{(i)}} \approx \sqrt{\det \left| U^{ij}  \right|} \, .
\ee

We can also apply the same procedure to the topological fluctuations and find the corresponding measure:
\be
  U^{\widetilde{\mathcal{A}}} \equiv \langle Z_\mu^{\widetilde{\mathcal{A}}} \big| Z_\mu^{\widetilde{\mathcal{A}}} \rangle\, .
\ee
As we will show now, the measure due to the topological fluctuations are constant, and therefore can be dropped from the path integral.

\begin{itemize}
\item 3D case:
\begin{align}
\left(Z^{\widetilde{G}} \right)_\mu^a  & = \left( \frac{\delta_{\mu a}}{|x - x_0|} - \frac{(x - x_0)_\mu  (x - x_0)_a}{|x - x_0|^3}   \right) \, , \nonumber\\
 \left(Z^{\widetilde{H}} \right)_\mu^a  & =  \frac{\epsilon_{\mu a i} (x - x_0)_i}{|x - x_0|^2}
\end{align}

\be
\Rightarrow \quad U^{\widetilde{G} \widetilde{G}} = \frac{1}{g^2} \int d^3 x \, \left(Z^{\widetilde{G}} \right)_\mu^a  \left(Z^{\widetilde{G}} \right)_\mu^a = \frac{8 \pi L }{g^2}\, ,
\ee
where $L$ is the size of the system.  The result is the same for $U^{\widetilde{H} \widetilde{H}} $. Therefore,
\be
[\textrm{Jac}]_{\widetilde{\mathcal{A}}} =  \frac{8 \pi L }{g^2} \, .
\ee

\item 4D case:
\be
\left(Z^{\widetilde{p}} \right)_\mu^a = 2 \eta_{a\mu\nu}  \frac{(x - x_0)_\nu}{|x- x_0|^2}
\ee

\be
\Rightarrow \quad U^{\widetilde{p} \widetilde{p}} = \frac{1}{g^2} \int d^4 x \, \left(Z^{\widetilde{p}} \right)_\mu^a  \left(Z^{\widetilde{p}} \right)_\mu^a = \frac{12 \pi^2 L^2}{g^2} \, ,
\ee
and
\be
[\textrm{Jac}]_{\widetilde{\mathcal{A}}} =  \frac{2 \sqrt{3} \pi L }{g} \, .
\ee
\end{itemize}

In practice, $[\textrm{Jac}]_{\gamma^{(i)}}$ is very difficult to compute. Here we take the 4D case as an example to explain how to do it. $A^\textrm{top}_\mu$ in 4D reads
\be
A_{\mu, a}^\textrm{top} = 2 \frac{p}{\tau} \eta_{a \mu\nu}  \, (x - x_0)_\nu \, ,
\ee
where $\tau = (x - x_0)^2$ and
\be
p = p(\tau; \, c_\rho ; \, c_1, c_2, c_3 ...) \, .
\ee
Here $p$ preserves the topological charge, and $c_\rho$ is the parameter that also leaves the action invariant, while $c_1$, $c_2$, $\cdots$   depend on topological modes and do change the action.  The definition of $c_1$, $c_2$, $\cdots$ will be given later in this appendix. Approximately,
\be
p(\tau; \, c_\rho ; \, c_1, c_2, c_3 ...) \approx p_0 (\tau; \, c_\rho) + \widetilde{p} (\tau,  c_1 , c_2 , ...) \, ,
\ee
where $p_0$ clearly corresponds to the classical solution. For instance,
\be
p_0 = \frac{\tau}{\tau + c_\rho \ell_0^2} \quad  \textrm{for instanton}\, , \quad \quad p_0 = \frac{c_\rho \ell_0^2}{\tau + c_\rho \ell_0^2} \quad \textrm{for anti-instanton}\, ,
\ee
where $\ell_0 \equiv \ell_{\textrm{cl}}$ is the averaged length scale of the classical background. For convenience, let us define
\be
\rho^2 \equiv c_\rho \ell_0^2\, .
\ee
  We consider the anti-instanton background as an example to explain how to calculate $[\textrm{Jac}]_{\gamma^{(i)}}$.  The results for different pseudo zero modes are listed in the following:
\begin{itemize}
\item Translational pseudo zero modes ($\gamma^{(\nu)} = x_0^\nu$):
\begin{align}
 \frac{\partial A_\mu^{\textrm{top}}}{\partial x_0^\nu} & =  - \partial_\nu A_\mu^{\textrm{top}} \, ,\nonumber\\
  \Lambda_{\nu} & = 2\, \frac{F_t }{\tau} \, \eta_{a\nu\rho} x_\rho  \, ,\nonumber\\
  Z_\mu^{(\nu)} & = 4 T^a \left(   x_\mu \eta_{a \nu\rho}  x_\rho \left[ \left(  \frac{F_t}{\tau} \right)' + \frac{p F_t}{\tau^2} \right]  -    x_\nu \eta_{a\mu\rho}  x_\rho\left[ \left(  \frac{p}{\tau} \right)' + \frac{p F_t}{\tau^2} \right] +  \eta_{a\mu\nu}  \left[ \frac{p F_t}{\tau} - \frac{F_t}{2\tau} - \frac{p}{2\tau}   \right]\right) \, ,\nonumber\\
  U_{\mu\nu} &  = \frac{12 \pi^2  }{g^2} \delta_{\mu\nu} \int_0^\infty    d \tau \,\,  \frac{1}{\tau } \Bigg(  \left(F_t^2 + p^2   \right) + 2 F_t p (F_t p - F_t - p) \nonumber\\
{} & \quad\quad\quad\quad\quad\quad\quad\quad - \frac{ \tau}{2} \left[ \left(  F_t - p \right)^2\right]' + \tau^2 \left[  \left({F_t}' \right)^2  + \left( p'  \right)^2 \right]  \Bigg)\, ,
\end{align}
where
\be
F_t = F_t (\tau; \, c_\rho; \, c_1, c_2 , \cdots) \, ,
\ee
satisfying
\be\label{eq:Fcondtion4Deq1}
- F_t + 2 F_t p + p^2 - 2 F_t p^2 + \tau {F_t}'  + \tau^2 {F_t}''  = 0 \, .
\ee

\item Gauge orientation pseudo zero modes  ($\gamma^{(\alpha)} = \varphi^\alpha$):
\begin{align}
 \frac{\partial A_\mu^{\textrm{top}}}{\partial \varphi^\alpha}|_{\varphi^\alpha=0} & = \delta_{\alpha a}  [A_\mu^{\textrm{top}}, T^a ] \, ,\nonumber\\
   \Lambda_{\alpha}  & =  \delta_{\alpha a} \left[  F_g- 1 \right] T^a ,\nonumber\\
  Z_\mu^{\alpha} & =   \delta_{\alpha a} \left( \left[ \partial_\mu F_g  \right] \delta_{a c} - F_g \epsilon_{a b c} A_{\mu, b}^{\textrm{top}} \right) T_c   \, ,\nonumber\\
  U_{\alpha \beta} (\varphi) & =e_\alpha\,^a (\varphi)\, e_\beta\,^a (\varphi)\,   \frac{4 \pi^2}{g^2}  \int_0^\infty    d \tau \,\,  \left(   \tau^2 \left[  \partial_\tau F_g \right]^2 + 2 F_g^2 p^2  \right)\, ,
\end{align}
where 
\be
F_g = F_g (\tau; \, c_\rho; \, c_1, c_2 , \cdots)
\ee
satisfies
\be\label{eq:Fcondtion4Deq2}
2 \tau \partial_\tau F_g + \tau^2 \partial_\tau^2 F_g = 2 F_g p^2 \, ,
\ee
and $e_\alpha\,^a (\varphi)$ is the group vielbein.

\item Dilatational pseudo zero modes (size $\rho$):
\begin{align}
  Z_\mu^{\rho}  & =  \frac{\partial A_\mu^{\textrm{top}}}{\partial \rho}   =  \left(  \frac{\partial p}{\partial \rho} \right) \left( \frac{2}{\tau} \eta_{a \mu\nu} (x - x_0)_\nu T_a  \right)  \, ,\nonumber\\
   \Lambda_{\rho}  & =0\, ,\nonumber\\
  U^{\rho\rho} (\varphi) & = \frac{ 12 \pi^2}{g^2}   \int_0^\infty    d \tau \,\,   \left(  \frac{\partial p}{\partial \rho} \right)^2\, .
\end{align}
Unlike the other modes, here we have
\be
 \left(  \frac{\partial p}{\partial \rho} \right)  =  \left(  \frac{\partial p_0}{\partial \rho} \right) = \frac{2 \rho \tau}{\left( \tau + \rho^2  \right)^2}
\ee
\be
  \Rightarrow\quad [\textrm{Jac}]_{\rho}^{\textrm{top}} = [\textrm{Jac}]_{\rho}^{\textrm{cl}}  = \sqrt{U^{\rho\rho}}= \frac{4 \pi }{g}\, ,
\ee
which is a constant and hence irrelevant in our discussion. 
\end{itemize}

For all the modes induced by $\gamma^{i}$, we have
\be\la{eq:allnormsbygamma}
  U_{ij} =   \left( \begin{array}{cc}
      U_{\mu\nu} & {}   \\
     {} &  U_{\alpha\beta}(\varphi)
  \end{array} \right)\, ,
\ee
and
\be
[\textrm{Jac}]_{\gamma^{(i)}}^{\textrm{top}} \, \left( c_\rho; c_1, c_2 , \cdots  \right) = \sqrt{\textrm{det}\, U_{ij}} \, \left( c_\rho; c_1, c_2 , \cdots  \right) \, ,
\ee
which clearly depends on topological fluctuations $\widetilde{p}$.

To simplify the discussions, we change the variable by performing the following conformal transformation:
\be\la{eq:conformalcoordinatetransformationdef}
\zeta \equiv \frac{1}{2} \frac{\tau - \ell_0^2}{\tau + \ell_0^2}\, ,
\ee
such that
\be
\int_0^\infty \, d \tau \longrightarrow  \ell_0^2 \int_{- \frac{1}{2}}^{\frac{1}{2}} \, \frac{1}{\left( \frac{1}{2} - \zeta  \right)^2} d \zeta \, ,
\ee
and
\be
\frac{\partial}{\partial \tau} \longrightarrow \frac{1}{\ell_0^2} \left(\frac{1}{2} - \zeta   \right)^2  \frac{\partial}{\partial \zeta}\, .
\ee
Consequently, we have
\begin{align}
 U_{\mu\nu} &  = \frac{8 \pi^2  }{g^2} \, \delta_{\mu\nu} \, \Omega_1 (c_\rho,  c_1, c_2 , \cdots)\, , \nonumber\\
 U_{\alpha \beta} (\varphi) & = e_\alpha\,^a (\varphi)\, e_\beta\,^a (\varphi)\,   \frac{4 \pi^2}{g^2} \, \ell^2_0\, \Omega_2 (c_\rho,  c_1, c_2 , \cdots)\, ,
\end{align} 
where
\begin{align}
\Omega_1 (c_\rho,  c_1, c_2 , \cdots) & \equiv \frac{3}{2} \int_{- \frac{1}{2}}^{\frac{1}{2}}   d \zeta \,    \Bigg( \frac{ \left(F_t^2 + p^2   \right) + 2 F_t p (F_t p - F_t - p)}{\left( \frac{1}{2} - \zeta  \right)\left( \frac{1}{2} +  \zeta  \right)} - \frac{1}{2} \partial_\zeta \left(  F_t - p  \right)^2  \nonumber\\
{} & \qquad\quad\quad\quad\quad + \left( \frac{1}{2} + \zeta  \right)  \left( \frac{1}{2} - \zeta  \right)  \left[  \left(\partial_\zeta p \right)^2 + \left(\partial_\zeta F_t \right)^2  \right] \Bigg)  \, , \nonumber\\
\Omega_2 (c_\rho,  c_1, c_2 , \cdots)& \equiv \int_{- \frac{1}{2}}^{\frac{1}{2}}   d \zeta \,    \left[   \left( \frac{1}{2} + \zeta  \right)^2 \left(   \frac{\partial F_g}{\partial \zeta} \right)^2 + \frac{2 F_g^2 p^2}{ \left( \frac{1}{2} - \zeta  \right)^2 }  \right]\, ,
\end{align}
and
\be
p = \frac{c_\rho}{c_\rho + \frac{1 + 2 \zeta}{1 - 2 \zeta}} + \widetilde{p} (\zeta; \, c_1 , c_2, \cdots)\, .
\ee

Now we can expand $\widetilde{p}$ in a Fourier series and define $c_1$, $c_2$, $\cdots$ in the following way:
\be\label{eq:tildepconformaldecompose}
\widetilde{p} (\zeta; \, c_1 , c_2, \cdots) \equiv \sum_{n=1}^\infty \, c_n \sin \left[ n \pi \left(    \frac{1}{2} + \zeta  \right) \right] \, ,
\ee
where $c_1$, $c_2$, $\cdots$ are the coefficients of the Fourier series. Plugging the expression above back to Eq.~\eqref{eq:Fcondtion4Deq1} and Eq.~\eqref{eq:Fcondtion4Deq2}, we can solve for $F_t$ and $F_g$.

We can combine everything above, then Eq.~\eqref{eq:allnormsbygamma} reads: 
\be
  U_{ij} =  \frac{8 \pi^2}{g^2} \left( \begin{array}{ccc}
      \delta_{\mu\nu}  \, \Omega_1 (c_\rho,  c_1, c_2 , \cdots) & {}   \\
    {} & \frac{1}{2} g_{\alpha\beta} (\varphi)\, \ell_0^2 \,  \Omega_2 (c_\rho,  c_1, c_2 , \cdots)
  \end{array} \right)\, ,
\ee
Hence,
\be
 [\textrm{Jac}]_{\gamma^{(i)}}^{\textrm{top}} = \sqrt{\textrm{det}\, U_{ij}} = \frac{2^{9} \pi^7 \ell^3_0}{g^7}  \sqrt{\textrm{det} \, g_{\alpha\beta} (\varphi)}   \,   \left(\Omega_1 \right)^2     \left(\Omega_2 \right)^\frac{3}{2} \, ,
\ee
and the measure in the path integral is
\be
  \frac{2^{12}\, \pi^{10}}{g^8} \ell_0^3 \int \, d^4 x_0\, \int \,  \ell_0 d \sqrt{c_\rho} \,  \int      \frac{  \sqrt{\textrm{det} \, g_{\alpha\beta} (\varphi)} \, d^3 \varphi}{ 2 \pi^2} \,  \left(\Omega_1 \right)^2     \left(\Omega_2 \right)^\frac{3}{2} \, ,
\ee
where we have included the dilatational pseudo zero mode.

In the rest of this appendix, we would like to discuss two issues. The first one is the relation between the discussions in this appendix and the wave function renormalization used in Section~\ref{approx}. Since the Jacobian depends on the parameters $c_i$ and $\varphi$:
\be
[\textrm{Jac}]_{\gamma^{(i)}}^{\textrm{top}} = \frac{2^{9} \pi^7 \ell^3_0}{g^7}  \sqrt{\textrm{det} \, g_{\alpha\beta} (\varphi)}   \,   \left[\Omega_1 (c_\rho,  c_1, c_2 , \cdots)\right]^2     \left[\Omega_2 (c_\rho,  c_1, c_2 , \cdots)\right]^\frac{3}{2} \, ,
\ee
it is clear from this expression that $[\textrm{Jac}]_{\gamma^{(i)}}^{\textrm{top}}$ may depend on the dilatations and the tolopogical fluctuations, but is independent of $x_0$, and the only part that depends on $\varphi$  is $ \sqrt{\textrm{det} \, g_{\alpha\beta} (\varphi)}$. To get rid of the $\varphi$-dependence, we notice that
\be
\frac{[\textrm{Jac}]_{\gamma^{(i)}}^{\textrm{top}}}{[\textrm{Jac}]_{\gamma^{(i)}}^{\textrm{cl}}} = \frac{[\textrm{Jac}]_{\gamma^{(i)}}^{\textrm{top}}}{[\textrm{Jac}]_{\gamma^{(i)}}^{\textrm{cl}}}  \left( c_\rho, c_1, c_2, \cdots  \right) \, ,
\ee
is independent of $x_0$ and $\varphi$, where $[\textrm{Jac}]_{\gamma^{(i)}}^{\textrm{cl}}$ is the classical measure, which can be understood as the measure when the topological fluctuations are turned off. Consequently, we can write the path integral measure as
\be
\int [\textrm{Jac}]_{\gamma^{(i)}}^{\textrm{top}}    \mathcal{D} \widetilde{p} \longrightarrow [\textrm{Jac}]_{\gamma^{(i)}}^{\textrm{cl}}\,  \int \prod_{n=1}^\infty  d c_n \,\,  \frac{[\textrm{Jac}]_{\gamma^{(i)}}^{\textrm{top}}}{[\textrm{Jac}]_{\gamma^{(i)}}^{\textrm{cl}}}  \left( c_\rho, c_1, c_2, \cdots  \right) \, .
\ee
Then we can define
\be
 \int \prod_{n=1}^\infty  d c_n \,\,  \frac{[\textrm{Jac}]_{\gamma^{(i)}}^{\textrm{top}}}{[\textrm{Jac}]_{\gamma^{(i)}}^{\textrm{cl}}}  \left( c_\rho, c_1, c_2, \cdots  \right)  \longrightarrow  \int \prod_{n=1}^\infty  d c'_n \, ,
\ee
which is equivalent to the wave function renormalization of the topological fluctuations used in Section~\ref{approx}:
\be
\int  \frac{[\textrm{Jac}]_{\gamma^{(i)}}^{\textrm{top}}}{[\textrm{Jac}]_{\gamma^{(i)}}^{\textrm{cl}} }  \mathcal{D} \widetilde{\mathcal{A}} \longrightarrow \int   \mathcal{D} \widetilde{\mathcal{A}}' \, .
\ee

Another issue is about the length scale. We expand $\widetilde{p}$ in Eq.~\eqref{eq:tildepconformaldecompose} as
\begin{displaymath}
 \widetilde{p} = \sum_{n=1}^\infty \, c_n \sin \left[ n \pi \left(    \frac{1}{2} + \zeta  \right) \right] \, .
\end{displaymath}
We will use the first quarter wavelength to define the characteristic length. Suppose that the phase starts with $0$, i.e.,
\be
n \pi \left[\frac{1}{2}+ \left(\zeta|_{\tau = 0} = -\frac{1}{2} \right) \right]\, = 0 \, ,
\ee
where we used \eqref{eq:conformalcoordinatetransformationdef}:
\begin{displaymath}
  \zeta = \frac{1}{2} \frac{\tau - \ell_0^2}{\tau + \ell_0^2}\, .
\end{displaymath}
At one quarter wavelength the phase is $\pi/2$, then the first quarter wave ends at $\zeta = -1/2 + 1/(2 n)$, because
\be
n \pi \left[\frac{1}{2} + \left( -\frac{1}{2} + \frac{1}{2n}   \right) \right]  = \frac{\pi}{2}\, .
\ee
Hence, we have the corresponding change in $\tau$ for the phase change on one quarter wavelength:
\be
\tau \left(  \zeta=  -\frac{1}{2} + \frac{1}{2n} \right)  = \frac{1}{2 n - 1} \ell_0^2\, .
\ee
 We approximately define the length scale for each mode as
\be
\ell\,_{\textrm{scale}}^n = \sqrt{ \frac{1}{2 n - 1}} \,\, \ell_0\, .
\ee
As we can see, for $n=1$ the length scale is just $\ell_0$. For $n = \infty$, the length scale is almost zero, which corresponds to very high energy. Strictly speaking, we need to consider all the modes and then average over them, but it is difficult in practice. In this paper, we use a topological length scale $\ell_{\textrm{top}}$ to effectively describe the average of the length scales $\sqrt{ 1/(2 n -1)}\,\ell_0$ of all the topological modes. Though the precise relation between $\ell_{\textrm{top}}$ and $\ell_0$ is difficult to obtain, we know that
\be
\ell_{\textrm{top}} \, \leq \, \ell_0\, .
\ee

\section{\label{reSpinChain}An Example of Spin Chain}

As an example of separating the slowly varying modes and the rapidly varying modes, in this appendix we summarize the effective action of 1-dimensional quantum antiferromagnets. The discussion follows Ref.~\cite{Fradkin}, and is parallel to the main topic of this paper.

Consider a spin chain with $N$ ($N$: an even integer) sites with spin-$s$ degrees of freedom on each site. The real time action for the antiferromagnet is
\be
  S[\vec{n}] = s \sum_{j=1}^N S_{WZ} [\vec{n} (j)] - \int_0^T dx_0\, \sum_{j=1}^N J s^2 \vec{n} (j, x_0) \cdot \vec{n} (j+1, x_0)\, ,
\ee
where we assume the periodic boundary condition on the spin chain, and $S_{WZ}$ is the Wess-Zumino term, while $x_0$ and $J$ denote the time variable and the coupling constant. Making the replacement
\be
  \vec{n} (j) \, \to \, (-1)^j \vec{n} (j)\, ,
\ee
up to an additive constant one obtains
\be
  S[\vec{n}] = s \sum_{j=1}^N (-1)^j S_{WZ} [\vec{n} (j)] - \frac{J s^2}{2} \int_0^T dx_0\, \sum_{j=1}^N \left(\vec{n} (j, x_0) - \vec{n} (j+1, x_0) \right)^2\, .
\ee
Next, we split the spin field $\vec{n}$ into a slowly varying mode $\vec{m} (j)$ and a rapidly varying mode $\vec{l} (j)$:
\be
  \vec{n} (j) = \vec{m} (j) + (-1)^j a_0 \, \vec{l} (j)\, .
\ee
The constraints $\vec{n}^2 = \vec{m}^2 = 1$ leads to
\be
  \vec{m} \cdot \vec{l} = 0\, .
\ee
In the continuum limit, one finds that
\be
  \lim_{a_0 \to 0} s \sum_{j=1}^N (-1)^j S_{WZ} [\vec{n} (j)] \approx \frac{s}{2} \int d^2 x\, \vec{m} \cdot (\partial_0 \vec{m} \times \partial_1 \vec{m}) + s \int d^2 x\, \vec{l} \cdot (\vec{m} \times \partial_0 \vec{m}) \, ,
\ee
\be
  \lim_{a_0 \to 0} \frac{J s^2}{2} \int_0^T dx_0\, \sum_{j=1}^N \left(\vec{n} (j, x_0) - \vec{n} (j+1, x_0) \right)^2 \approx \frac{a_0 J s^2}{2} \int d^2 x\, \left((\partial_1 \vec{m})^2 + 4 \vec{l}^2 \right)\, ,
\ee
where $a_0$ is the lattice spacing. Hence, the effective Lagrangian becomes
\be
  \mathcal{L} (\vec{m}, \vec{l}) = -2 a_0 J s^2 \vec{l}^2 + s \vec{l} \cdot (\vec{m} \times \partial_0 \vec{m}) - \frac{a_0 J s^2}{2} (\partial_1 \vec{m})^2 + \frac{s}{2} \vec{m} \cdot (\partial_0 \vec{m} \times \partial_1 \vec{m})\, .
\ee
Integrating out the rapidly varying mode $\vec{l}$, we obtain an effective Lagrangian for the slowly varying mode $\vec{m}$:
\be
  \mathcal{L} (\vec{m}) = \frac{1}{2g} \left(\frac{1}{v_s} (\partial_0 \vec{m})^2 - v_s (\partial_1 \vec{m})^2 \right) + \frac{\theta}{8\pi} \epsilon_{\mu\nu} \vec{m} \cdot (\partial_\mu \vec{m} \times \partial_\nu \vec{m})\, ,
\ee
where
\be
  g \equiv \frac{2}{s}\, ,\quad v_s \equiv 2 a_0 J s\, ,\quad \theta \equiv 2 \pi s\, .
\ee

\section{Feynman's Path Integral}\label{FeynmanPathInt}

We discussed in Subsection~\ref{remarks} the difference between classical solutions, topological fluctuations and quantum fluctuations, and how they contribute to the path integral. As a concrete example, in this appendix we review the famous work by R.~Feynman on the path integral formulation of 1-dimensional non-relativistic quantum mechanics. We follow closely the original paper \cite{FeynmanQM} by R.~Feynman, and then discuss this formalism from some modern point of view.

For an initial state $\psi_{t'}$ at time $t'$ and a final state $\chi_{t''}$ at time $t'' > t'$ the transition amplitude is
\be\la{eq:randomwalk1}
  \langle \chi_{t''} | 1 | \psi_{t'} \rangle = \lim_{\epsilon \to 0} \int \cdots \int \frac{dx_0}{A} \cdots \frac{dx_{N-1}}{A} dx_N\, \chi^* (x'', t'')\, e^{i S / \hbar} \, \psi(x', t')\, ,
\ee
where $x_0 \equiv x'$, $x_N \equiv x''$, $N = (t'' - t')/ \epsilon$, and $A$ is a normalization constant. It can be justified that most of the contributions to Eq.~\eqref{eq:randomwalk1} comes from the Brownian motion (random walk) paths between $x_0$ and $x_N$. For a function $F$ of the coordinates $x_i$ with $t' < t_i < t''$, the transition element of $F$ between $\psi(x', t')$ and $\chi(x'', t'')$ is
\begin{align}
  \langle \chi_{t''} | F | \psi_{t'} \rangle & = \lim_{\epsilon \to 0} \int \cdots \int \frac{dx_0}{A} \cdots \frac{dx_{N-1}}{A} dx_N\, \chi^* (x'', t'') \, F(x_0, x_1, \cdots, x_N)\nonumber\\
  {} & \qquad\qquad\qquad\qquad\qquad\qquad\qquad \cdot \textrm{exp}\left[\frac{i}{\hbar} \sum_{i=0}^{N-1} S(x_{i+1}, x_i) \right]\, \psi(x', t')\, .
\end{align}
Therefore, we obtain after a partial integration
\be
  \bigg\langle \chi_{t''} \bigg| \frac{\partial F}{\partial x_k} \bigg| \psi_{t'} \bigg\rangle = -\frac{i}{\hbar} \bigg\langle \chi_{t''} \bigg| F \frac{\partial S}{\partial x_k} \bigg| \psi_{t'} \bigg\rangle\, ,
\ee
or formally, the equivalence for the operators
\be\label{eq:opequiv}
  -\frac{\hbar}{i} \frac{\partial F}{\partial x_k} \longleftrightarrow F \frac{\partial S}{\partial x_k}\, ,
\ee
where
\be\label{eq:intermediate-1}
  \frac{\partial S}{\partial x_k} = \frac{\partial S (x_{k+1}, x_k)}{\partial x_k} + \frac{\partial S (x_k, x_{k-1})}{\partial x_k}\, .
\ee

Now let us consider a particle of mass $m$ moving in one dimension under a potential $V(x)$ as an example. In this case,
\be
  S (x_{i+1}, x_i) = \epsilon\, L \left(\frac{x_{i+1} - x_i}{\epsilon}, x_{i+1} \right) = \frac{m \epsilon}{2} \left(\frac{x_{i+1} - x_i}{\epsilon} \right)^2 - \epsilon\, V(x_{i+1})\, .
\ee
Hence, for this example Eq.~\eqref{eq:intermediate-1} becomes
\begin{align}
  \frac{\partial S}{\partial x_k} & = \frac{\partial S (x_{k+1}, x_k)}{\partial x_k} + \frac{\partial S (x_k, x_{k-1})}{\partial x_k} \nonumber\\
  {} & = - m \frac{x_{k+1} - x_k}{\epsilon} + m \frac{x_k - x_{k-1}}{\epsilon} - \epsilon\, V'(x_k)\, .\label{eq:intermediate-2}
\end{align}
We should distinguish different cases for the operator $F$ in Eq.~\eqref{eq:opequiv}:
\begin{itemize}
  \item If $F$ does not depend on $x_k$, then Eq.~\eqref{eq:intermediate-2} simply becomes the Newton's law.

  \item If $F$ depends on $x_k$, for example $F = x_k$, then Eq.~\eqref{eq:opequiv} becomes
\be\label{eq:intermediate-3}
  \frac{\hbar}{i} \longleftrightarrow  m \left(\frac{x_{k+1} - x_k}{\epsilon} \right) x_k -  x_k m \left(\frac{x_k - x_{k-1}}{\epsilon} \right)\, ,
\ee
where we neglect terms of order $\epsilon$. In the language of time-ordered operators, the expression above is equivalent to
\be
  \bold{p} \, \bold{x} - \bold{x} \, \bold{p} = \frac{\hbar}{i}\, ,
\ee
which is the commutation relation between the operators $\bold{x}$ and $\bold{p}$ in the non-relativistic quantum mechanics. One can also shift $k \to k+1$ in the second term on the right-hand side of Eq.~\eqref{eq:intermediate-3}, which does not change the expression to the zeroth order in $\epsilon$, then Eq.~\eqref{eq:intermediate-3} reads
\be
  \left(\frac{x_{k+1} - x_k}{\epsilon} \right)^2 \longleftrightarrow  -\frac{\hbar}{i m \epsilon}\, .
\ee
From this expression one can learn that the ``velocity'' is of the order $\sqrt{\hbar / m \epsilon}$, which diverges as $\epsilon \to 0$. It implies that the paths are continuous but not differentiable.

\end{itemize}

Based on the discussions above, we can decompose the function space of all the possible paths in the following way:
\be
  V = V_{C^1} \oplus V_{\textrm{random}} \oplus \cdots\, ,
\ee
where $V_{C^1}$ stands for the space of all the $C^1$-differentiable paths, while $V_{\textrm{random}}$ denotes the space of all the Brownian motion (random walk) paths.  The space $V_{C^1}$ also includes a subspace consisting of the paths corresponding to all the classical solutions, which we call $V_{\textrm{sol}}$, i.e.
\be
  V_{C^1} \supset V_{\textrm{sol}}\, .
\ee
It can be proven \cite{math-1, math-2} that most continuous functions are nowhere differentiable. Therefore, it really suffices to capture the relevant quantum physics by considering only the continuous but nowhere differentiable paths, i.e. the paths from $V_{\textrm{random}}$. As shown in the original paper by R.~Feynman \cite{FeynmanQM} and also demonstrated above in this appendix, taking into account the paths from Brownian motion (random walk) indeed leads to the canonical quantization condition of quantum mechanics.

In the classical limit $\hbar \to 0$, all the quantum modes including all the paths in $V_{\textrm{random}}$ and $V_{C^1} \backslash V_{\textrm{sol}}$ are suppressed, hence the paths in $V_{\textrm{sol}}$ will give us the classical physics in this limit. On the other hand, when $\hbar$ cannot be neglected, $V_{\textrm{random}}$ dominates the configuration space, and the paths in $V_{\textrm{random}}$ provide the relevant physics in this case, which leads to the quantum mechanics. However, to probe the IR regime of a quantum theory a special limit turns out to be relevant, in which one suppresses most quantum modes while still keeps some lowest quantum modes, then both $V_{\textrm{sol}}$ and $V_{C^1} \backslash V_{\textrm{sol}}$ will be important.

\section{Fundamental Gap of the Gross-Pitaevskii Equation}\label{GP}

As we have seen in Section~\ref{LE} and \ref{MG}, the mass gap problem of the Yang-Mills theory can be mapped to the mass gap problem of a special kind of defocusing nonlinear Schr\"odinger equation or defocusing Gross-Pitaevskii equation. The mass gap problem of another related nonlinear Schr\"odinger equation was recently studied in Ref.~\cite{Singapore-2}. In this appendix, we summarize some results of Ref.~\cite{Singapore-2}, which are relevant to our discussions in Section~\ref{MG}.

In Ref.~\cite{Singapore-2}, the authors considered the following dimensionless nonlinear Schr\"odinger equation in $D$-dimensions ($D = 1,\, 2,\, 3$):
\be\label{eq:SingaporeEq}
  \left[-\frac{1}{2} \Delta + V(x) + \beta |\phi (x)|^{2 \sigma} \right] \phi(x) = \mu\, \phi(x)\, ,\quad x \in \Omega \subset \mathbb{R}^D\, .
\ee
The eigenvalue $\mu$ is given by
\be
  \mu (\phi) = \widetilde{E} (\phi) + \frac{\sigma \beta}{\sigma + 1} \int_\Omega dx\, |\phi(x)|^{2 \sigma + 2}\, ,
\ee
where $\widetilde{E} (\phi)$ is defined as
\be
  \widetilde{E} (\phi) = \int_\Omega dx\, \left[\frac{1}{2} |\nabla \phi(x)|^2 + V(x) |\phi(x)|^2 + \frac{\beta}{\sigma + 1} |\phi(x)|^{2\sigma + 2} \right]\, .
\ee
Ref.~\cite{Singapore-2} has discussed different boundary conditions, including
\begin{itemize}
  \item Periodic boundary condition:
\be\la{eq:periodicboundarycondition}
  \phi (x) \textrm{ is periodic on } \partial \Omega\, .
\ee

  \item Dirichlet boundary condition:
\be
  \phi (x) \big|_{\partial \Omega} = 0\, .
\ee

  \item Homogeneous Neumann boundary condition:
\be
  \partial_{\bf n} \phi \big|_{\partial \Omega} = 0\, .
\ee

\end{itemize}
Because in this paper we are mostly interested in the mass gap problem, we would like to focus on the trivial vacuum background
\be
  \psi (|x| = 0) = \psi (|x| = \infty) = \pm 1\, ,
\ee
where the relation between $\psi$ and $\phi$ will be given later, and this boundary condition for $\psi$  corresponds to the periodic boundary condition for $\phi$ (see Eq.~\eqref{eq:periodicboundarycondition}).

The authors of Ref.~\cite{Singapore-2} have studied for various boundary conditions the existence of the fundamental gap in a finite system, i.e. the energy difference between the first excited state and the ground state. To illustrate different boundary conditions, we will present the analytical results of the 1D case in Appendix~\ref{Toy}. For now, let us focus on the periodic boundary condition and recall a theorem from Ref.~\cite{Singapore-2}.
\begin{flushleft}
  {\bf Theorem} For the following Gross-Pitaevskii equation:
\end{flushleft}
\be
  \left[-\frac{1}{2} \Delta + V(x) + \beta |\phi(x)|^2 \right] \phi(x) = \mu\, \phi(x)\, ,\quad x\in \Omega\, ,\quad \phi(x) \textrm{ periodic on } \partial \Omega\, .
\ee
When $\Omega = \prod_{j=1}^D (0,\, L_j)$ ($D = 1,\, 2,\, 3$) satisfying $L_1 = \textrm{max} \{L_1,\, \cdots,\, L_D \}$ and $V(x) \equiv 0$ for $x \in \Omega$, the fundamental gaps $\delta_{\widetilde{E}} (\beta)$ and $\delta_\mu (\beta)$ are increasing functions for $\beta \geq 0$ and they have the following asymptotics:
\be
  \delta_{\widetilde{E}} (\beta) = \Bigg\{
  \begin{array}{ll}
    \frac{2 \pi^2}{L_1^2} + \frac{A_0^2}{4} \beta + o (\beta)\, , & 0 \leq \beta \ll 1\, ,\\
    \frac{8 A_0}{3 L_1} \beta^{1/2} + \frac{8}{L_1^2} + o (1)\, , \quad& \beta \gg 1\, ;
  \end{array}
\ee
\be
  \delta_\mu (\beta) = \Bigg\{
  \begin{array}{ll}
    \frac{2 \pi^2}{L_1^2} + \frac{A_0^2}{2} \beta + o (\beta)\, , & 0 \leq \beta \ll 1\, ,\\
    \frac{4 A_0}{L_1} \beta^{1/2} + \frac{8}{L_1^2} + o (1)\, , \quad& \beta \gg 1\, ,
  \end{array}
\ee
where $A_0$ is related to the volume of the system $V$:
\be
  A_0 = \frac{1}{\sqrt{\prod_{j=1}^D L_j}} = \frac{1}{\sqrt{V}}\, ,
\ee
and for $\beta \geq 0$
\be
  \delta_{\widetilde{E}} (\beta) \equiv \widetilde{E} (\phi_1^\beta) - \widetilde{E} (\phi_g^\beta) > 0\, ,\quad \delta_\mu (\beta) \equiv \mu (\phi_1^\beta) - \mu (\phi_g^\beta) > 0
\ee
are the energy differences between the first excited state and the ground state.

In particular, when $\beta \gg 1$ the first excited state is
\be\la{eq:1ksD}
  \phi_1^\beta \approx \sqrt{\frac{\mu_1}{\beta}}\, \left[1 + \textrm{tanh} \left(\sqrt{\mu_1} \left(\frac{L_1}{4} - x_1 \right) \right) - \textrm{tanh} \left(\sqrt{\mu_1} \left(\frac{3L_1}{4} - x_1 \right) \right) \right]\, ,
\ee
and
\begin{align}
  \mu_1 (\beta) & = A_0^2 \, \beta + \frac{4 A_0}{L_1} \beta^{1/2} + \frac{8}{L_1^2} + o(1)\, ,\\
  \widetilde{E}_1 (\beta) & = \frac{A_0^2}{2} \beta + \frac{8 A_0}{3 L_1} \beta^{1/2} + \frac{8}{L_1^2} + o(1)\, .
\end{align}

{\flushleft $\square$}

Although Ref.~\cite{Singapore-2} has only discussed 1-, 2- and 3-dimensions. The same idea can be applied to 4-dimensions, and similar results hold.

In order to apply the results of Ref.~\cite{Singapore-2} to our case, we have to map Eq.~\eqref{eq:SingaporeEq} from Ref.~\cite{Singapore-2} with $V(x) \equiv 0$ and $\sigma = 1$ to the equation that we obtained in Section~\ref{LE}:
\begin{align}
  {} & \quad \left(-\frac{1}{2} \Delta + \beta |\phi|^2 \right) \phi = \mu \phi \label{eq:SingaporeEq-2}\\
  \Leftrightarrow & \quad \Delta \phi - 2 \beta |\phi|^2 \phi + 2 \mu \phi = 0 \\
  \Leftrightarrow & \quad \Delta \psi - 2 m^2 (|\psi|^2 - 1) \psi = 0\, ,\label{eq:OurEq}
\end{align}
where in the last step we set
\be\label{eq:mapNLS}
  m^2 = \mu\, ,\quad \psi = \sqrt{\frac{\beta}{\mu}} \phi\, .
\ee
Hence, if one can find a solution to Eq.~\eqref{eq:SingaporeEq-2}, there is a corresponding solution to Eq.~\eqref{eq:OurEq}.

We also need to be careful about the boundary conditions. In fact, if we map Eq.~\eqref{eq:1ksD}, which is a solution to Eq.~\eqref{eq:SingaporeEq-2}, into a solution to Eq.~\eqref{eq:OurEq}, we find the boundary values of the new solution satisfy
\be
\psi\big|_{\textrm{Boundaries}} \approx 1 \, ,
\ee
where we used $m_D^2 \gg 1$ and the fact that $L_1$ is of the order of $1$ (see Eq.~\eqref{eq:rescaleSD}). This boundary condition is exactly the one that we used for the trivial vacuum background in Section~\ref{LE} and \ref{MG}. Hence, the results of Ref.~\cite{Singapore-2} are consistent with our results.

Also, for $\sigma=1$ the definitions of the energy $\widetilde{E}$ and the chemical potential $\mu$ from Ref.~\cite{Singapore-2} become
\begin{align}
  \widetilde{E} (\phi, \beta) & = \int_\Omega dx \left[\frac{1}{2} |\nabla \phi|^2 + \frac{\beta}{2} |\phi|^4 \right]\, ,\label{eq:SingaporeE}\\
  \mu (\phi, \beta) & = \widetilde{E}(\phi, \beta) + \frac{\beta}{2} \int_\Omega dx\, |\phi|^4\, .\label{eq:Singaporemu}
\end{align}
They obey the relation
\be
  \widetilde{E}(\phi, 2\beta) = \mu (\phi, \beta) \geq \widetilde{E} (\phi, \beta)\, .
\ee
In our case, the definition of the energy $E$ is
\be
  E = \int_\Omega dx \left[|\nabla \psi|^2 + \mu (|\psi|^2 - 1)^2 \right]\, .
\ee
After some steps, one can show that
\be\label{eq:NewE}
  E = \mu V - 2 \beta \left(1 - \frac{\widetilde{E}}{\mu} \right)\, .
\ee

\section{1D Nonlinear Schr\"odinger Equation as a Toy Model}\label{Toy}

We have discussed in Section~\ref{LE} and \ref{MG}, that the mass gap problem of the quantum Yang-Mills theory becomes a mass gap problem of a certain kind of defocusing nonlinear Schr\"odinger equation with a cubic interaction. For dimensions higher than one, the solutions do not have analytical expressions, which makes the problem hard to analyze. However, for the 1D case, the analytical solution is known. Although the 1D equation does not have an origin from the quantum Yang-Mills theory. As a toy model, it still sheds some light on the higher-dimensional cases. Hence, we analyze the 1D case analytically in this appendix, and we believe the 1D results should capture the qualitative features of the higher-dimensional cases of interest, especially the 3D and the 4D cases.

Let us recall the problem that we have seen in the main text. As we discussed in Subsection~\ref{sec:sec6remark}, for a general $D$-dimensional space up to some overall constant factors the effective action of the topological fluctuations can be brought into the expression:
\be\label{eq:app1Deffaction}
  S = \int d^D x\, \left[(\partial_\mu \Phi)^2 + m^2 (\Phi^2 - 1)^2 \right]\, ,
\ee
where $\Phi$ is a real function, and $m$ is a real parameter. In Euclidean space, the action itself can also be viewed as the energy of the configuration, which is non-negative definite. If we focus on the spherically symmetric configurations, the effective action above leads to the equation of motion:
\be
  \Delta \Phi - 2 m^2 \Phi (\Phi^2 - 1) = 0\, ,
\ee
where
\be
  \Delta \Phi = \partial_{|x|}^2 \Phi + \frac{D-1}{|x|} \, \partial_{|x|} \Phi\, .
\ee
For the 1D case the equation above simplifies to
\be\label{eq:app1DNLS}
  \partial_{|x|}^2 \Phi - 2 m^2 \Phi (\Phi^2 - 1) = 0\, ,
\ee
which has some trivial solutions
\be
  \Phi = \pm 1\quad \textrm{or}\quad \Phi = 0\, .
\ee
Among them $\Phi = \pm 1$ have zero energy, and they correspond to the pure gauge solution and the trivial vacuum solution, while the energy of $\Phi = 0$ is proportional to the volume of the system, i.e. divergent before regularization, hence it does not correspond to the vacuum. As the ground state of the system, $\Phi = \pm 1$ are more preferable at the low energy.

Now let us discuss the boundary condition for Eq.~\eqref{eq:app1DNLS}. At the boundaries ($x = 0, \, \infty$), $\Phi$ should return to the constant values corresponding to the classical solutions discussed above. Therefore, there are 5 possible boundary conditions for the spherically symmetric configurations:
\begin{enumerate}
\item
\be\label{eq:1DNLSbc-1}
  \Phi (|x| = 0) = 1\, ,\quad \Phi (|x| = \infty) = 1\, ;
\ee

\item
\be\label{eq:1DNLSbc-2}
  \Phi (|x| = 0) = -1\, ,\quad \Phi (|x| = \infty) = -1\, ;
\ee

\item
\be\label{eq:1DNLSbc-3}
  \Phi (|x| = 0) = 0\, ,\quad \Phi (|x| = \infty) = 0\, ;
\ee

\item
\be\label{eq:1DNLSbc-4}
  \Phi (|x| = 0) = -1\, ,\quad \Phi (|x| = \infty) = 1\, ;
\ee

\item
\be\label{eq:1DNLSbc-5}
  \Phi (|x| = 0) = 1\, ,\quad \Phi (|x| = \infty) = -1\, .
\ee
\end{enumerate}

Here we want to emphasize that, if we use a complex field $\Psi$ instead and rewrite Eq.~\eqref{eq:app1Deffaction} as
\be 
  S = \int d^D x\, \left[ \, \big| \partial_\mu \Psi\big| ^2 + m^2 \left(\big| \Psi\big|^2 - 1\right)^2 \,\right]\, ,
\ee
the solutions corresponding to the last two boundary conditions  do not show up because the new effective action for $\Psi$ has a rigid $U(1)$ symmetry, which makes the vacua $\Psi = \pm 1$ equivalent.

These boundary conditions also agree with the ones discussed in Ref.~\cite{Singapore-2}. The first one and the second one correspond to the periodic boundary condition. The third one corresponds to the Dirichlet boundary condition, while the fourth one and the fifth one correspond to the homogeneous Neumann boundary condition.

In the following, let us analyze the low-energy spectrum, i.e. the solutions to Eq.~\eqref{eq:app1DNLS} with lowest energies, under different boundary conditions \eqref{eq:1DNLSbc-1} $\sim$ \eqref{eq:1DNLSbc-5}.
\begin{itemize}
\item $\Phi (|x| = 0) = 1 \quad \textrm{and} \quad \Phi (|x| = \infty) = 1$:

For this boundary condition, the lowest energy state, i.e. the ground state, is just the one of the trivial vacua given by
\be
  \Phi = 1\, ,
\ee
which has zero energy. The first excited state is composed of an anti-kink kink pair, which has the expression:
\be
  \Phi =   \textrm{tanh} \left[ m \left(   a - |x| \right) \right] + \textrm{tanh} \left[ m \left(|x| - b\right) \right] + 1\, ,
\ee
where $a \gg m^{-1}$ and $b-a \gg m^{-1}$. The ground state is displayed in Fig.~\ref{app:figSol-1}~(left), and the first excited state is in Fig.~\ref{app:figSol-1}~(right).

\begin{figure}[!htb]
\hfill
 \minipage{141.5mm}
\minipage{0.49 \linewidth}
\begin{center}
\includegraphics[height=36mm]{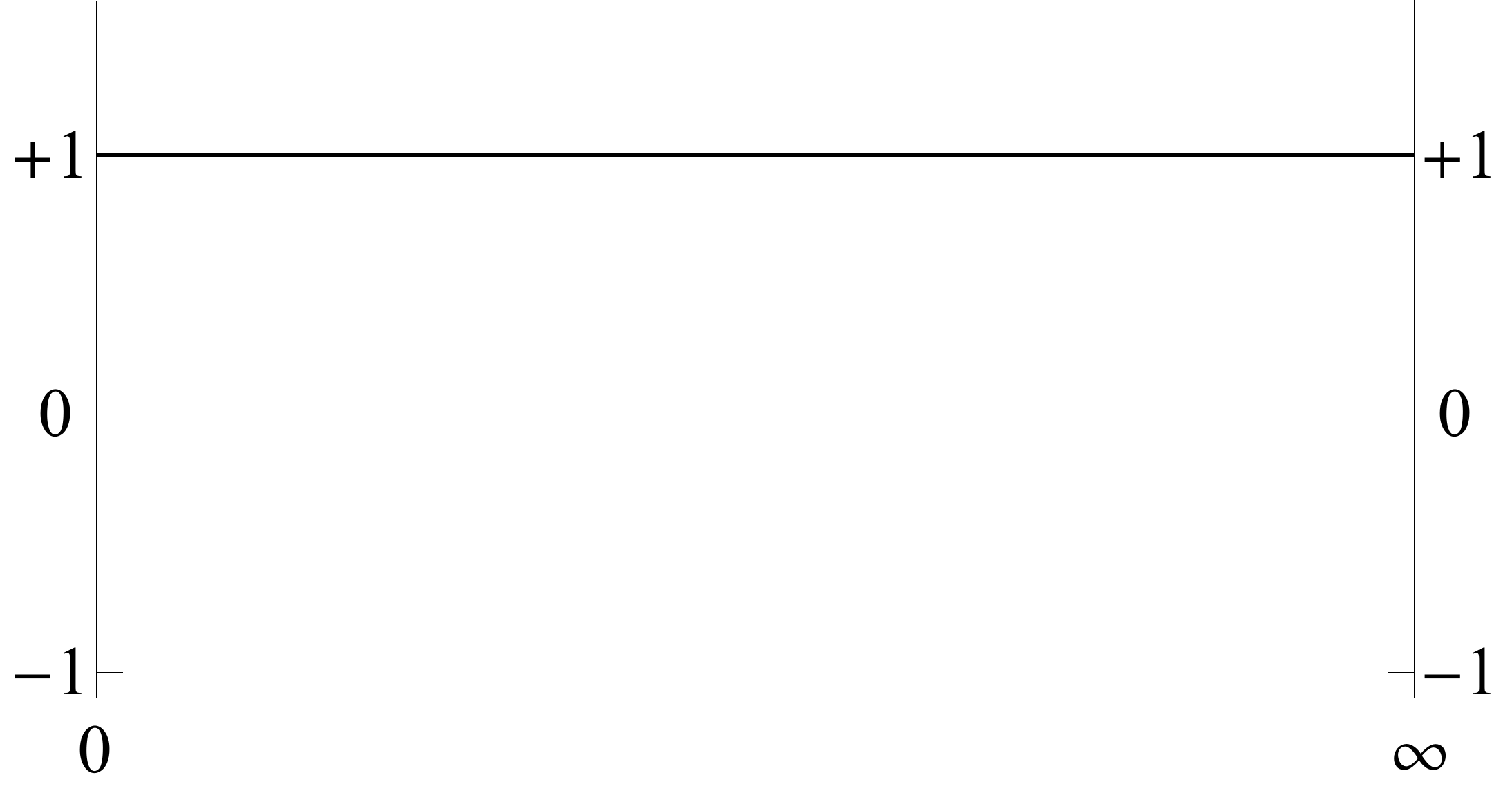}
\end{center}
\endminipage
\minipage{0.02\linewidth}
\ \
\endminipage
\minipage{0.49 \linewidth}
\begin{center}
\includegraphics[height=36mm]{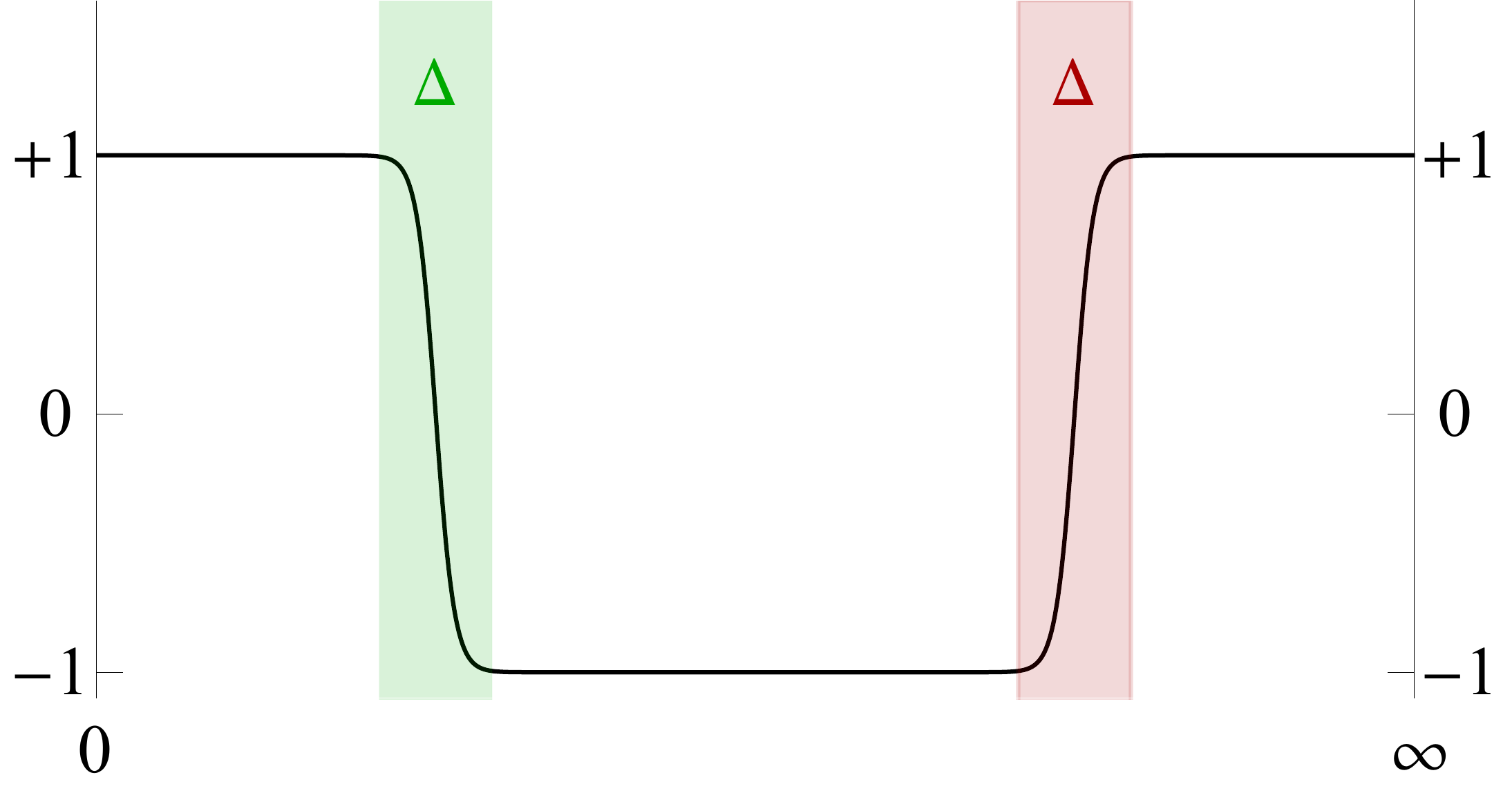}
\end{center}
\endminipage
  \caption{The 1D exact solutions with $\Phi (|x| = 0) = \Phi (|x| = \infty) = 1$. Left:~The ground state. Right:~The first excited state. }\label{app:figSol-1}
\endminipage
\end{figure}

The energy of the first excited state is twice of the energy of a kink or an anti-kink, i.e. $2 \Delta$, where $\Delta$ is defined by
\be
 \Delta = \int_{-\infty}^\infty dx\, \Big[(\partial_{x} \Phi)^2 + m^2 (\Phi^2 - 1)^2 \Big]_{\Phi = \textrm{tanh} (m x)} = \frac{8 m}{3}  \, .
\ee
One can also consider higher excited states, which correspond to more anti-kink kink pairs, and their energies are just multiples of the energy of an anti-kink kink pair. The spectrum with this boundary condition includes $\{0,\, 2 \Delta,\, 4 \Delta,\, \cdots \}$.

\item $\Phi (|x| = 0) = -1 \quad \textrm{and} \quad \Phi (|x| = \infty) = -1$:

For this boundary condition, the lowest energy state, i.e. the ground state, is another trivial vacuum given by
\be
  \Phi = -1\, ,
\ee
which has zero energy. The first excited state is composed of a kink anti-kink pair, which has the expression:
\be
  \Phi = \textrm{tanh} \left[ m \left( |x| - a \right) \right] + \textrm{tanh} \left[ m \left(  b - |x| \right) \right]- 1\, ,
\ee
where $a \gg m^{-1}$ and $b-a \gg m^{-1}$. The ground state is displayed in Fig.~\ref{app:figSol-2}~(left), and  the first excited state is in Fig.~\ref{app:figSol-2}~(right).

\begin{figure}[!htb]
\hfill
 \minipage{141.5mm}
\minipage{0.49 \linewidth}
\begin{center}
\includegraphics[height=36mm]{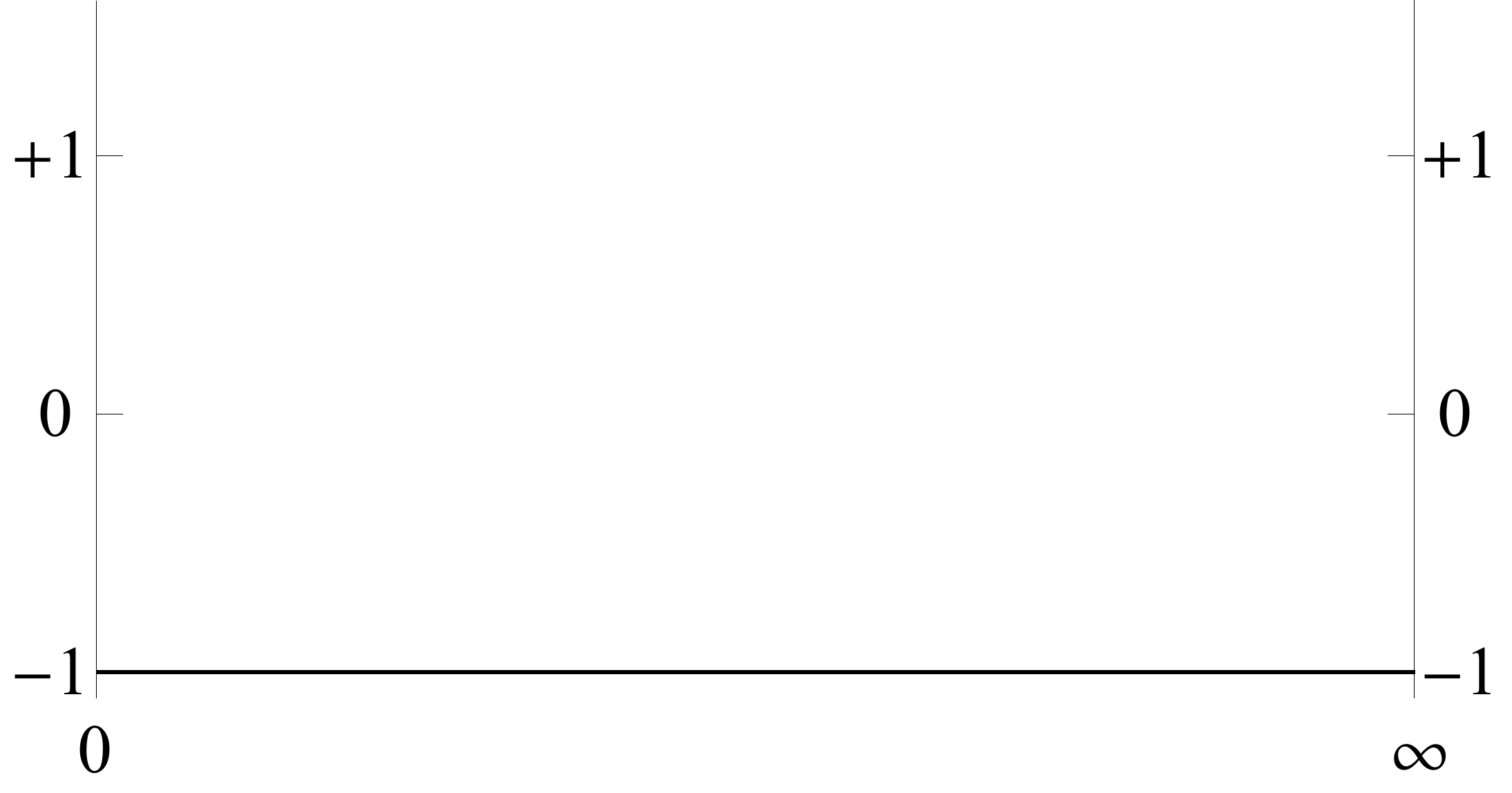}
\end{center}
\endminipage
\minipage{0.02\linewidth}
\ \
\endminipage
\minipage{0.49 \linewidth}
\begin{center}
\includegraphics[height=36mm]{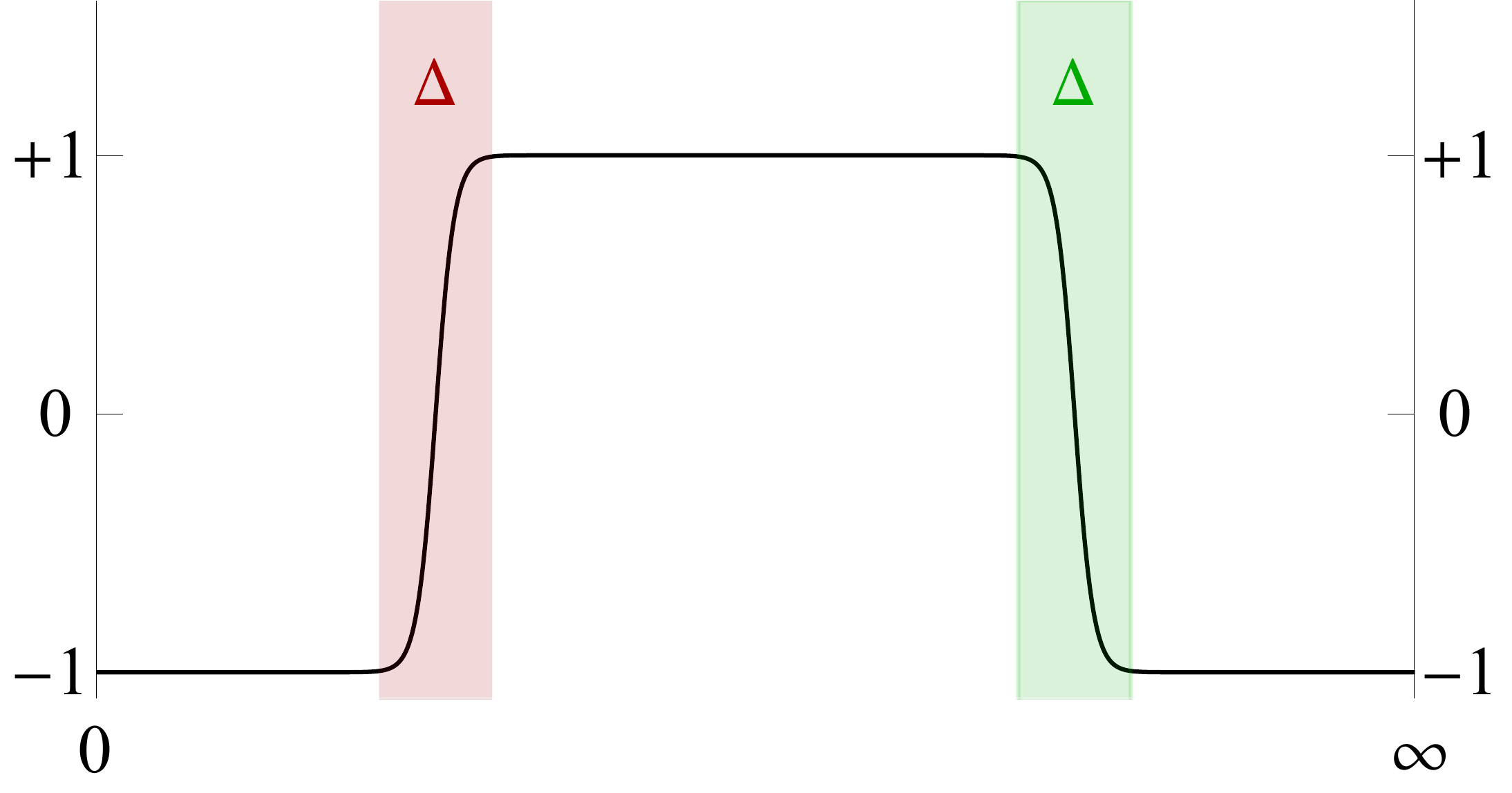}
\end{center}
\endminipage
  \caption{The 1D exact solutions with $\Phi (|x| = 0) = \Phi (|x| = \infty) = -1$. Left:~The ground state. Right:~The first excited state. }\label{app:figSol-2}
\endminipage
\end{figure}

The energy of the first excited state is $2 \Delta$. One can also consider higher excited states, which correspond to more kink anti-kink pairs, and their energies are just multiples of the energy of a kink anti-kink pair. The spectrum with this boundary condition includes $\{0,\, 2 \Delta,\, 4 \Delta,\, \cdots \}$.

\item $\Phi (|x| = 0) = 0 \quad \textrm{and} \quad \Phi (|x| = \infty) = 0$:

For this boundary condition, as we discussed before, the solution $\Phi = 0$ has divergent energy before regularization. Instead, the lowest energy state with this boundary condition is given by
\be
  \Phi = \pm \left\{\textrm{tanh} \left[ m |x| \right] +  \textrm{tanh} \left[ m \left(  |x|_\infty - |x| \right) \right] - 1 \right\} \, ,
\ee
where $|x|_\infty \to \infty$. These configurations corresponds to a half kink at $|x| = 0$ and a half anti-kink at $|x| = \infty$ or vice versa. There are also excited states, which correspond to inserting more kinks or anti-kinks between $|x| = 0$ and $\infty$. One of the lowest energy states is displayed in Fig.~\ref{app:figSol-3}~(left), and one of the first excited states is displayed  in Fig.~\ref{app:figSol-3}~(right).

\begin{figure}[!htb]
\hfill
 \minipage{141.5mm}
\minipage{0.49 \linewidth}
\begin{center}
\includegraphics[height=36mm]{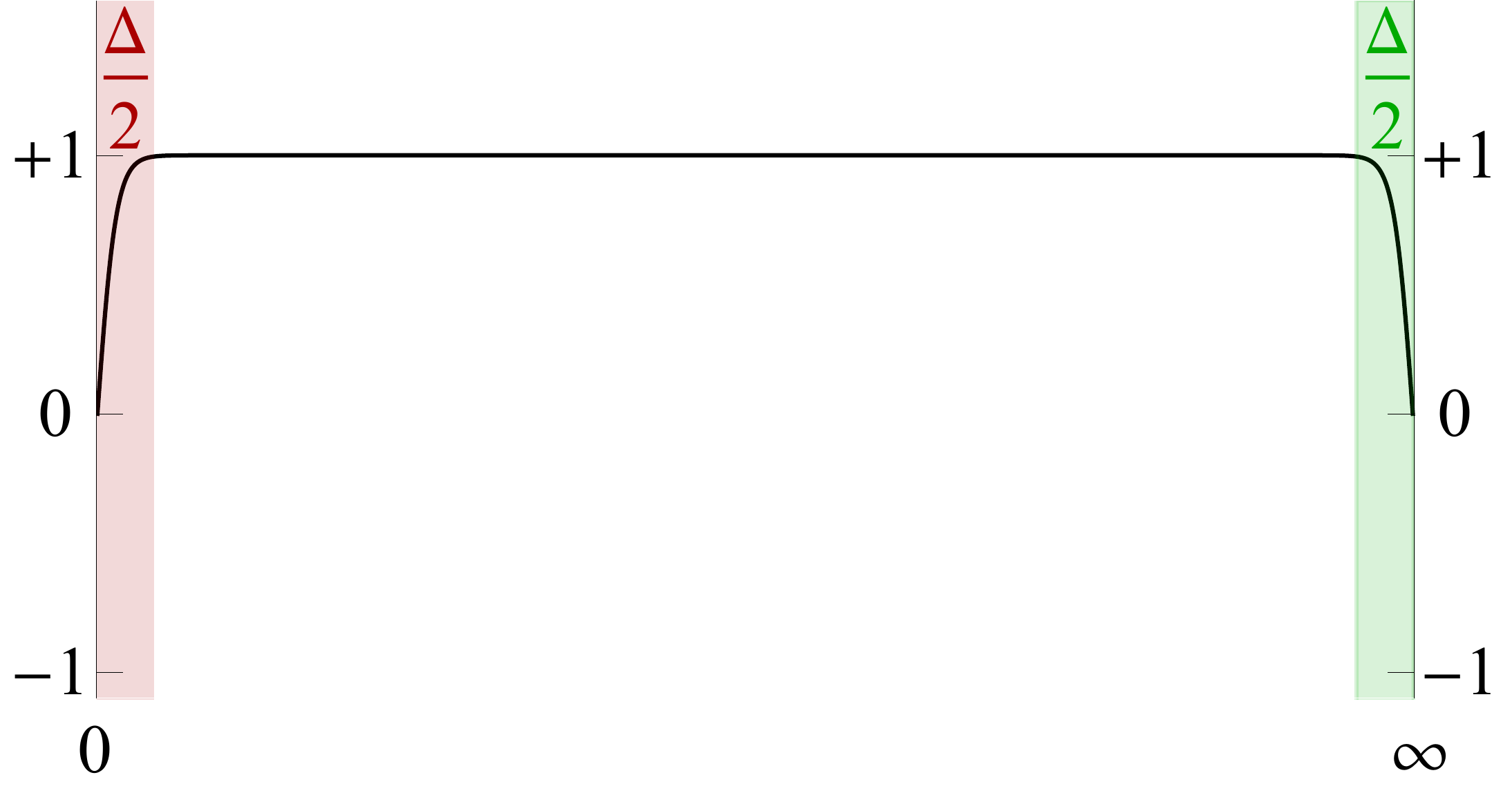}
\end{center}
\endminipage
\minipage{0.02\linewidth}
\ \
\endminipage
\minipage{0.49 \linewidth}
\begin{center}
\includegraphics[height=36mm]{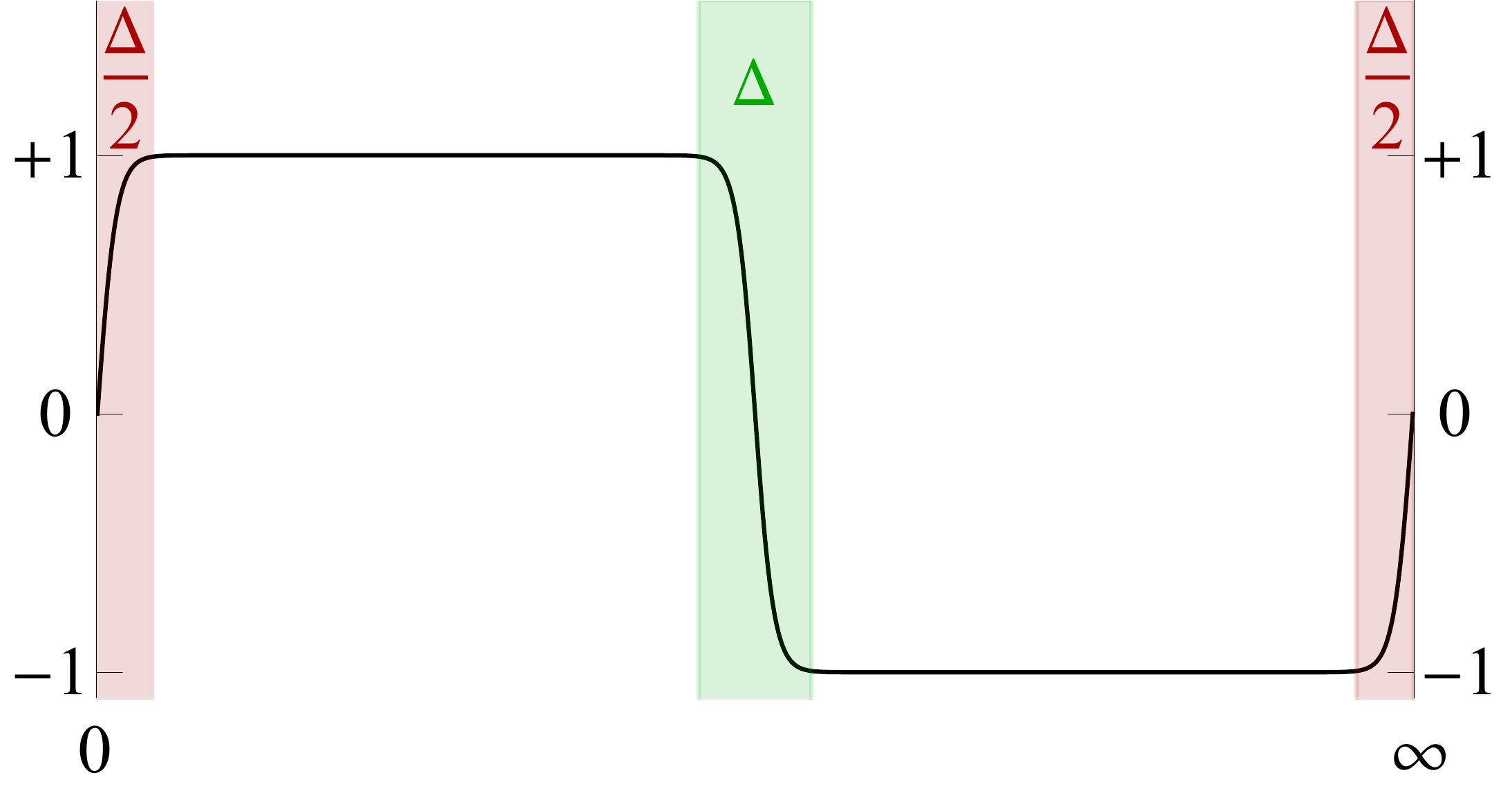}
\end{center}
\endminipage
  \caption{The 1D exact solutions with $\Phi (|x| = 0) = \Phi (|x| = \infty) = 0$. Left:~One of the lowest energy states. Right:~One of the first excited states.}\label{app:figSol-3}
\endminipage
\end{figure}

The energy of the lowest energy state with this boundary condition is the same as a kink or an anti-kink, i.e. $\Delta$. Hence, the spectrum with this boundary condition includes $\{\Delta,\, 2 \Delta,\, 3 \Delta,\, \cdots \}$.

\item $\Phi (|x| = 0) = -1 \quad \textrm{and} \quad \Phi (|x| = \infty) = 1$:

For this boundary condition, the lowest energy state is just a kink solution given by
\be
  \Phi = \textrm{tanh} \left[ m \left( |x| - a \right)  \right] \, ,
\ee
where $a \gg m^{-1}$. The excited states correspond to inserting more kink anti-kink or anti-kink kink pairs. The lowest energy state is displayed in Fig.~\ref{app:figSol-4}~(left), and  the first excited states is displayed in Fig.~\ref{app:figSol-4}~(right).

\begin{figure}[!htb]
\hfill
 \minipage{141.5mm}
\minipage{0.49 \linewidth}
\begin{center}
\includegraphics[height=36mm]{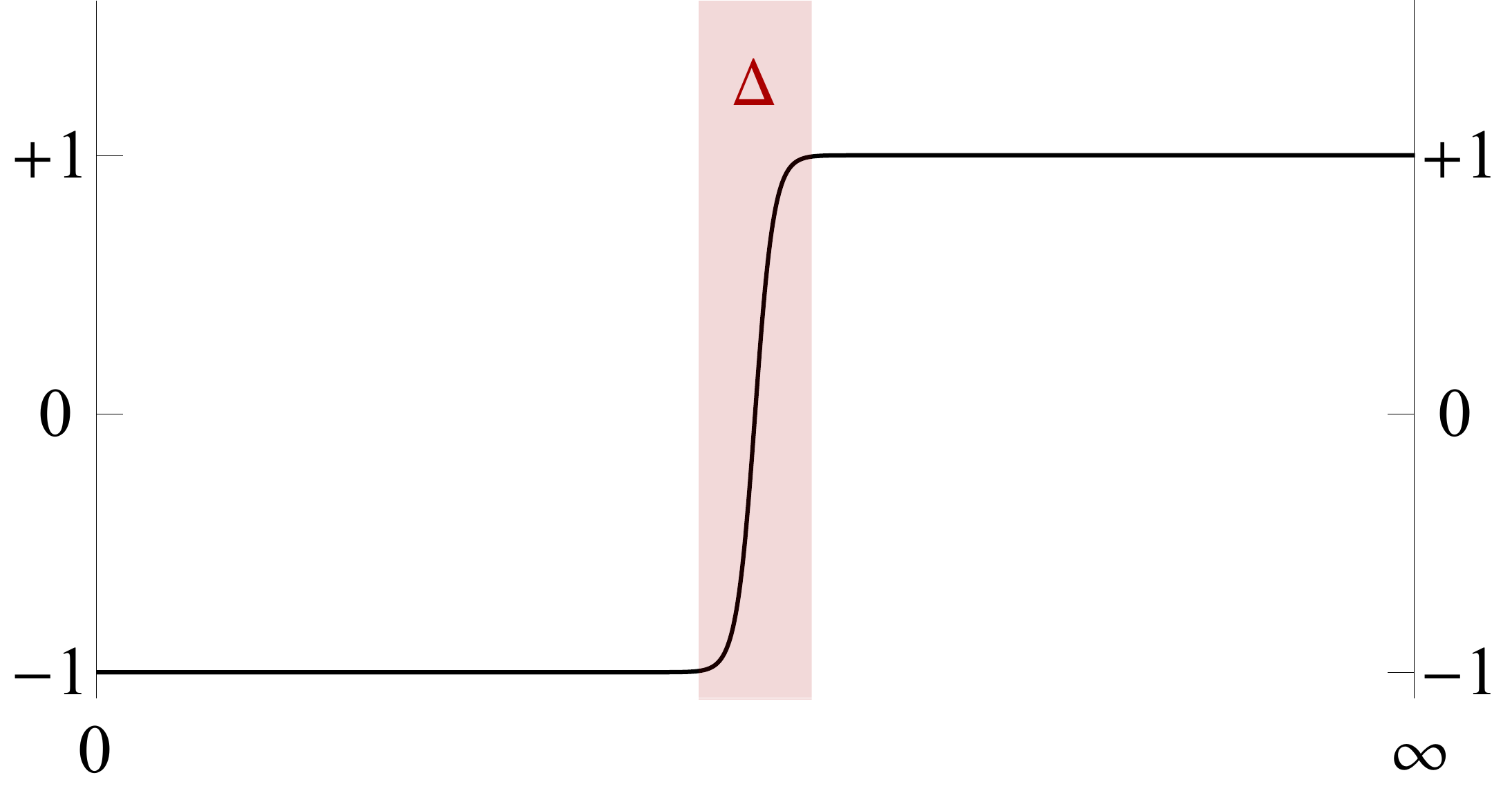}
\end{center}
\endminipage
\minipage{0.02\linewidth}
\ \
\endminipage
\minipage{0.49 \linewidth}
\begin{center}
\includegraphics[height=36mm]{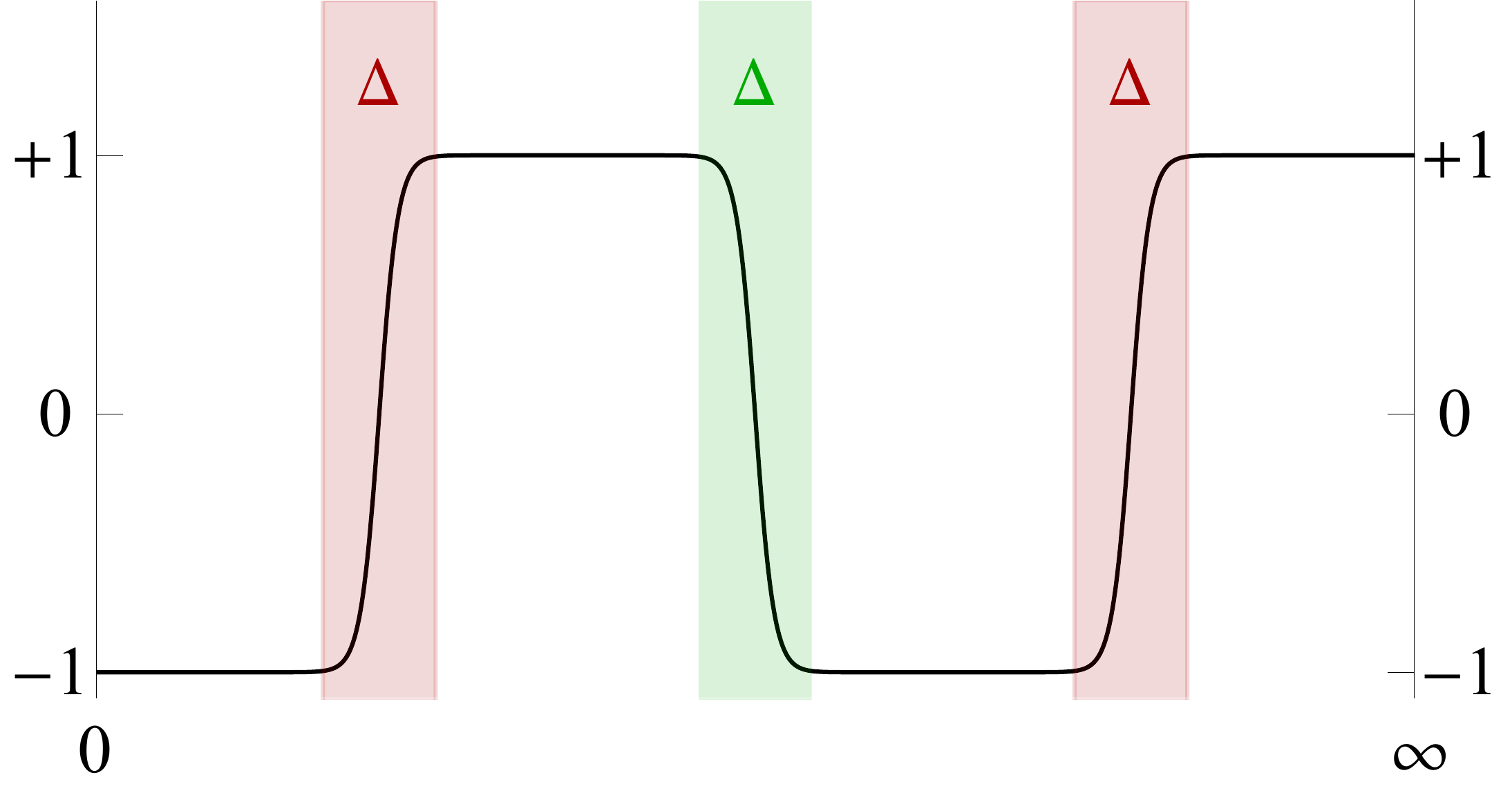}
\end{center}
\endminipage
  \caption{The 1D exact solutions with $\Phi (|x| = 0) = -1$ and $\Phi (|x| = \infty) = 1$. Left:~The lowest energy state. Right:~The first excited state.}\label{app:figSol-4}
\endminipage
\end{figure}

The energy of the lowest energy state with this boundary condition is $\Delta$. Hence, the spectrum with this boundary condition includes $\{\Delta,\, 3 \Delta,\, 5 \Delta,\, \cdots \}$.

\item $\Phi (|x| = 0) = 1 \quad \textrm{and} \quad \Phi (|x| = \infty) = -1$:

For this boundary condition, the lowest energy state is just a kink solution given by
\be
  \Phi =   \textrm{tanh} \left[ m \left( |x| - a \right)  \right] \, ,
\ee
where $a \gg m^{-1}$. The excited states correspond to inserting more kink anti-kink or anti-kink kink pairs. The lowest energy state is displayed in Fig.~\ref{app:figSol-5}~(left), and  the first excited states is displayed in Fig.~\ref{app:figSol-5}~(right).

\begin{figure}[!htb]
\hfill
 \minipage{141.5mm}
\minipage{0.49\linewidth}
\begin{center}
\includegraphics[height=36mm]{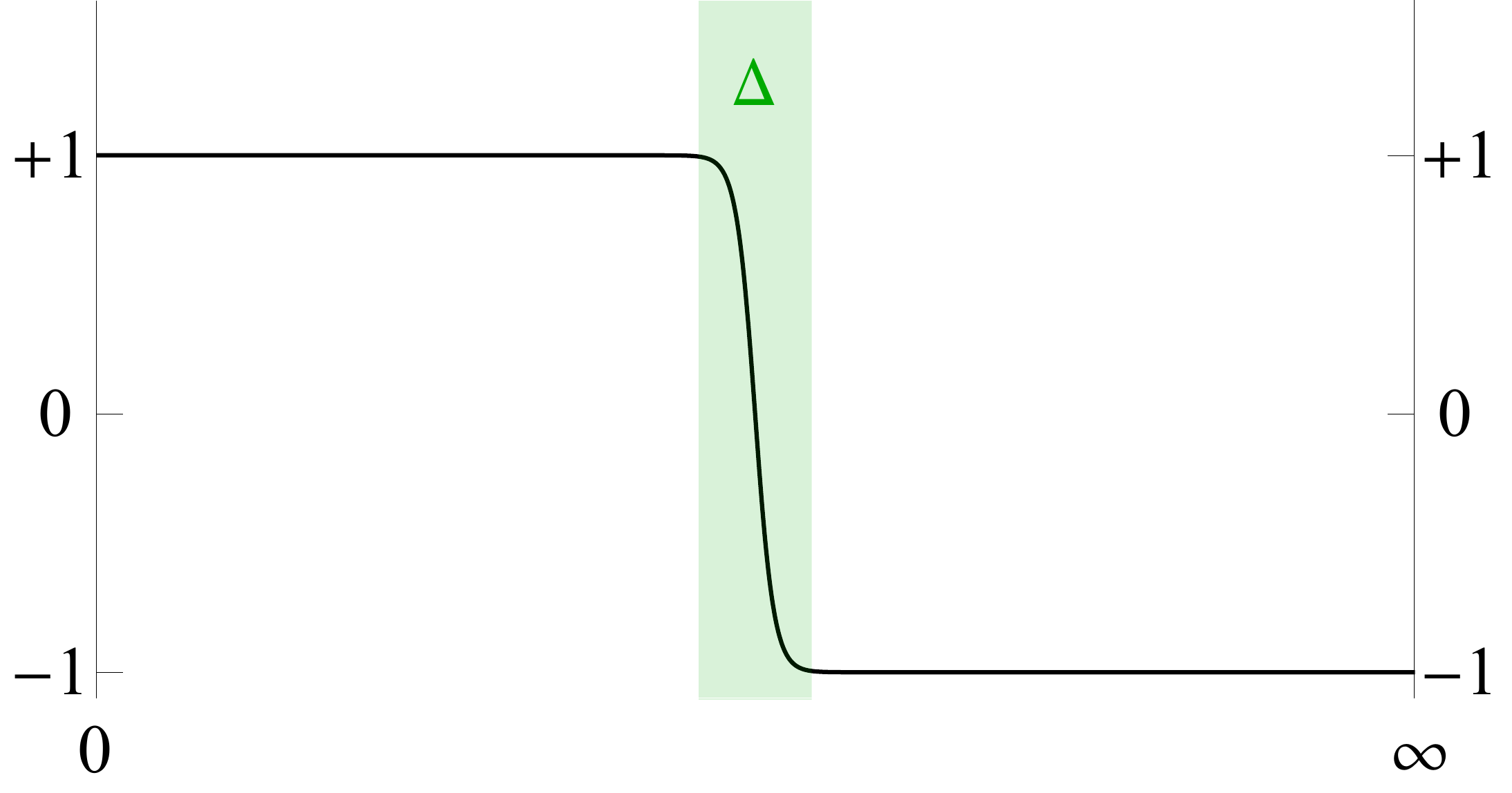}
\end{center}
\endminipage
\minipage{0.02\linewidth}
\ \
\endminipage
\minipage{0.49\linewidth}
\begin{center}
\includegraphics[height=36mm]{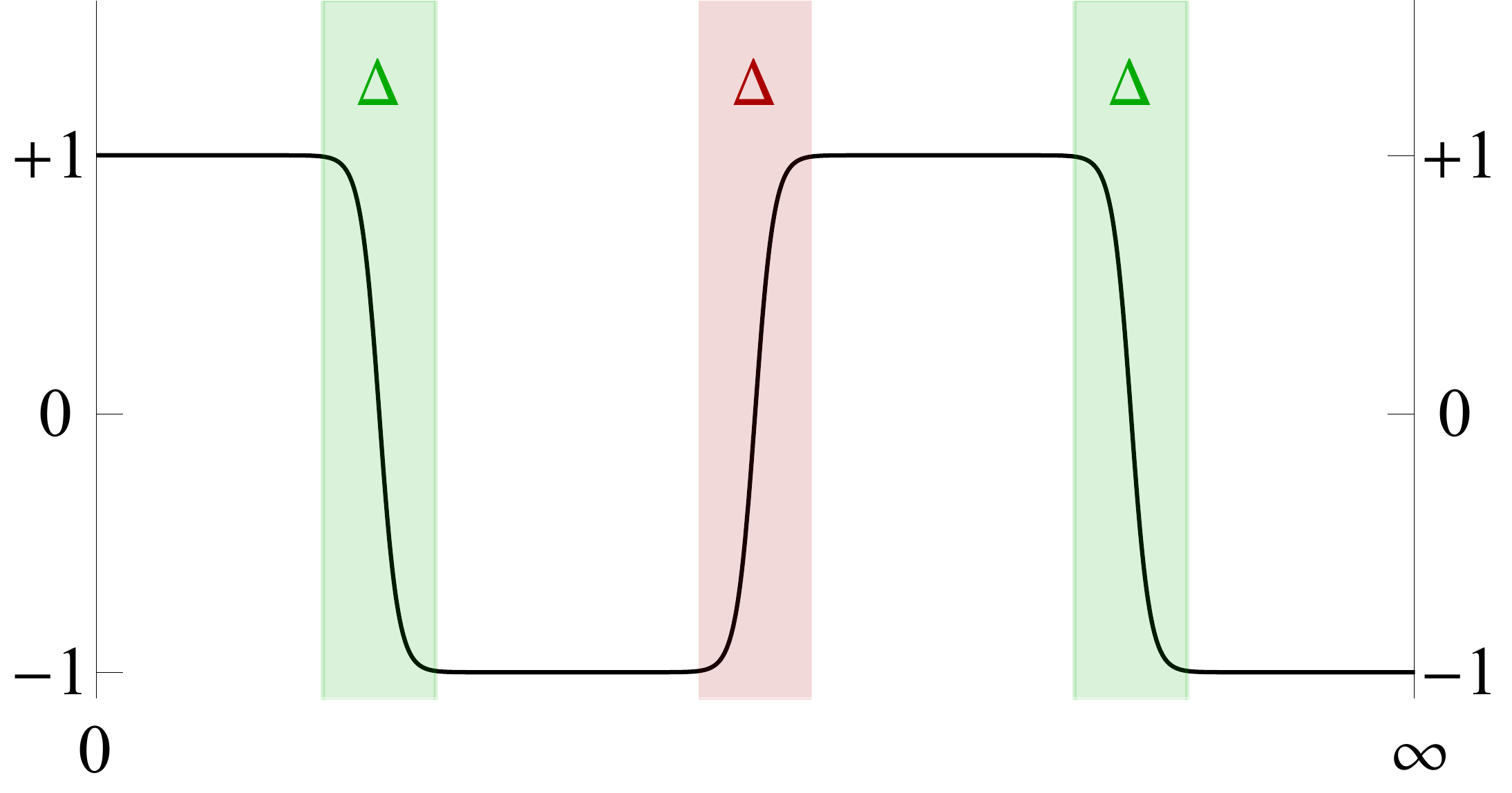}
\end{center}
\endminipage
  \caption{The 1D exact solutions with $\Phi (|x| = 0) = 1$ and $\Phi (|x| = \infty) = -1$. Left:~The lowest energy state. Right:~The first excited state.}\label{app:figSol-5}
\endminipage
\end{figure}

The energy of the lowest energy state with this boundary condition is $\Delta$. Hence, the spectrum with this boundary condition includes $\{\Delta,\, 3 \Delta,\, 5 \Delta,\, \cdots \}$.

\end{itemize}

The complete low-energy spectrum of the theory should be the union of the spectrum from different boundary conditions, which include

\begin{itemize}
\item For $\Phi (|x| = 0) = \Phi (|x| = \infty) = \pm 1$:
\begin{center}
  $0,\, 2\Delta,\, 4\Delta,\, \cdots$
\end{center}

\item For $\Phi (|x| = 0) = \Phi (|x| = \infty) = 0$:
\begin{center}
  $\Delta,\, 2\Delta,\, 3\Delta,\, \cdots$
\end{center}

\item For $\Phi (|x| = 0) = - \Phi (|x| = \infty) = \pm 1$:
\begin{center}
  $\Delta,\, 3\Delta,\, 5\Delta,\, \cdots$
\end{center}

\end{itemize}
From these results, we can conclude the existence of the mass gap at semi-classical level. In principle, we can even predict the energy eigenvalues of the excited states and their multiplicities. Of course, the analysis in this appendix works only for a special kind of 1D defocusing nonlinear Schr\"odinger equation with a cubic interaction, which can only be qualitatively true for higher dimensions.

Because in this paper we would like to study the existence of the mass gap, the trivial vacuum background is the most relevant background. For the 3D and the 4D case discussed in the text, we only focus on the trivial vacuum background, which corresponds to the first and the second boundary condition in this appendix for the 1D case.

\bibliographystyle{utphys}
\bibliography{YMEuclideanRef}

\end{document}